\documentclass[letters,a4paper,fleqn,usenatbib]{mnras}

\usepackage{newtxtext,newtxmath}

\usepackage[T1]{fontenc}
\usepackage{ae,aecompl}

\usepackage{graphicx}	
\usepackage{amsmath}	
\usepackage{hyperref}
\usepackage{pdflscape}
\usepackage{wasysym}
\usepackage{siunitx}
\usepackage{mathtools}
\usepackage{color}
\usepackage{gensymb}

\title[Taurid observations]{An observational synthesis of the Taurid meteor complex}

\author[A. Egal et al.]{
A. Egal,$^{1,2,3}$\thanks{E-mail: aegal@uwo.ca (AE)}
P. G. Brown,$^{1,2}$
P. Wiegert$^{1,2}$
Y. Kipreos$^{1,2}$
\\
$^{1}$Department of Physics and Astronomy, The University of Western Ontario, London, Ontario N6A 3K7, Canada\\
$^{2}$ Institute for Earth and Space Exploration (IESX), The University of Western Ontario, London, Ontario N6A 3K7, Canada\\
$^{3}$IMCCE, Observatoire de Paris, PSL Research University, CNRS, Sorbonne Universit\'{e}s, UPMC Univ. Paris 06, Univ. Lille, France\\
}

\date{Accepted 2022 February 10; Received 2022 February 9; in original form 2022 January 3}

\pubyear{2021}

\hypersetup{draft}

\begin{document}
\label{firstpage}
\pagerange{\pageref{firstpage}--\pageref{lastpage}}
\maketitle

\begin{abstract}
We provide an overview of the observational properties of the four major Taurid showers, namely the Northern and Southern Taurids (\#017 NTA and \#002 STA), the $\beta$ Taurids (\#173 BTA) and the $\zeta$ Perseids (\#172 ZPE). Analysing more than two decades of meteor observations from visual, optical and radar measurements we present the Taurids average activity, annual variations in strength, radiant drift and orbital variations as a function of solar longitude and particle size. 
The Taurid showers are detected over several weeks in the spring and autumn, but their annual activity level is generally low (less than 15 visual meteors per hour). We find the STA to be predominant in autumn, while its twin the ZPE dominates over the BTA in spring. Due to their long duration, the position of each shower's radiant and orbital elements are variable with time.  Optical measurements have previously recorded enhanced STA activity and increased fireball rates caused by the return of a swarm of meteoroids trapped in the 7:2 mean motion resonance with Jupiter. However, we find no presence of the swarm in radar data, suggesting that small meteoroids are removed from the resonance faster than fireball-producing meteoroids. We also find the STA to be enriched in smaller particles early in their activity period.
The differences we identify in our analysis between the showers at different particle sizes provide strong observational constraints to future dynamical modelling of the Taurid Meteoroid Complex. 

\end{abstract}




\section{Introduction}

The Taurid meteor shower is among the most enigmatic of all the major showers. It is related to a wide meteoroid complex, with a structure showing both North and South branches and approaching Earth's orbit before and after perihelion. These meteoroid branches are the source of the North and South Taurids of autumn and the daytime $\beta$ Taurids and $\zeta$ Perseid of the spring. The meteoroid orbits suggest that comet 2P/Encke is the immediate parent, a connection first proposed more than 80 years ago by \cite{Whipple1940}. 

The long duration of the showers, the broadness of the Taurid radiants, and the linkage of at least four showers \citep[and potentially many more such as the N. and S. $\chi$ Orionids, cf.][]{Jenniskens2016} together with the suggestion that the Taurids are responsible for a significant fraction of the broader Helion and Anti-Helion sporadic radiant sources \citep{Stohl1986} are all indicators that the stream complex is old. The detection of meter-sized Taurid meteoroids \citep{Spurny2017}, a unique feature among all the major showers, further hints at the complex origin and history of the stream. Finally, the shower shows periodic high fireball activity, linked to the 7:2 mean motion resonance (MMR) with Jupiter \citep{Asher1998}, suggestive of a denser 'swarm' of larger meteoroids centred at the 7:2 MMR (hereafter termed the {\it Taurid Swarm Complex} or TSC).

The formation and evolution of the Taurids has been a subject of intense debate. The linkage to comet 2P/Encke was originally proposed as a parent - sibling association \citep{Whipple1940,Whipple1952}. Later, it was suggested that 2P/Encke is simply the largest member of the Taurid complex \citep{Clube1984}, which was produced by the breakup of a larger progenitor comet some 20 ka ago. In this scenario, successive fragmentation/splitting events of the original progenitor contributed to create the {\it Taurid Meteoroid Complex} (TMC), the TSC, and large near-Earth asteroids within the complex. Over the past decades, an increasing number of Apollo-type asteroids have been proposed as the parent body of minor Taurid showers \citep[e.g.,][]{Madiedo2013}, of the TSC resonant swarm \citep[][]{Spurny2017,Olech2017,Devillepoix2021}, or linked to the asteroidal Taurid complex \citep[see review in][]{Egal2021}. 

The formation and evolution of the TMC has been examined in several numerical studies which followed the early analysis of \cite{Whipple1940} and \cite{Whipple1952}. Most of these models are based on the assumption that the Taurid meteoroids are injected into the stream from comet 2P/Encke, and attempt to reproduce the orbital elements and radiant distribution of the associated meteor showers \citep[e.g.,][]{Babadzhanov1990,Asher1991,Steel1991,Tomko2019}. However the duration and activity levels of the individual showers --- which provide important clues to the formation and evolution of the Taurid complex --- have not been used in any modelling efforts to date. 

A possible explanation for this deficiency relates to the challenges in observing even the major Taurid showers. Their low activity and the long duration make them difficult to distinguish from the sporadic background, in particular for visual observers. In addition, the daytime $\beta$ Taurids and $\zeta$ Perseids are difficult to observe with optical instruments, requiring dedicated radar instruments. Recently, \cite{Dewsnap2021} presented the first long-term analysis of the daytime Taurids recorded by the Canadian Meteor Orbit Radar (CMOR). 

This work aims to analyze, summarize and synthesize the characteristics of the four major Taurid showers, to serve as an observational basis for any future model of the Taurid meteoroid complex. We examine in particular the activity, duration, radiants and orbital elements of the Northern and Southern Taurids (\#017 NTA and \#002 STA), the $\beta$ Taurids (\#173 BTA) and the $\zeta$ Perseids (\#172 ZPE) as measured over the past two decades. When available, the observational properties of each shower are compared across visual, photographic, video and radar databases, as well as with previous published analyses. A special effort has been made to include an Appendix with a suite of figures (available as an online supplementary material) that can serve as direct measurements to validate future dynamical models of the complex. 
In the following section, we start by reviewing the different sources of observations of the Taurid showers available in the literature, and provide details about our analysis methodology. 

\section{Observations}

\subsection{Visual observations}

The identification of the N. and S. Taurid meteor showers (IAU designations: NTA and STA) presents a challenge for visual observers due to their low activity levels that make them difficult to distinguish from the sporadic background. As a consequence, long-term visual observations of Taurid activity are limited to a small number of literature sources. In addition, distinguishing between the broad radiants of the northern and southern branches is demanding for observers. As a result, many publications using visual data \citep[e.g.,][]{Roggemans1989,Dubietis2006} amalgamate the NTA and STA as a single shower.  Historic records suggest that the Taurids  may have been particularly active a millennium ago \citep{Ahn2003}. However, due to the shower timing in the fall months there is possible confusion between outbursts of NTA/STA and storms/outbursts from the Leonids or the Orionids, further complicating interpretation of the ancient visual record. 

The most comprehensive and uniform data set of visual observations of the NTA/STA showers is provided by the IMO Visual Meteor Data Base\footnote{\url{https://www.imo.net/members/imo_live_shower}, accessed in 2021 August} (VMDB). Activity profiles of the two showers are available in the VMDB dating back to 1988. Complementary observations of the Taurids have been collected by the Dutch Meteor Society \citep[DMS, cf.][]{Johannink2006} and the Nippon Meteor Society (NMS). In particular, the NMS have monitored  Taurid fireballs rates consistently since 1972. Their records include references to remarkable fireball activity dating back to 1934 \citep{Asher1998}. 

Additional activity profiles of the NTA and STA were also published for specific years, generally corresponding to the return of a swarm of Taurid meteoroids trapped in the 7:2 Mean Motion Resonance (MMR) with Jupiter (cf. Section \ref{sec:resonance}). This includes the 1988 \citep{Roggemans1989}, 1998 \citep{Arlt2000,Dubietis2007} and 2005 \citep{Dubietis2006,Dubietis2007,Miskotte2006,Johannink2006} apparitions of the showers.  A compilation of the activity profiles computed from visual observations of the Taurids during these resonant years is presented in Appendix A, Figures A.1.1 and A.1.2. The reported maximum activity time of these showers, as well as estimates of their total duration, are summarized in Table \ref{table:shower_times}. 

In this work, we have collected all the activity profiles of the NTA and STA available in the VMDB since 1988. The showers' intensity are characterized by the zenithal hourly rate (ZHR), which represents the number of meteors a single observer would observe with the radiant overhead per hour under standard reference conditions \citep{Rendtel2017}. The ZHR profile of each NTA and STA apparition since 2002 is provided in Appendix A2.

Such profiles provide valuable information about the general activity of the showers, including their duration, average intensity level and location of the broad maxima. However, the ZHR estimates were obtained using a fixed set of parameters over the whole activity period (e.g., population index, limiting sensitivity) that can affect the shape of the profiles computed, and the main maxima's location and strength. The profiles presented in Appendix A2 are therefore not well suited to search for shower details, such as the precise timing and intensity of the Taurids peak of activity.  
	
Obtaining a reliable, time-resolved ZHR profile for each NTA and STA apparition would require a careful analysis of the visual data, correcting  (among other things) for variations of the limiting magnitude (due to moonlight conditions), number and quality of individual reports, and variations of the population index over the activity period  \citep{Egal2020b}. Such analysis, complicated by the long duration and low meteor rates recorded for the NTA and STA, is beyond the scope of the present work. In our study, the visual activity profiles of the IMO VMDB are used to investigate the general characteristics of the nighttime showers but not to explore the fine structure of the stream (including the detection of the Taurids resonant swarm).

As the typical meteor limiting magnitude from visual data for the STA and NTA is in the range of +3 to +4, these observations are appropriate to meteoroid masses of >10$^{-5}$ kg, using the mass-magnitude-velocity relationship from \citet{Verniani1973}. This roughly corresponds to particles with diameters of order >2 mm.

\begin{table}
	\centering
	\begin{tabular}{ccrrrc}
		Shower & Year & $SL_{max}$ & $SL_b$ & $SL_e$ & source \\
		\hline
		& & & & & \\[-0.25cm]
		ZPE & -    & 76.7   & 70.7 & 86.7  & [C]\\
		& -    & 78     & 59   & 90    & [S]\\
		& 1950 &        & 70.6 & 85    & [As]\\
		& 1967 & 76.42  &      &       & [Pe]\\
		& 2000 & 79.8   & -    & -     & [P]\\
		& 2001 & 76.5   & -    & -     & [P]\\
		& 2002 & 77.2   & -    & -     & [P]\\
		& 2003 & 78.8   & -    & -     & [P]\\
		& 2003 & 76.81  &      &       & [Pe]\\
		&      & 78.81  &      &       & [Pe]\\
		& 2004 & 81.5   & -    & -     & [P]\\
		& 2005 & (81.2) & -    & -     & [P]\\
		& -    & 77.0   & 56   & 92    & [De]\\
		\hline
		& & & & & \\[-0.25cm]
		BTA & -    & 96.7   & 91.7 & 103.7 & [C]\\
		& -    & 95     & 81   & 104  & [S]\\
		& 1950 &        & 94.5 & 102  & [As]\\
		& 1997 & 94.7   & -    & -    & [P]\\
		& 1999 & 96.4   & -    & -    & [P]\\
		& 2003 & 94.1   & -    & -    & [P]\\
		& 2003 & 94.13  & -    & -    & [Pe]\\
		& 2004 & 95.9   & -    & -    & [P]\\
		& 2005 & (98.4) & -    & -    & [P]\\
		& -    & 95.0   & 85   & 103  & [De]\\
		
		\hline
		& & & & & \\[-0.25cm]
		STA & -    & 220.7  & 172.7 & 244.7 & [C]\\
		& 1988 & 219.5  & 172.7 & 235   & [R]\\
		& 1998 & 217.0  & 195   & 245   & [A]\\
		& 2003 & 221.1  & -     & -     & [Pe]\\
		&      & 223.08 & -     & -     & [Pe]\\
		
		\hline
		& & & & & \\[-0.25cm]
		NTA & -    & 230.7 & 176.7 & 249.7 & [C]\\
		& 1988 & 231.3 &       & 247.8 & [R]\\
		& 1998 & -     & 195   & 245   & [A]\\
		& 2003 & 231.9 & -     & -     & [Pe]\\
		\hline
		& & & & & \\[-0.25cm]
		NSTA & -    & 226  & 180   & 245   & [Du]\\
		& 2005 & 219.6 & 185   & 245   & [Du]\\ 
	\end{tabular}
	\caption{\label{table:shower_times}Timing of activity maximum ($SL_{max}$), activity beginning ($SL_b$) and end ($SL_e$, in solar longitude) from various literature sources for each of the four major Taurid streams in various years. Here we use the three letter IAU code for each shower: ZPE - $\zeta$ Perseids, BTA - $\beta$ Taurids, STA - S. Taurids, NTA - N. Taurids and NSTA (references which merge activity of both NTA/STA showers when reporting).  All solar longitudes are J2000. References are As: \protect\cite{Aspinall1951}, C: \protect\cite{Cook1973}, S: computed from dates in \protect\cite{Sekanina1973} for observation year 1965, R: \protect\cite{Roggemans1989}, Pe: \protect\cite{Pecina2004}, Du: \protect\cite{Dubietis2006}, P: \protect\cite{Porubcan2007}, De: \protect\cite{Dewsnap2021}, A: \protect\cite{Arlt2000}. Uncertain times of maximum are given in parenthesis; in these cases the real maximum was probably missed because of bad observing conditions. }
\end{table}

\subsection{Photographic and video observations}

An important source of Taurid observations are photographic and video records of the nighttime showers. From the analysis of a few photographic Taurids, \cite{Whipple1940} and later \cite{Whipple1952} identified comet 2P/Encke as the probable parent body of the NTA and STA. Since then, the radiant structure and orbital characteristics of the NTA and STA have been often investigated using the photographic meteor measurements contained in the IAU Meteor Data Center (MDC) database \citep[e.g.,][]{Steel1991,Porubcan2002,Svoren2011,Kanuchova2012,Kanuchova2014,Tomko2019}. 

The recent maturation of video networks dedicated to meteor observations has greatly improved systematic measurement of the activity of the nighttime Taurids. Surveys such as SonotaCo \citep{SonotaCo2009} or CAMS \citep[Cameras for Allsky Meteor Surveillance project, ][]{Jenniskens2016} release regular catalogues of meteoroids orbits, that are publicly available online. In our analysis, we used the last release of the CAMS catalogue (v3.0) to estimate the orbital elements distribution of the NTA and STA  (cf. Section \ref{sec:orbelts} and Appendix C).  To define the activity variations of these showers, we used the data provided by the IMO Video Meteor Network \citep{Molau2009}, which we have shown to be suitable for longer term analysis of shower rates in prior works \citep[e.g.,][]{Egal2020b}. 

The IMO Video Meteor Network (hereafter VMN)  is comprised of about 130 video cameras capable of recording meteors down to a magnitude of 3.0 $\pm$ 0.8. This limiting meteor magnitude corresponds to roughly 10$^{-4}$< m <10$^{-5}$ kg limiting masses and particles sizes of several mm. Reports of the VMN observations are regularly published in WGN, the Journal of the IMO, and a dedicated web interface\footnote{\url{https://meteorflux.org}} allows analysis of shower flux profiles of specific meteor showers recorded by the network since 2011.  The software allows different filters (time bin, limiting magnitude, population index) to be applied when computing the flux and equivalent ZHR profile of a given shower. 

In this study, we computed the activity of the NTA and STA observed by the VMN using a time bin of one day and a population index of 2.3 \citep{rendtel2014}. The selection of comparatively large time bins was chosen to limit the short-term fluctuations in the flux profiles due to the low number of Taurid meteors. The annual activity profile of each shower since 2011 using the VMN and compared with VMDB measurements is given in supplementary Appendix A2.

\subsection{Previous radar observations} \label{sec:radio_obs}

Radio observations provide essential measurements of the Taurid showers activity, in particular for the daytime BTA and ZPE that are difficult to detect with optical instruments due to the radiant's proximity to the sun.  The showers were first detected using specular backscattering radio techniques at the Jodrell Bank Experimental Station \citep{Aspinall1951,Lovell1954}. The semi-major axis, eccentricity and perihelion distances of the orbits measured were found to be very similar to the present-day orbital elements of comet 2P/Encke. Later, additional orbital data of the daytime Taurids were obtained as part of the Harvard Radio Meteor Project, which produced lower estimates of the streams' semi-major axis, and a reduced eccentricity for the ZPE \citep[][cf. Table \ref{tab:orbelts}]{Sekanina1976}. Radar observations performed in the southern Hemisphere in 1969  \citep{Gartrell1975} suggested that the ZPE are a probable continuation of the STA.

\cite{Steel1991} compiled and analysed the orbital correlations among the several hundred Taurid observations contained in the IAU Meteor Data Center. Meteors were identified as a Taurid member using a modified Southworth and Hawkins D-criterion \citep{Southworth1963} centered on a perihelion distance, eccentricity and inclination of 0.375 AU, 0.82 and 4$\degree$ respectively. The authors identified 170 photographic and 87 radar meteors of the nighttime Taurids with D$<$0.15, and only 56 orbits of the daytime showers. Their main conclusions are presented and discussed in Section \ref{sec:orbital_correlation}.

Radio measurements of the BTA and ZPE between 2000 and 2005 were also performed using the Ond\v{r}ejov backscatter meteor radar \citep{Pecina2004,Porubcan2007} and the  Budrio-Lecce-Modra (BLM) forward-scatter system \citep{Porubcan2007}. The time of maximum activity of the ZPE and BTA was found to vary from year-to-year by a few days (cf. Table \ref{table:shower_times}), which was interpreted as evidence for a filamentary structure of the streams. In addition, differences between the ZPE observed in 1967 and 2003 highlighted the possibility of a long-term variability of the shower's activity \citep{Pecina2004}.

\subsection{CMOR measurements}

\subsubsection{Activity}

Since 2002, the Canadian Meteor Orbit Radar (CMOR) has been providing consistent single-station and orbital multi-frequency observations of the NTA, STA, BTA and ZPE \citep{Brown2008,Brown2010}. Recently, \cite{Dewsnap2021} measured the activity, orbital characteristics and mass distribution of the ZPE and BTA recorded by CMOR including measurements up to 2020. Estimates of the showers' period of activity, peak time, average radiant and orbital elements determined by \cite{Dewsnap2021} and other authors are summarized in Tables \ref{table:shower_times}, \ref{table:radiant_drift} and \ref{tab:orbelts}. For the Taurids, CMOR measurements probe to an effective limiting meteor magnitude of +7 to +8, corresponding to meteoroids with mass $\sim$ 10$^{-7}$ kg (diameters of $\sim$ 0.5 mm).

For this work, we directly computed the flux of the NTA, STA, BTA and ZPE meteor showers using the 29.85 and 38.15 MHz data, following the procedure described in \cite{Egal2020b}. Details of the data processing can be found in \cite{CB2015}. Using optical measurements of the NTA and STA as our baseline, we assumed for this work a constant mass index $s$ of 1.90, corresponding to the population index 2.3 commonly adopted for the NTA/STA showers (cf. Section \ref{sec:s}). The mass index was computed from the population index $r$ using the relation $s=1+2.5\text{log}_{10}r$ \citep{McKinley1961}. Building on the results of \cite{Dewsnap2021}, we adopted a mass index of 1.87 for the BTA and 1.81 for the ZPE. 

\subsubsection{Wavelet analysis}  \label{sec:wavelet}

In addition, we performed a wavelet analysis of the CMOR measurements to identify and locate the Taurid meteor shower radiant as a function of solar longitude. A 3D wavelet transform is applied to the set of measured velocity vectors of all meteor orbits detected by CMOR from 2002-2020 (approximately 20 million orbits) binned in one degree increments of solar longitude. From the resulting wavelet coefficients, a search is then performed for local enhancements in the number of meteors with a particular radiant and velocity relative to the sporadic background. Shower meteors are characterized by a significant clustering of their drift-corrected radiant and geocentric velocity at a particular time, identified by high values of the wavelet strength coefficient Wc. 

In the first stage of analysis, local maxima in the Wc value in a particular solar longitude bin of 1$\degree$ are identified. Local maxima are defined by Wc coefficients higher than 3$\sigma$ above the annual median Wc$_m$ value at the same (radiant,velocity) location, a value we term x$_\sigma$. In the second analysis stage, possible linkages between local maxima through time are identified. The procedure results in a chain of temporally linked maxima in radiant-geocentric velocity space, characterizing the time-evolution of a specific meteor shower radiant in the radar data. Here we used a wavelet angular probe size of 4$\degree$ and a velocity probe size of 10\% of the geocentric velocity. More details about the method and the different filters considered can be found in \cite{Brown2010}. 

\subsubsection{Convex Hull Approach} \label{sec:convex_hull}

The wavelet analysis is particularly suitable for investigating the orbital evolution of the core of a meteor shower, since it presents less dispersion than individual measurements of the meteoroids' orbit. However, the dispersion of a meteor shower also provides important clues about the age and evolution of a meteoroid stream. Though photographic measurements of the nighttime Taurids allow an assessment of the orbital scatter of these showers (for multi-mm sized particles), only a few observations of the daytime Taurids dispersion are available in the literature \citep{Steel1991}.

We therefore extracted individual meteor orbits associated with the NTA, STA, BTA and ZPE from the CMOR database. Care was taken to employ a new and more robust meteor identification method than the D-criteria technique adopted in \cite{Steel1991}, termed the Convex Hull Approach. We applied the Convex Hull method to the meteors' velocity vector space ($\lambda-\lambda_\odot$, $\beta$, $V_g$), with ($\lambda-\lambda_\odot$, $\beta$) the sun-centered ecliptic coordinates of the radiant and $V_g$ the meteors' geocentric velocity. 

We first calculate the number density matrix of the shower by dividing the ($\lambda-\lambda_\odot$, $\beta$, $V_g$) phase space into cubes of equal size called voxels. We then measure the number of meteor radiants in each voxel when the shower is not active to estimate the average sporadic background density matrix. Next, the number density matrix of the shower is subtracted from the number density matrix of the average sporadic background. Any voxel that contains a number of meteors less than three standard deviations above the background is rejected. Additionally, any voxel with less than four meteors is not considered significant and is therefore also rejected. From the remaining voxels, a set of shower meteors is extracted. 

Once the showers' radiants are separated from the sporadic background, we analyze the distribution of the meteors ($\lambda-\lambda_\odot$, $\beta$, $V_g$) centered on the wavelet-generated radiant. Only meteors in radiant, velocity voxels populated more than three standard deviations above the background and centered around the wavelet radiant are identified as shower members. This provides a separation of the shower's members with the sporadic background with a confidence level of 95\%.

Via this procedure, we identified 1229 NTA, 10 263 STA, 1901 BTA and 4862 ZPE individual orbits in the CMOR database between 2011 and 2021. The parameters used for the Convex Hull extraction are detailed in Appendix C2.2. A full description of this method and its application to the Daytime Sextantids meteor shower will be presented in a separate paper.  

\section{Activity}

\subsection{General considerations} \label{sec:general_considerations}

The activity profiles of the NTA, STA, BTA and ZPE as measured from single-station echo data with the CMOR 29 MHz and 38 MHz radar systems since 2002 are presented in Appendix A2. Gaps in the radar profiles indicate periods for which the records are missing (because of instrumental issues or lack of reliable measurements), and not to a null ZHR estimate. When available, activity profiles computed from the IMO VMN measurements or provided by the VMDB are also presented for comparison. In the case of the VMDB, only ZHR estimates computed for solar longitude bins containing at least five meteors have been retained. In addition, video and radar rates have been arbitrarily scaled to match the average activity level of the nighttime showers determined from visual observations (cf. Section \ref{sec:average_activity}). 

For each shower and each measurement system considered, we can see from plots in Appendix A2 significant variability of the meteor rates measured from year to year. The number and location of activity peaks changes each year. This is an artifact largely caused by the low Taurid meteor rates. In contrast, the annual activity period of the showers remains similar each year, ranging between solar longitudes (SL) 197$\degree$ and 255$\degree$ for the NTA, 170$\degree$ and 245$\degree$ for the STA,  84$\degree$ and 106$\degree$ for the BTA, and  55$\degree$ and 94$\degree$ for the ZPE. 

The annual activity level of the NTA is generally below 10 meteors per hour, for an average maximum ZHR of about 5. The STA displays slightly higher activity than the NTA in the visual range, with episodically high rates exceeding 5 meteors per hour. However, the average activity level of the shower as measured with video and visual systems remains below a limit of 6 meteors per hour. Equivalent STA rates recorded by the radar are generally similar to the visual rates after 212$\degree$ SL; however, we observe a significant enhancement of the STA activity in the radar range around 195-205$\degree$ SL, especially when compared with the VMN rates (e.g., in 2007, 2011, 2016 or 2020). Average radar rates during this period of activity are close to an equivalent of 9-10 meteors per hour, about twice the level estimated from optical measurements. 

The equivalent ZHR profiles of the BTA and ZPE computed from CMOR 29 and 38 MHz data are generally in good agreement between the two systems, except for some specific years (e.g., in 2004, 2010 and 2020) when the discrepancy of the absolute meteor rates differ by a factor $<$2 due to instrumental effects. The maximum ZHR  of the BTA vary by year from 6 to 16 meteors per hour with an average maximum annual level lying around 8 meteors per hour. Maximum ZHR rates of the ZPE vary between 15 and 30 meteors per hour, but the highest activity of the shower is on average closer to 8-12 meteors per hour. 

The main difference between the two radar systems is the lower power of 38 MHz and its shorter wavelength which makes it less sensitive (on average) than the 29 MHz CMOR system. However, the 29 MHz system has undergone changes (in 2009) to its transmitter and cabling, while the 38 MHz system hardware has remained unchanged since 2002, providing a longer, more reliable baseline for comparison. For this reason, we expect the 38 MHz data to provide more reliable measurements of the showers' fluxes and activity profiles; however, the results from the 29 MHz system are presented in this work for comparison.

The ZHR profiles of the autumn Taurids during a year when the Earth passes close to the center of the 7:2 MMR and encounters meteoroids trapped in that resonance \citep[][termed a {\it swarm year}, cf. Figures A1.1 and A1.2]{Asher1998}, generally display a clear primary maximum of activity, with rates exceeding 15 meteors per hour. The timing of the peak activity in solar longitude for a return of the resonant meteoroid swarm are remarkably consistent: 219.5$\degree$, 220$\degree$, 219.5$\degree$ and 220$\degree$ SL were determined respectively for the 1988, 1998, 2005 and 2015 apparitions \citep{Roggemans1989,Arlt2000,Dubietis2007, Olech2017}. These maxima of activity were found to be related to the southern branch (STA), with a possible secondary peak around 229-231$\degree$ SL corresponding to the epochs of highest NTA rates reported in Table \ref{table:shower_times}.  

In a recent analysis, \cite{Devillepoix2021} compared the measurements of the resonant return in 2015 as observed by CAMS and the Desert Fireball Network (DFN) in Australia. De-biased rates of the resonant swarm (TSC) measured by CAMS were compared with the remaining STA and NTA observed in 2015. In Appendix A Figure A1.2, we computed the ZHR of the TSC, STA and NTA showers using the rates (normalized to the sporadic background) presented in Figure 3 of \cite{Devillepoix2021}, assuming a constant population index of 2.3. We added an offset of 1 meteor per hour to better match the level of average NTA and STA activity recorded by the VMN (cf. Section \ref{sec:average_activity}). 

From Figure A1.2, we note that the main peak of activity as observed by CAMS/DFN in 2015 was caused by resonant meteoroids  around SL 221-223$\degree$, corresponding to the timing of the secondary maximum noted for the STA in many years. Possible increased TSC rates were also measured around 229-231$\degree$ SL, in agreement with what was observed during previous returns of the Taurid swarm.  

\subsection{Average activity} \label{sec:average_activity}

\begin{figure}
	\includegraphics[width=.49\textwidth]{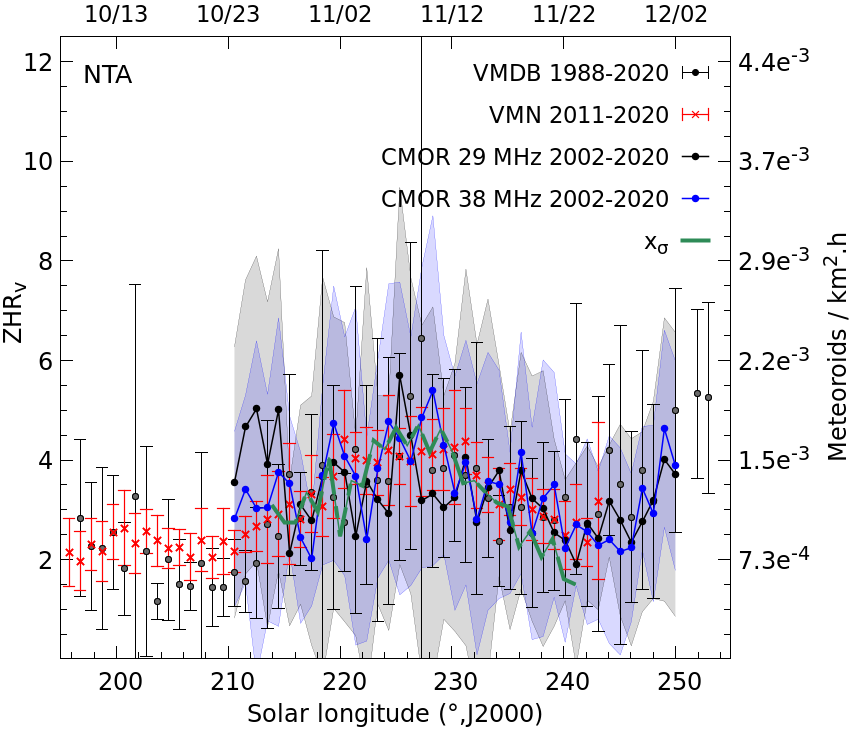}\\[0.2cm]
	\includegraphics[width=.49\textwidth]{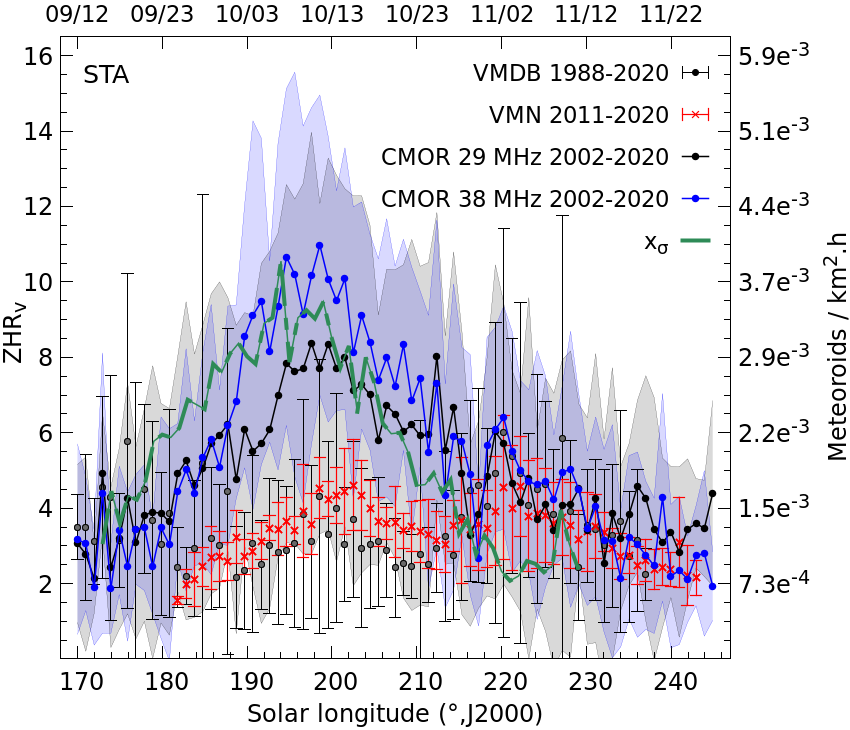}
	\caption{\label{fig:average_NSTA} The average activity (ZHR$_v$) of the NTA (top panel) and STA (bottom panel) as measured from single station echoes with the CMOR 29 MHz system (black) and the 38 MHz system (blue) between 2002 and 2020, the VMN (red) between 2011 and 2020, and visual observations from the IMO from 1988 to 2020 (grey). Error bars reflect the standard deviation in measurements across all years. The profiles are compared with the coefficient x$_\sigma$ (in green), representing the number of standard deviations by which the CMOR-based wavelet strength Wc of radiant overdensity exceeds the median wavelet coefficient Wc$_m$ through the year. This coefficient is arbitrarily scaled to the average ZHR$_v$ profile to emphasize the shape/relative intensity (see text for details). The equivalent absolute flux to a limiting magnitude of +6.5 assuming a mass index of 1.9 is shown as well. The dates corresponding to each solar longitude label in the abscissa are provided at the top of each graph.}
\end{figure}

\begin{figure}
	\includegraphics[width=.49\textwidth]{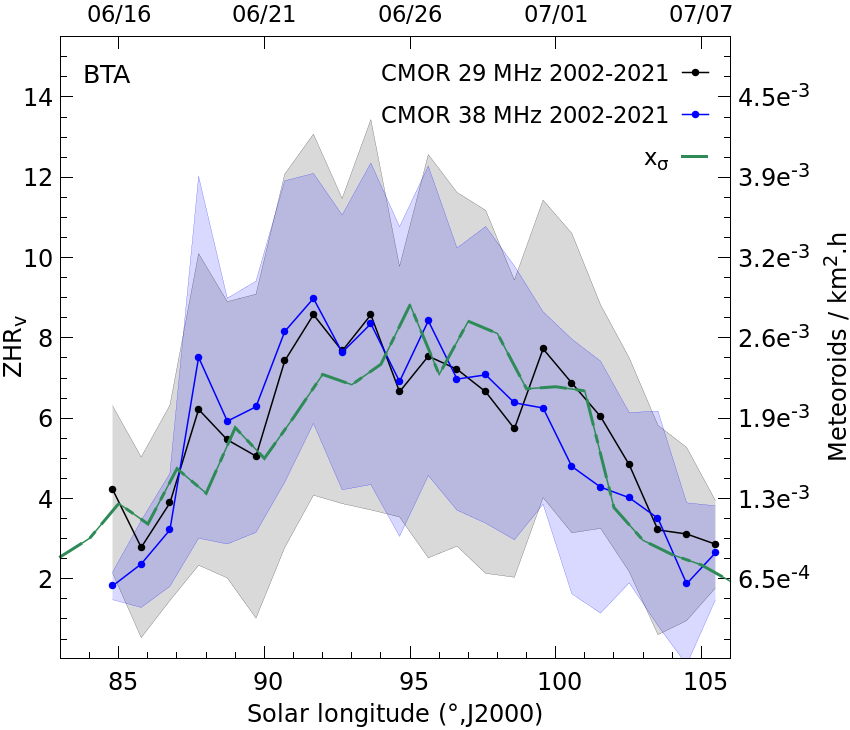}\\[0.2cm]
	\includegraphics[width=.49\textwidth]{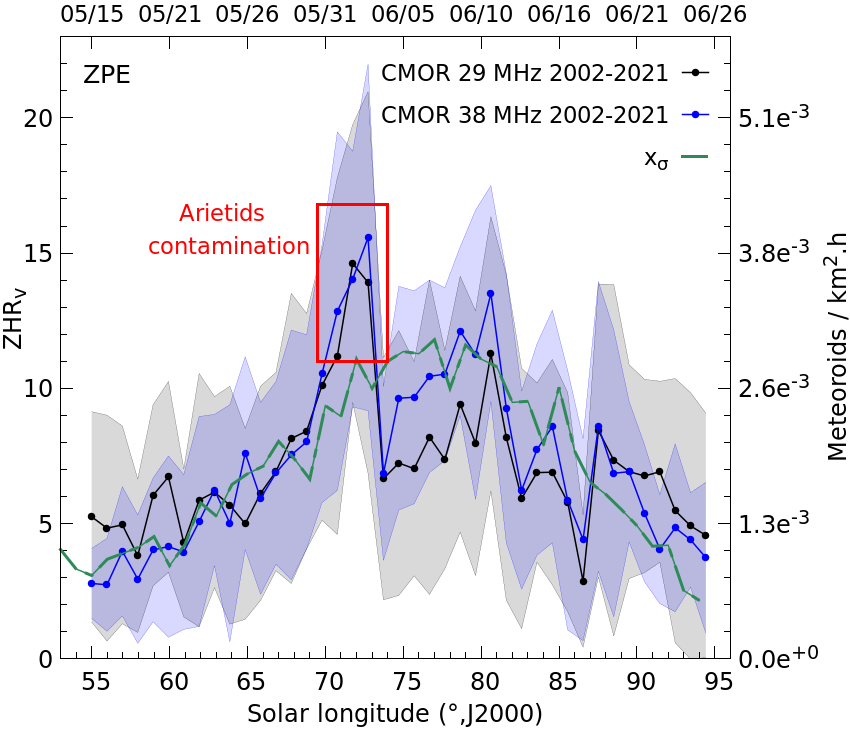}
	\caption{\label{fig:average_BZ} The average activity (ZHR$_v$) of the BTA (top panel) and ZPE (bottom panel) as measured with the CMOR 29 MHz system (black) and the 38 MHz system (blue) between 2002 and 2020. The profiles are compared with the coefficient x$_\sigma$ (in green), arbitrarily scaled to the average ZHR$_v$ profile to emphasize the shape/relative intensity as in Figure \ref{fig:average_NSTA}. The equivalent absolute flux to a limiting magnitude of +6.5 assuming a mass index of 1.87 (BTA) and 1.81 (ZPE) is shown as well. Note the local maximum in single station flux for the ZPE near 72-73 SL is due to contamination from the nearby (and much stronger) Daytime Arietid radiant. This contamination does not affect the wavelet-determined relative intensity. The dates corresponding to each solar longitude label in the abscissa are provided at the top of each graph.}
\end{figure}

Figure \ref{fig:average_NSTA} presents the average activity profile of the NTA and STA as measured with the CMOR 29 MHz and 38 MHz systems, the VMN and the VMDB. Each technique has observations of the two showers since 2002, 2011 and 1988 respectively. The average activity profile of the BTA and ZPE, computed from CMOR measurements between 2002 and 2021 is presented in Figure \ref{fig:average_BZ}. A fixed bin window of 1$\degree$ in solar longitude was used for the average profile computation, following the methodology detailed in \cite{Egal2020b}. 

Examining the original absolute activity profiles derived from visual, video and radar measurements, we observed a systematic initial offset in the ZHR-equivalent levels. We found the VMN activity to be consistently 1.5 to 2 times smaller than the VMDB estimates, while the radar levels were systematically higher than the visual data. These differences were observed for each apparition of the NTA and STA for which activity profiles were available, and are therefore likely due to systematic biases, particularly as the VMN and VMDB sample the same particle sizes. In the case of differences between the radar and optical estimates, some of this difference may also be due to changes in the population index with meteoroid size.

To focus on the relative intensity changes with SL between the different networks, we scale the average CMOR and VMN profiles of the NTA to the VMDB activity curve. A similar procedure was done  in \cite{Egal2020b}, where  the radar and video rates were adjusted to the visual ZHR results to make the visual data the comparison standard as visual measurements have the longest continuous period of measurement. This does not imply that we expect visual observations to be more accurate, especially in the case of the Taurids and without a careful processing of the observations quality over time, but rather represent a convenient choice for a standard measure. 

To highlight this distinction in the rest of this work, we adopt the notation ZHR$_v$ to emphasize that the instrumental fluxes are equivalent ZHR estimates scaled to the visual activity level. The resulting ZHR$_v$ profiles of the VMN, CMOR 29 MHz and 38 MHz presented in Figures \ref{fig:average_NSTA}, \ref{fig:average_BZ} and in supplementary Appendix A2 were obtained by multiplying the original ZHR by 1.7, 0.23 and 0.45 respectively. 

Once rescaled, we find the shape of the average profiles of the NTA and STA computed from visual (VMDB) and video (VMN) records to be in good agreement. Average visual rates range from 1 to 6 meteors per hour, during 50$\degree$ to 70$\degree$ in solar longitude. The activity profiles of the NTA and STA are not symmetric in solar longitude. The STA is more active in October and early November, while the NTA activity increases in middle and late November. 

The onset of  NTA activity is observed at 195$\degree$ SL, with a slight increase of the ZHR$_v$ around 201$\degree$ SL. However, most of the NTA activity is confined to the interval 210-250$\degree$, with highest rates occurring between 220$\degree$ and 232$\degree$ SL. From the individual profiles presented in Supplementary Appendix A2, we observe that the location and magnitude of peak activity of the NTA varies from year to year; no pronounced absolute maximum is visible in the average profile. 

Neither the average visual profile nor the CMOR measurements show a definite maximum activity time of the NTA (see Figure \ref{fig:average_NSTA}), but rather display a broad activity plateau. The equivalent radar ZHR$_v$ follows the general trend of the optical profile but with higher variability in ZHR$_v$ estimates. This is probably a consequence of the low shower meteor rates which are comparable to the sporadic background. 

The bottom panel of  Figure \ref{fig:average_NSTA} shows the average activity profile of the STA computed from optical and radar observations. The shower is active between 170$\degree$ and 245$\degree$ in solar longitude, and shows two distinct periods of enhanced activity. The first ZHR$_v$ peak occurs around 197-198$\degree$, and is at least twice as strong in the radar profile than in the visual data. Excluding this first peak of activity (between SL 187$\degree$ and 213$\degree$), the shape of the VMN, VMDB and CMOR (29 MHz and 38 MHz) profiles are in good agreement. We suggest that the marked difference between optical and radar rates at this first peak reflects a real difference in the meteoroid population at sub-mm and multi-mm sizes rather than calibration differences or observational biases. We find the STA to be enriched in fainter, radar-sized (sub-mm) meteoroids early in their activity period. The second activity peak occurs near 220$\degree$ SL, corresponding to the beginning of the maximum NTA activity, and reaches a similar ZHR$_v$ level of 5 to 6 meteors per hour for each technique. 

Figure \ref{fig:average_BZ} presents the average activity profile of the BTA (top panel) and ZPE (bottom panel) as measured by CMOR between 2002 and 2021. As expected, our results are in good agreement with those recently published by \cite{Dewsnap2021}; we find the BTA and ZPE profiles to be asymmetric as a function of solar longitude, with a slow rise and more sudden decrease in activity. The BTA are active between 85 and 103$\degree$ SL, with equivalent ZHR$_v$ rates of 7-9 meteors per hour recorded between 91 and 96$\degree$ SL. The ZPE are active between 55 and 95$\degree$, with a broad maximum ZHR$_v$ of 10-14 measured between 76.5 and 81.5$\degree$ SL in the 38 MHz data.

From the ZPE profile of Figure \ref{fig:average_BZ}, we observe a local maximum of the ZHR$_v$ between 69 and 73$\degree$ SL. This increased activity is due to contamination from the daytime Arietids meteor shower. The daytime Arietids, also discovered with the Jodrell Bank radar in 1947, are frequently recorded along with the ZPE \citep{Porubcan2007,Dewsnap2021}. The ZPE are difficult to distinguish from the Arietids, having radiants less than 15$\degree$ apart, speeds within 12 km/s and similar times of maximum activity. For this reason, the activity profile of the ZPE is easily contaminated by the Arietids, resulting in an artificial apparent increase of the ZPE meteor rates around SL 72$\degree$ \citep{Dewsnap2021}.

Since the average activity profile of the Taurids might be affected by the return of resonant meteoroids, we computed the average activity level of each shower (and for each system) excluding known swarm years \citep{Asher1998}.  The NTA and STA profiles were computed without the 1988, 1991, 1995, 1998, 2005 and 2015 apparitions, while the years 2002, 2012 and 2019 were not considered during the BTA and ZPE processing. However, we found no noticeable difference from the average profile computed using all years. This implies that while the resonant swarm increases the proportion of very bright Taurids, the enhancement in overall rates is not significant. We therefore conclude that the ZHR$_v$ profiles of Figures \ref{fig:average_NSTA} and \ref{fig:average_BZ} are representative of the annual Taurid activity. 

Using the average profiles of Figures \ref{fig:average_NSTA} and \ref{fig:average_BZ}, we estimate that the maximum annual ZHR$_v$ of the NTA, STA, BTA and ZPE are about 6, 11, 9 and 14 meteors per hour respectively. Note, the ZHR$_v$ of each shower has been obtained using different best-fit values for mass indices (1.9 for the NTA and STA, 1.87 for the BTA and 1.81 for the ZPE), that slightly affect the estimated activity level. However, we use the same mass index values for all instruments for a given shower to allow intercomparison.

To quantify the real relative strength of the four Taurid showers, we present on a secondary vertical axis the flux of meteoroids per km$^2$ and per hour recorded by each network. We remind the reader the values presented in Figures \ref{fig:average_NSTA} and \ref{fig:average_BZ} were obtained from the original VMN and CMOR 29 MHz and 38 MHz fluxes after being multiplied by scaling coefficients of 1.7, 0.23 and 0.45 respectively.  We note that our flux estimates of the ZPE and BTA from the CMOR 38 MHz data (not multiplied by the scaling coefficients) are consistent with the values presented in \cite{Dewsnap2021}. 

We observe that the strongest absolute fluxes (non-scaled, in km$^{-2}$.h$^{-1}$) measured by CMOR are for the STA ($\sim4.0\times 10^{-3}$), followed by the ZPE ($\sim3.5\times 10^{-3}$), the BTA ($\sim2.7\times 10^{-3}-2.9\times 10^{-3}$) and finally the NTA ($\sim2.0\times 10^{-3}$) for Taurids with absolute magnitude brighter than +6.5. Our result confirms the observation of \cite{Steel1991}, that the northern Taurid branch is predominant at the daytime intersection with Earth, while the southern branch is dominant for the nighttime showers. 

\subsection{Fireballs} \label{sec:fireballs}

The autumn Taurids are well known for producing bright fireballs \citep{Hindley1972,Bone1991}. Observational records indicate it might have been one of the most important sources of fireballs around the 11th century \citep{Beech2004}. Most of the observed Taurid fireballs seem to be related to the STA and not the NTA in video records, another reason for the apparent difference in relative activity/strength between the two nighttime showers \citep{Roggemans1989,Jenniskens2006}. 

At the present epoch, the proportion of Taurid fireballs is found to be similar to other major showers like the Perseids or the Geminids \citep{Dubietis2007}, with an annual ratio of fireballs over other visual meteors close to 1\% on average. However, enhanced fireball rates have been reported for each return of the Taurid resonant swarm since 1988 \citep[e.g.,][]{Asher1998,Arlt2000,Beech2004,Dubietis2007}. During swarm years, the proportion of fireballs may approach  2.4-4.6\% of the number of visual meteors close to the time of Taurid maximum activity \citep{Dubietis2007}.

The observation of bright fireballs within the shower led to the suggestion that the Taurids might produce meteorites  \citep{Brown2013,Madiedo2014}. In a recent study, \cite{Borovicka2020} analyzed the physical properties of 16 Taurid fireballs, most of them belonging to the resonant swarm presented in \cite{Spurny2017}. From this sample of meteoroids of diverse masses, ranging from 8 g to 650 kg, the authors found that most of the Taurids lose 90\% of their mass when reaching a dynamic pressure of 0.05 MPa. The meteoroids were found to have generally low strength (below 0.01 MPa) with a trend of weaker material at larger sizes (more than 10 cm in size). Stronger material, potentially compatible with meteorite strengths, was found in some meteoroids, but only at smaller sizes. For these reasons, \cite{Borovicka2020} concluded it is unlikely that the Taurids produce meteorites.

\section{Physical properties}

\subsection{Bulk Density}

Dynamical and spectral analysis of the Taurids has revealed significant heterogeneity among meteoroids of different masses.
Taurid meteoroids were found to be generally weak and have cometary affinities, but with some stronger material present as small inclusions of possibly carbonaceous chondrite-like components \citep{Matlovic2017}. Estimates of Taurid meteoroid bulk density vary greatly; from 400 kg.m$^{-3}$ \citep{BR2002} to 1600 kg.m$^{-3}$ \citep{Babadzhanov2009} or even as high as 2300-2800 kg.m$^{-3}$ \citep{Konovalova2003}. 

\cite{Verniani1967} derived a mean particle density of 2500 kg.m$^{-3}$ for the STA and 2700 kg.m$^{-3}$ for the NTA for photographic-sized (> tens of grams) particles. \cite{Matlovic2017} estimated a mineralogical grain density of between 1300 and 2500 kg.m$^{-3}$, which exceeds the bulk density because of the meteoroids' porous structure. In their sample, the lowest mineralogical densities were measured for the most massive bodies. The authors suggested that the diversity of the observed Taurids is due to a different evolution of the meteoroids' surfaces, or reflects a real heterogeneity of their parent body. They also found that spectra and physical properties of cm-sized Taurids were most consistent with cometary material. This was supported by normal to enhanced sodium levels, albeit with significant diversity in chemistry/strength as evidenced by a wide range in meteoroid strengths as evidenced by indices such as K$_b$ and P$_E$. 

In a subsequent analysis, \cite{Borovicka2020} derived a wide range of bulk densities from 200 to 2000 kg.m$^{-3}$. Most meteoroids were found to have a density below 1000 kg.m$^{-3}$ and a low strength compatible with a cometary origin. The most notable trend, confirmed by the results of \cite{Matlovic2017}, was that the larger the Taurid meteoroid the more fragile it was. Stronger material existed as small inclusions in these large bodies or as separate small (cm-sized) meteoroids \citep{Spurny2017}. Thus the overall picture from Taurid physical measurements is of material likely cometary in origin, but with a notable anti-correlation of strength with size.

\subsection{Size distribution} \label{sec:s}

Estimates of the NTA and STA population indices, as measured from long-term visual observations, typically lie around 2.3 \citep{Jenniskens1994,Rendtel1995, rendtel2014}. This value was adopted by the IMO to compute the ZHR profiles of the two showers since 1988. However, it has been shown that the population index $r$ of the nighttime Taurids varies with solar longitude, and also from year to year. 

By measuring the population index of the NTA and STA from visual observations between 1988 and 1998, \cite{Arlt2000} identified low $r$-values between 211$\degree$ and 225$\degree$ SL. The lowest $r$-estimates ranged from 1.8 to 2.4, and were especially small during the swarm years. Later, \cite{Dubietis2006} and \cite{Devillepoix2021} confirmed this result, measuring a population index of 1.9 and 1.75 during the return of the resonant swarm in 2005 and 2015, respectively. 

Radio measurements of the BTA and ZPE show similar variability in the showers' mass index $s$ \citep{Pecina2004,Porubcan2007,Dewsnap2021}. From observations of the showers between 2000 and 2005, \cite{Porubcan2007} estimated that $s$ ranged from 1.98$\pm$0.04 and 2.28$\pm$0.21 for the ZPE, and from 2.10$\pm$0.02 to 2.37$\pm$0.09 for the BTA. In addition, observations of the ZPE in 2004 and 2005 in Ondrejov displayed evident hints of mass segregation, with the maximum of activity of faint radar meteors appearing two days prior to the maximum measured for overdense echoes \citep{Pecina2004,Porubcan2007}. Similar variation with mass was observed for the BTA and the STA, but not for the NTA \citep{Pecina2004}. However, as these estimates are based solely on the number of overdense echoes and their duration,  distribution, significant sporadic contamination is probable. Moreover, overdense durations are heavily influenced by chemistry effects which tends to obscure the true mass distribution \citep{Baggaley2002}. Thus these are likely upper limits to the true mass index. 

More recently, \cite{Dewsnap2021} measured the mass index of the daytime Taurids using CMOR observations since 2011. They used the underdense echo amplitude distribution of radar echoes occurring on the shower echo line to isolate shower members, though some sporadic contamination along the echo line was still present. 
The ZPE mass index ranged from 1.73 to 1.85 between 2011 and 2019, and from 1.80 to 1.91 for the BTA. No correlation between the swarm returns and epochs of low $s$-values were identified in CMOR data. The mass index was also found to vary over the showers' period of activity, with values ranging between 1.71 and  1.89 for the ZPE, and between 1.76 and 2.0 for the BTA. These authors estimated  a mass distribution index near the time of maximum of 1.81 $\pm$ 0.05 for the ZPE and 1.87 $\pm$ 0.05 for the BTA. 

\section{Radiants}

\subsection{Structure}

Early photographic observations performed by \cite{Whipple1940} and \cite{Whipple1952} first revealed that the nighttime Taurid radiants are split into two distinct branches, called the northern and southern branches, located symmetrically relative to the ecliptic. The two branches were created by planetary perturbations acting over a long timescale of a few thousand years, representing a full cycle (or more) in the precession of the argument of perihelion for the mean orbits of the NTA/STA at their time of maximum \citep{Asher1991}. Particles with slightly different semi-major axis precess differentially and result in multiple points of intersection of the stream at the Earth \citep{Babadzhanov1990}. The existence of N and S branches is generally interpreted as a sign of old age (cf. Section \ref{sec:discussion}).

Over time, small clusters of meteor radiants located in the constellations of Taurus, Aries, Cetus and Pisces (among others), have variously been claimed to be related to the Taurid Meteoroid Complex (TMC). \cite{Denning1928} identified 13 possible active radiants associated with the TMC, and additional associations of meteoroid streams with the Taurids have been explored by different authors \citep[e.g.,][]{OS1988,Steel1991,Babadzhanov1990,Stohl1990,Babadzhanov2001,Porubcan2002,Porubcan2006,Babadzhanov2008,Bucek2014,Jenniskens2016}.

Searching for Taurid sub-streams in the IAU MDC database of photographic meteor orbits, Porub\u{c}an and co-workers suggested that the Southern and Northern Piscids, the Southern Arietids, both branches of the $\chi$ Orionids, and Omicron Orionids are related to the TMC \citep{Stohl1990,Porubcan1992,Porubcan2002,Porubcan2006,Bucek2014}. The authors found the main Taurid filaments extracted from the IAU MDC to be connected with comet 2P/Encke, while additional sub-streams may be related to near-Earth asteroids like 2003 QC10, 2004 TG10, 2005 UY6, 2005 TF50 or 2007 RU17. The radiant distribution and ephemeris of these streams computed from IMO VMN records between 1995 and 2004 is presented in \cite{TC2005}. 

\cite{Svoren2011} selected 84 NTA and 143 STA orbits from the IAU MDC, that were further analyzed by \cite{Kanuchova2012} and \cite{Kanuchova2014}. The authors identified several subgroups within each branch, including the NTA, the STA and 9 additional minor meteor showers. Their associations suggest that several of these showers could be genetically linked with the 2P/Encke meteor complex, but that additional parent bodies are necessary to explain the whole extent of the TMC. Complicating this interpretation is the heavy contamination from the sporadic background as the STA/NTA are embedded in the anti-helion sporadic source.

Processing the meteor observations collected by CAMS between 2010 and 2013, \cite{Jenniskens2016} noticed a discontinuity in the drift of the NTA and STA radiants. The authors observed that the drift-corrected radiants, after remaining stationary for a few days, jumped to different positions during the whole activity period of the shower. They suggested that presuming this phenomenon is not due to observational selection effects, it may reflect the existence of the several subcomponents of the Taurid streams explored in previous works. On this basis, \cite{Jenniskens2016} identified 19 subcomponents of the nighttime Taurids, most with a typical duration of 3 to 10 days. They also identified sub-components of the complex still present in January, highlighting the possibility of late meteor activity linked to the TMC.

\begin{figure*}
	\includegraphics[width=.48\textwidth]{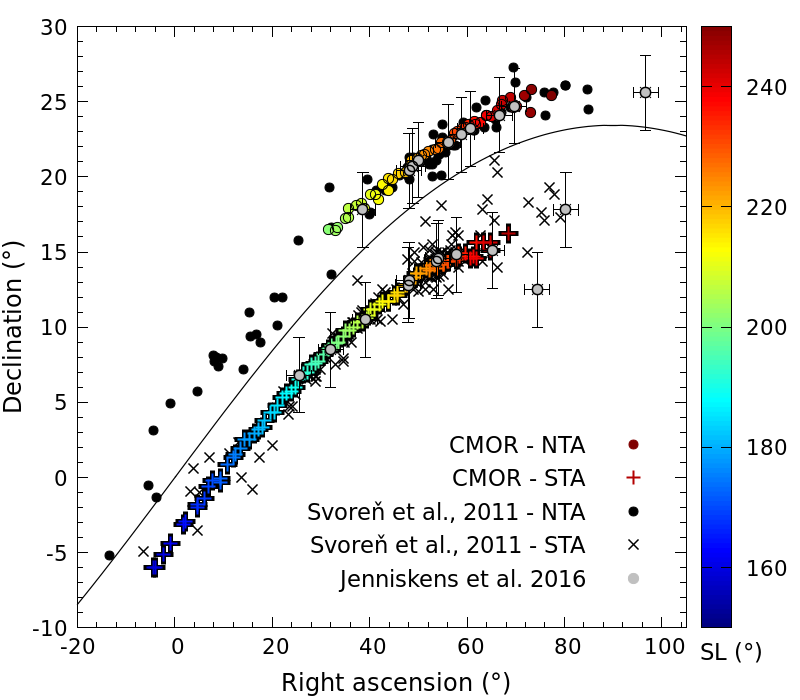}
	\includegraphics[width=.48\textwidth]{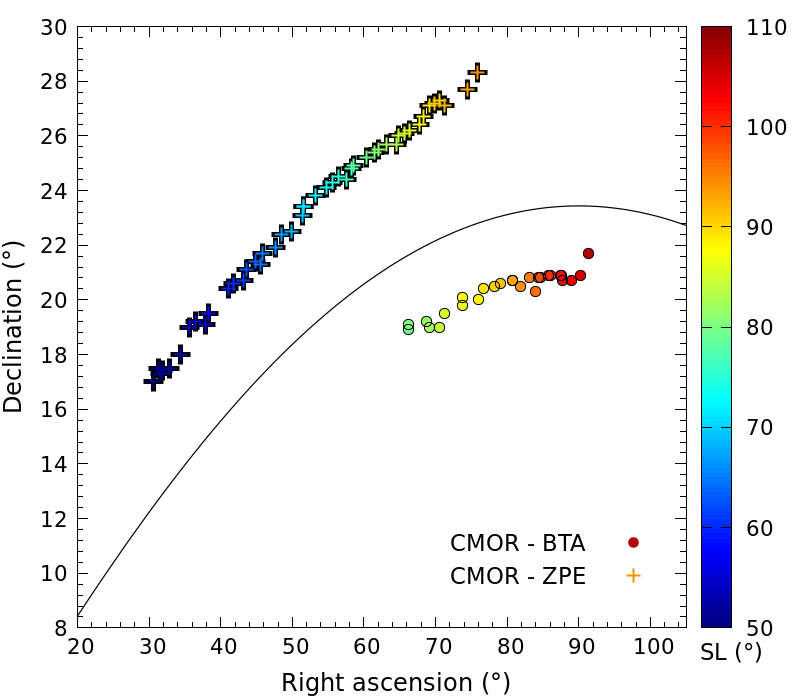}
	\caption{Geocentric right ascension and declination (J2000.0) of the NTA, STA (left panel) and BTA, ZPE (right panel) radiants as a function of solar longitude (color code). The average shower radiant location in one degree SL bins determined from the wavelet analysis of 29 MHz CMOR data is identified with circles (NTA \& BTA) and crosses (STA \& ZPE). The radiants of the NTA and BTA lie above the ecliptic (represented by the solid black line), while the STA and ZPE are located below the ecliptic line. Radiants of the STA and NTA meteors as selected by  \protect\cite{Svoren2011} in the IAU MDC are represented by black circles and crosses. The radiants of the Taurid sub-components identified by \protect\cite{Jenniskens2016} are illustrated by the grey circles with errorbars.}
	\label{fig:radiants}
\end{figure*}

In this work, we restrict our analysis to the properties of the four main Taurid showers, that is the NTA, STA, BTA and ZPE, widely accepted as being the core of the TMC. Figure \ref{fig:radiants} shows each shower's apparent radiant location as function of the solar longitude. The evolution of the sun-centred ecliptic longitude and latitude of the radiants as a function of the solar longitude is presented in Appendix B1. As expected, the nighttime and daytime showers are separated into two components, one above (NTA \& ZPE) and one below (STA \& BTA) the ecliptic plane. The colored points in Figure \ref{fig:radiants} represent the average radiant location (in one degree SL bins) of the showers determined from the  wavelet analysis of CMOR data. The average NTA and STA radiants are compared with the Taurids selected from the IAU MDC by \cite{Svoren2011}, and with the 19 sub-components of the showers identified by \cite{Jenniskens2016}. We see that there is generally good agreement in the radiant location of the NTA and STA as determined from photographic, video and radar measurements and between different studies. 

\begin{table*}
	\begin{tabular}{rrrrrr}
		& NTA\hspace{0.5cm}\mbox{ } & STA\hspace{0.5cm}\mbox{ } & BTA\hspace{0.5cm}\mbox{ } & ZPE\hspace{0.5cm}\mbox{ } & \\
		\cline{2-5} \\[-0.25cm]
		\hline
		& & & &\\[-0.3cm]
		R.A. & 52.981+0.834*L & 30.786+0.794*L & 82.510+0.829*L & 58.143+0.959*L & [C] \\
		& 54.553+0.936*L & 31.907+0.848*L & -- & -- & [S]\\
		& 54.515+0.906*L & 32.228+0.806*L & -- & -- & [J]\\
		& 54.276+0.875*L & 32.860+0.760*L & -- & -- & [Cd] \\
		\cline{2-6} \\[-0.25cm]
		& 58.60+0.80*L & 48.70+0.73*L & -- & -- & [P] \\
		& 58.0+0.76*L & 50.0+0.79*L & -- &--  & [R] \\
		& 58.43+0.86*L & 33.57+0.77*L &--  &--  & [T]\\
		& -139.21+0.86*SL & -120.27+0.77*SL & -- & -- & [T5] \\
    	\cline{2-6} \\[-0.25cm]
		& 54.13+0.80*L & 31.91+0.73*L & -- & -- & [P*] \\
		& 54.2+0.76*L & 29.5+0.79*L & -- &--  & [R*] \\
		& 54.13+0.86*L & 13.55+0.77*L &--  &--  & [T*]\\
		& 54.29+0.86*L & 31.42+0.77*L & -- & -- & [T5*] \\
		\hline
		& & & &\\[-0.3cm]
		Dec. & -0.0029*L$^{2}$+0.195*L+21.891   & -0.0022*L$^{2}$+0.259*L+8.322 & 20.704+0.038*L & -0.0028*L$^{2}$+0.208*L+24.768 & [C] \\
		& -0.0019*L$^{2}$+0.207*L+21.883   & -0.0020*L$^{2}$+0.286*L+8.384 & -- & -- & [S]\\
		& -0.0020*L$^{2}$+0.171*L+21.171 & -0.0016*L$^{2}$+0.239*L+8.842 & -- & -- & [J]\\
		& -0.0015*L$^{2}$+0.185*L+22.501 & -0.0019*L$^{2}$+0.240*L+9.728 & -- & -- & [Cd] \\
		\cline{2-6} \\[-0.25cm]
		& 21.6-0.16*L & 13.00-0.18*L & -- &--  & [P] \\
		& 22.0-0.10*L & 13.0-0.15*L & -- & -- & [R]\\
		& 22.50-0.20*L  & 14.30-0.18*L &-- & -- & [T]\\
		& -0.0013*SL$^{2}$+0.76*SL-87.08 & -0.0026*SL$^{2}$+1.34*SL-153.08 & -- & -- & [T5] \\
    	\cline{2-6} \\[-0.25cm]
		& 20.8-0.16*L & 17.14-0.18*L & -- &--  & [P*] \\
		& 22.5-0.10*L & 16.9-0.15*L & -- & -- & [R*]\\
		& 23.50-0.20*L  & 18.98-0.18*L &-- & -- & [T*]\\
		& -0.0013*L$^{2}$+0.175*L+18.107 & -0.0026*L$^{2}$+0.316*L+9.997 & -- & -- & [T5*] \\
		\hline
		& & & &\\[-0.3cm]
		$\lambda-\lambda_\odot$ & 190.928-0.205*L  & -0.0023*L$^{2}$-0.159*L+194.900 & 347.970-0.224*L & -0.0033*L$^{2}$-0.111*L+344.412 & [C] \\
		& 192.027-0.064*L  & -0.0023*L$^{2}$-0.077*L+195.976 & -- & -- & [S]\\
		& 192.114-0.235*L  & -0.0008*L$^{2}$-0.231*L+196.278 & -- & -- & [J]\\
		& 191.323-0.130*L  & -0.0016*L$^{2}$-0.174*L+196.219 & -- & -- & [Cd] \\
		\cline{2-6} \\[-0.25cm]
		$\lambda$ & 53.61+0.760*L & 51.89+0.744*L & -- & -- & [P] \\
		& 61.16+0.83*L & 53.66+0.78*L & -- & -- & [T] \\
    	\cline{2-6} \\[-0.25cm]
		& 57.41+0.760*L & 34.78+0.744*L & -- & -- & [P*] \\
		& 57.01+0.83*L & 33.38+0.78*L & -- & -- & [T*] \\
		\hline
		& & & &\\[-0.3cm]
		$\beta$ & 2.639         & -4.156-0.029*L & -2.502-0.001*L & 4.445+0.0202*L & [C] \\
		& 2.562-0.005*L & -4.335-0.016*L & -- & -- & [S]\\
		& 2.479-0.003*L & -4.064-0.032*L & -- & -- & [J]\\
		& 2.681-0.013*L & -4.346-0.027*L & -- & -- & [Cd] \\
		\cline{2-6} \\[-0.25cm]
		& 2.71-0.0136*L  & -4.72-0.0233*L & -- & -- & [P] \\
		& 2.27-0.005*L & -4.33-0.039*L  & -- & -- & [T]\\
    	\cline{2-6} \\[-0.25cm]
		& 2.77-0.0136*L  & -4.18-0.0233*L & -- & -- & [P*] \\
		& 2.29-0.005*L & -3.32-0.039*L  & -- & -- & [T*]\\
		\hline
		& & & &\\[-0.3cm]
		L & SL-225$\degree$ & SL-197$\degree$ & SL-95$\degree$ & SL-77$\degree$ &  [C,S,J,Cd,*]\\
		 & SL-220$\degree$ & SL-220$\degree$ & -- & -- &  [P]\\
		 & SL-230$\degree$ & SL-223$\degree$ & -- & -- &  [R,T,T5]\\
		\hline
		\hline
	\end{tabular}
	\caption{\label{table:radiant_drift} Estimates of the radiant drifts of the NTA, STA, BTA and ZPE from different literature sources. The linear and quadratic functions of the table describe the evolution of the apparent (R.A., dec.) and ecliptic ($\lambda,\beta$) coordinates of the radiant with the solar longitude $L$, centred on the showers' maximum activity time determined in Section \ref{sec:average_activity}. $\lambda$ refers to the  geocentric ecliptic longitude and $\lambda-\lambda_\odot$ to the sun-centered geocentric ecliptic longitude. The radiant drifts were computed using C: CMOR measurements (wavelet analysis), S: the selection of photographic meteors of \protect\cite{Svoren2011}, J: the showers' sub-components identified by \protect\cite{Jenniskens2016} and Cd: the NTA and STA contained in the CAMS database between 2010 and 2016. Estimates published in the literature by P: \protect\cite{Porubcan2002}, R: \protect\cite{Rendtel1995} and (T,T5): \protect\cite{TC2005}, modified to correct for the updated $L$ values (*), are shown for comparison. See text for details.}
\end{table*}

\subsection{Radiant drift}\label{sec:radiant_drift}

The radiant drift of the four major Taurid showers using the wavelet analysis of CMOR data, the photographic meteors selected by \cite{Svoren2011}, the sub-components determined by \cite{Jenniskens2016} and any meteor referenced as a NTA or a STA in the CAMS database between 2010 and 2016 (i.e., about 1800 NTA and 2000 STA) are summarized in Table \ref{table:radiant_drift}. The evolution of the radiants' apparent right ascension and declination (R.A, Dec.) and sun-centered ecliptic longitude and latitude ($\lambda-\lambda_\odot$,$\beta$) is plotted in Appendix B2.

Because of the long duration of the Taurid stream, the motion of the showers' radiants in equatorial coordinates do not follow a perfect linear trend. The meteors' declination, in particular, are not well represented by a linear function over more than a few degrees in solar longitude \citep{TC2005}. As a consequence, estimates of the radiants drift available in the literature are only valid for a limited interval around a solar longitude of reference $L_m$, whose value depends on the data set considered \citep[e.g.,][]{Rendtel1995,Porubcan2002}. 

Since this work focuses on the general characteristics of the major Taurid showers, the choice was made to fit the radiants' motion over the longest possible period of activity. Where appropriate, we fit a quadratic function to the radiant coordinates instead of decomposing its evolution in linear fits over small time bins. Such an approach has also been adopted by \cite{TC2005}, in order to reproduce the NTA and STA average radiants over 5$\degree$ solar longitude intervals. We defined the reference solar longitude $L_m$ of each shower as the approximate epoch of maximum activity determined in Section \ref{sec:average_activity}, that is around 225$\degree$ for the NTA, 197$\degree$ for the STA, 95$\degree$ for the BTA and 77$\degree$ for the ZPE. 

The radiant drift of each shower measured from CMOR [C], \cite{Svoren2011} [S], \cite{Jenniskens2016} [J] and CAMS [Cd] data is provided in Table \ref{table:radiant_drift}. In Appendix B2 we note that the radiant's apparent right ascension and ecliptic latitude are well represented by a linear function, while the apparent declination and the sun-centered ecliptic longitudes are best approximated with a quadratic function. The apparent radiant drifts estimated from the four data sets are in good agreement. A higher discrepancy is noted when considering the ecliptic coordinates, because of the radiants' more clustered shape. In particular, the NTA contained in CAMS database and selected by \cite{Svoren2011} display several clumps in solar longitude that may reflect a difference in geocentric velocity \citep[][cf. Appendix B2]{TC2005}. However, the drift measured remain similar for each source of observation (cf. Table \ref{table:radiant_drift}).  

The radiant drift estimates published by different authors \citep{Porubcan2002,Rendtel1995,TC2005} are presented in Table \ref{table:radiant_drift} for comparison (see [P,R,T,T5]). In the table, references to [T] and [T5] both relate to the work of \cite{TC2005}, but using a time resolution of 1$\degree$ or 5$\degree$ in solar longitude. However, since the $L_m$ values adopted by these authors differ from the ones used in this work, these results can not be directly compared with our measured drifts. To simplify the comparison, we present in Table \ref{table:radiant_drift} a modified version of [P], [R], [T]  and [T5] coefficients (noted [P*,R*,T*,T5*]), transposed to our selected $L_m$ values. However, these modified coefficients may only be valid for small differences in $L_m$ (like for the NTA, with $\Delta L_m = $5$\degree$), and are probably not adapted to the STA ($\Delta L_m \sim $25$\degree$). When comparable, we find our results to be consistent with the estimates of \cite{Porubcan2002}, \cite{Rendtel1995} and \cite{TC2005}.

  \begin{table*}
    \begin{tabular}{cllllllll}
      Shower & q (AU) & a (AU) & e & i ($\degree$) & $\omega$ ($\degree$) & $\Omega$ ($\degree$) & Source \\[0.1cm]
      \hline
      NTA &0.369$\pm$0.071 & 2.14$\pm$0.26 & 0.826$\pm$0.036 & 3.1$\pm$1.3 & 293.0$\pm$8.7 & 227.3$\pm$34.2 & [P$_1$]\\ 
      & 0.350$\pm$0.053 & 2.12$\pm$0.25 & 0.832$\pm$0.033 & 3.1$\pm$1.1 & 294.9$\pm$6.5 & 226.2$\pm$10.2  & [P$_2$] \\
      & 0.352$\pm$0.066 & -- & 0.833$\pm$0.040 & 3.1$\pm$1.4 & 294.9$\pm$8.0 & 216.3$\pm$25.0 & [S,K] \\
      & 0.355$\pm$0.029 & 2.13$\pm$0.06 & 0.829$\pm$0.022 & 3.0$\pm$1.3 & 294.6$\pm$3.1 & 220.6$\pm$1.9  & [J]\\
      & 0.383$\pm$0.032 & 2.18$\pm$0.10 & 0.825$\pm$0.010 & 2.9$\pm$1.3 & 290.9$\pm$4.1 & 227.5$\pm$8.6 &   CMOR \\
      \hline
      STA & 0.370$\pm$0.092 & 2.03$\pm$0.33 & 0.815$\pm$0.046 & 5.5$\pm$1.2 & 113.6$\pm$11.5 & 38.3$\pm$27.9 & [P$_1$]\\ 
      & 0.352$\pm$0.058 & 2.07$\pm$0.32 & 0.825$\pm$0.039 & 5.4$\pm$1.1 & 115.4$\pm$7.2 & 37.3$\pm$11.1 & [P$_2$] \\
      & 0.347$\pm$0.064 & -- & 0.826$\pm$0.455 & 5.4$\pm$1.5 & 116.4$\pm$8.2 & 32.9$\pm$18.9 & [S,K] \\
      & 0.353$\pm$0.029 & 1.95$\pm$0.06 & 0.798$\pm$0.022 & 5.3$\pm$1.3 & 116.6$\pm$3.1 & 34.4$\pm$1.9 & [J]\\
      & 0.342$\pm$0.055 & 1.82$\pm$0.16 & 0.813$\pm$0.014 & 5.0$\pm$1.2 & 117.9$\pm$7.3 & 24.5$\pm$16.6 & CMOR \\
      \hline
      BTA & 0.34 & 2.2 & 0.85 & 6 & 224 & 278.1 & [L] \\
      & 0.274 & 1.653 & 0.834 & 0.3  & 52.3   & 102.7 & [Se] \\ 
      & 0.383 & 1.94  & 0.802 & 3.5  & 246.47 & 274.0 & [B] \\  
      & 0.3737$\pm$0.0004 & 1.793$\pm$0.004 & 0.785$\pm$0.0006 & 2.74$\pm$0.01 & 243.87$\pm$0.04 & 275.46$\pm$0.02 & [D] \\
      & 0.359$\pm$0.021 & 1.91$\pm$0.04 & 0.812$\pm$0.011 & 2.9$\pm$0.2 & 243.7$\pm$2.3 & 275.0$\pm$4.2 &  CMOR \\
      \hline
      ZPE & 0.35  & 2.33  & 0.85   & 8.0  & 61 & 77.0 &  [L] \\
      & 0.365 & 1.492 & 0.755  & 6.5  & 60.5 & 81.5 & [Se] \\
      & 0.331 & 1.65  & 0.800  & 3.9  & 58.74 & 74.0 & [B] \\
      & 0.324$\pm$0.0002 & 1.426$\pm$0.002 & 0.7809$\pm$0.0003 & 5.37$\pm$0.01 & 56.35$\pm$0.02 & 77.51$\pm$0.01 & [D] \\
      & 0.315$\pm$0.019 & 1.61$\pm$0.05 & 0.804$\pm$0.008 & 5.4$\pm$0.5 & 56.8$\pm$2.4 & 73.5$\pm$10.5 & CMOR \\
      \hline  
      2P/Encke & 0.336 & 2.215 & 0.848 & 11.781 & 186.547 & 334.568 & JPL K204/19 \\
     \hline
    \end{tabular}
  \caption{\label{tab:orbelts} Mean orbital elements of the NTA, STA, BTA and ZPE computed from CMOR measurements (wavelet analysis). The streams perihelion distance (q), semi-major axis (a), eccentricity (e), inclination (i), perihelion argument ($\omega$) and longitude of the ascending node ($\Omega$) are compared with the mean orbits published by P$_1$: \protect\cite{Porubcan1992}, P$_2$: \protect\cite{Porubcan2002}, S: \protect\cite{Svoren2011}, K: \protect\cite{Kanuchova2014}, J: \protect\cite{Jenniskens2016}, L: \protect\cite{Lovell1954}, Se: \protect\cite{Sekanina1976}, B: \protect\cite{Brown2010}, D: \protect\cite{Dewsnap2021}. The orbit of comet 2P/Encke in 2015 provided by the JPL is indicated at the end of the Table.}
  \end{table*}
  
\section{Orbital elements} \label{sec:orbelts}

\subsection{Mean orbit} \label{sec:mean_orbit}

Along with the activity profiles and the radiants, the Taurid meteors' orbital elements are an important source of information when analyzing a shower. In particular, specific correlations and time-variations of the orbital elements can help discriminate between different scenarios of meteoroid stream formation \citep[e.g.,][]{Steel1991}. As was done in Section \ref{sec:radiant_drift}, we examine the orbital distribution of the four major Taurid streams measured by CMOR (wavelet analysis), CAMS \citep[database and][]{Jenniskens2016} and contained in the IAU MDC \citep{Svoren2011}.

The average orbital elements of the NTA, STA, BTA and ZPE measured by different authors are summarized in Table \ref{tab:orbelts}. The orbit of comet 2P/Encke in 2015, provided by the JPL\footnote{\url{https://ssd.jpl.nasa.gov}, accessed in October 2021},  is also presented for comparison. We see generally good agreement between the mean orbits measured for the nighttime Taurids, while estimates of the daytime streams orbital elements are more disparate. These discrepancies may be due to the paucity of observations of the daytime Taurids, and/or to biases in the processing of radar meteors. One such bias is the fact that many past radar measurements of the daytime streams have been restricted to small intervals in solar longitude and/or sparse sampling. 

We note for example that the mean orbit of the daytime Taurids computed from our wavelet analysis of CMOR's data differs from the results of \cite{Dewsnap2021}, also based on CMOR measurements. These differences are mainly due to a different correction applied for the meteors' deceleration in the atmosphere. Our analysis, which used the original deceleration correction for CMOR \citep{Brown2005}, resulted in slightly higher geocentric velocities compared to \citet{Dewsnap2021} who used a more recent correction determined by \citet{Froncisz2020}. We consider the mean orbits determined by \cite{Dewsnap2021} for the BTA and ZPE to be more accurate given their use of an updated deceleration correction model. However, since the present wavelet analysis provides the only consistent source of measurements of both the daytime and nighttime Taurids, our computed mean orbits are also presented in Table \ref{tab:orbelts} for comparison. 

It is noteworthy that for the STA and NTA orbits, the semi-major axis of the mean orbits measured optically or by radar do not differ within uncertainty. Indeed, none of the mean orbital elements differ significantly between optical and radar measurements within error bounds. 

\subsection{Time variations} \label{sec:orbital_drift}

The determination of a mean orbit of the Taurid streams is complicated by the notable evolution of the meteoroids' orbit during the year. Appendix C1 details the evolution of the meteoroids' perihelion distance ($q$), semi-major axis ($a$), eccentricity ($e$), inclination ($i$), perihelion argument ($\omega$), longitude of perihelion ($\Pi$), aphelion distance ($Q$) and geocentric velocity ($V_g$) with solar longitude for each Taurid shower and each data set. Figures in Appendix C1 would remain similar if plotted against the meteoroids' longitude of the ascending node ($\Omega$), since $\Omega$ increases linearly with the solar longitude. The best line fit (or quadratic fit) of each orbital element is indicated by the colored lines in the figures of Appendix C1, and the corresponding coefficients are summarized in Table 5.

Appendix C2.1 presents a similar set of Figures for the NTA and the STA, but plotted against the density of radiants measured by CAMS. We see that the core of the stream (highest density areas) in CAMS data have solar longitude ranges which straddle the radar wavelet-determined periods of maximum activity - the blue circles span the dense optical core of each shower. To analyze the daytime shower variations, Figures C2.2.1 to C2.2.4 present a comparison between the wavelet analysis and individual orbits extracted from the  CMOR database using the Convex Hull approach described in Section \ref{sec:convex_hull}.

We observe that the density distribution of the radar-sized meteors with time is generally well characterized by the wavelet result. The wavelet orbits and the densest concentration of meteors identified with the Convex Hull method present a similar evolution across time. However, the slight shift in orbital elements between these two data sets can be explained by the different deceleration corrections applied for the meteors' deceleration in the atmosphere. As detailed in Section \ref{sec:mean_orbit}, the wavelet analysis considered CMOR original deceleration correction \citep{Brown2005} while the Convex Hull involved orbits computed with the new deceleration correction of  \citet{Froncisz2020}. Since the wavelet approach allows characterizing the evolution of the meteoroid streams' core with a higher accuracy than the scattered individual measurements, we will analyze the orbital variations of radar-sized meteoroids using the wavelet results in the rest of this section. 

In Appendices C1 \& C2, we see significant orbital dispersion of the meteors selected by \cite{Svoren2011} or contained in the CAMS database with time, with a complex structure that can not be reduced to a simple linear dependence. The evident clustering of some orbital elements with time also prevents an easy estimate of the streams' mean orbit drift. This is particularly true for the NTA, for which two overlapping structures are visible at L $<-10\degree$ and  -60$\degree<$ L $<$ 80$\degree$ (with L=SL-225$\degree$). These structures are similar to the ones reported by \cite{Steel1991}, who identified two groups of NTA with $\Omega$ below and above 205$\degree$ in photographic data (see top panel of Figure \ref{fig:qW}).

A similar but smaller separation of orbital elements may also exist for the STA, with some CAMS meteors detected at L $<-40\degree$ (L=SL-197$\degree$). Because of the large scatter of orbital elements reported by CAMS and the IAU MDC, the fits to these data should be considered as unreliable and serve only to guide the eye. For the NTA, we restricted the time period for the fit to L= SL-225$\degree$ $>-30\degree$, in order to capture most of the second cluster of meteors. 

From Figure C1.1, we observe a positive correlation of $q$, $a$, $Q$ and $\Pi$ with time, and a negative evolution of $i$, $\omega$, $V_g$ and possibly $e$ with SL for the NTA (L$>-30\degree$). Similar variations are observed for the STA, except for the inclination for which a negative trend is not clear. For the BTA and the ZPE, we observe a slight negative dependence of $q$, $a$, $\omega$ and a positive trend in $e$, $i$, $\Pi$ and $V_g$. A negative correlation of $Q$ is visible for the ZPE, while no clear trend in aphelion distance is observed for the ZPE. As expected, the variations of the daytime Taurids orbital elements in our analysis are in good agreement with \cite{Dewsnap2021}, despite the different deceleration corrections applied to the data. 

We summarize in Table C1.4 the drifts measured for each data set. As stated previously, these coefficients must be considered with caution, especially in the case of the CAMS database or the meteors selected by \cite{Svoren2011}. They are presented here as an approximation of the streams' orbital drift for each branch, and do not describe the evolution of the meteoroids osculating elements as a function of solar longitude with accuracy. The mean orbit variations of the NTA and STA computed by \cite{Porubcan2002} are also provided for comparison.

\subsection{Orbital element correlations}\label{sec:orbital_correlation}

\citet{Steel1991} analysed and interpreted the broad correlations among then available orbits (several hundred) for the Taurid complex. Among several other findings, they suggested that optical and radar observations then available showed trends such that:

\begin{enumerate}
    \item Semi major axis and perihelion distance increased with solar longitude (and $\Omega$) for the NTA/STA.
    \item The NTA appear to show a change in slope and inflection in $q$ as a function of nodal longitude for optical orbits near $\Omega$=205$^\circ$.
    \item Radar measured orbits had smaller semi-major axis than those measured optically. 
    \item Eccentricity showed no trend with solar longitude, but that $a$ and $e$ were positively correlated.
    \item The two daytime showers had very different inclination ranges, with the BTA having higher inclinations than the ZPE.
    \item The STA inclinations showed an anti-correlation with nodal longitude (i.e., also with SL) and occurred at higher inclination than the NTA. The NTA showed a positive correlation of inclination with nodal longitude, but only in radar measurements. 
\end{enumerate} 

\begin{figure}
\centering
	\includegraphics[width=0.48\textwidth]{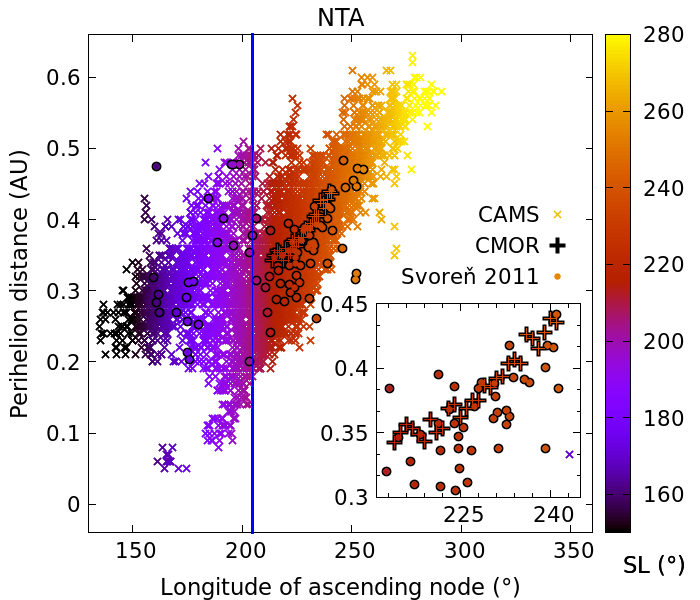}\\[0.1cm]
	\includegraphics[width=0.48\textwidth]{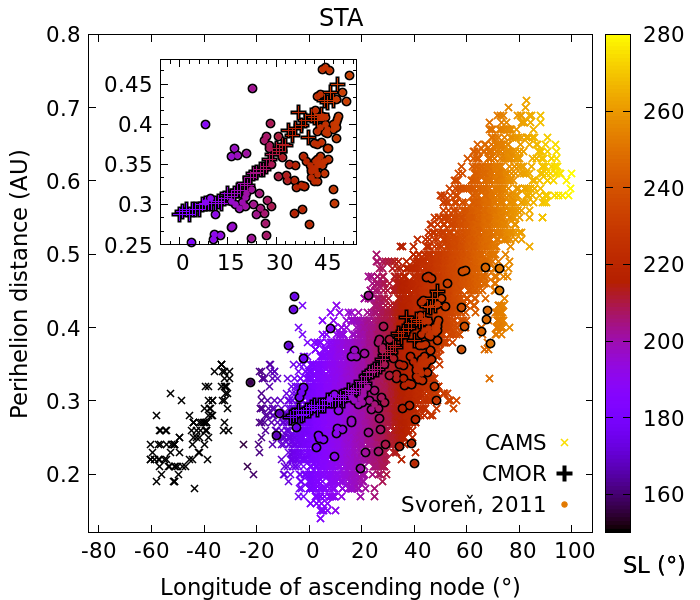}\\[0.1cm]
	\includegraphics[width=0.48\textwidth]{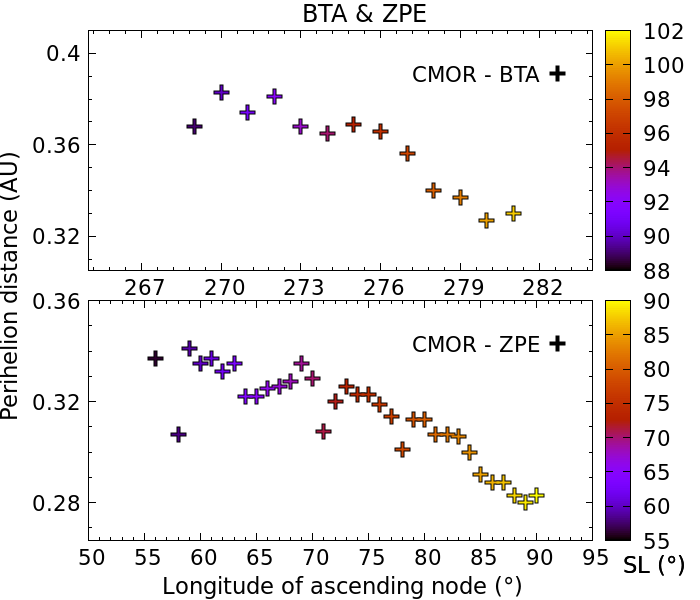}\\[0.1cm]
	\caption{\label{fig:qW} Perihelion distance and longitude of ascending node of the NTA (top), STA (middle), BTA and ZPE (bottom) as measured by CAMS, CMOR (wavelet analysis) or selected by \protect\cite{Svoren2011} from the IAU MDC database. The vertical blue line on the NTA plot indicate the location of $\Omega=205\degree$.}
\end{figure}

\begin{figure}
	\includegraphics[width=0.48\textwidth]{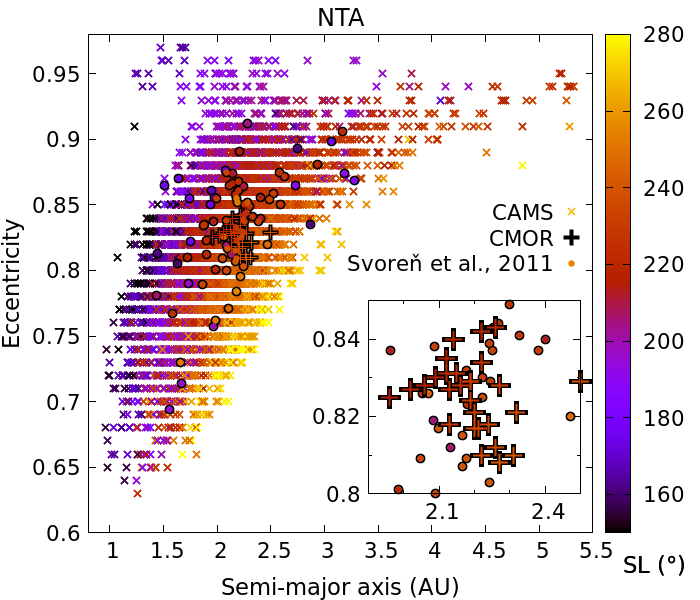}\\[0.1cm]
	\includegraphics[width=0.48\textwidth]{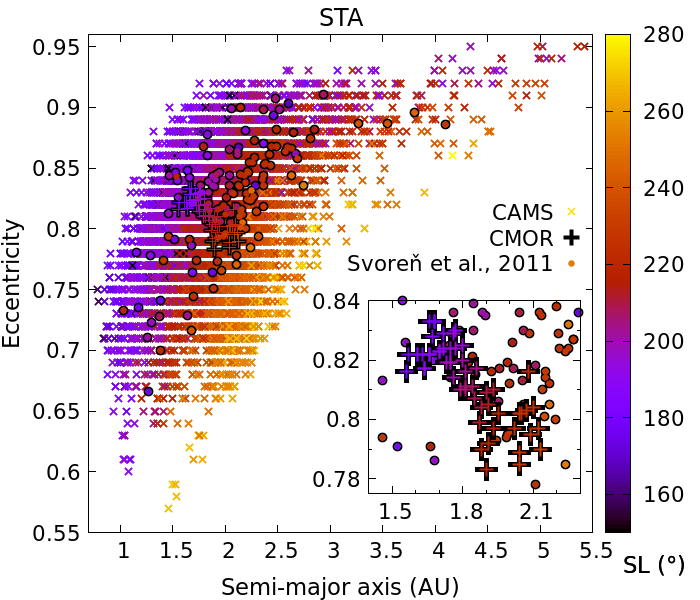}\\[0.1cm]
	\includegraphics[width=0.48\textwidth]{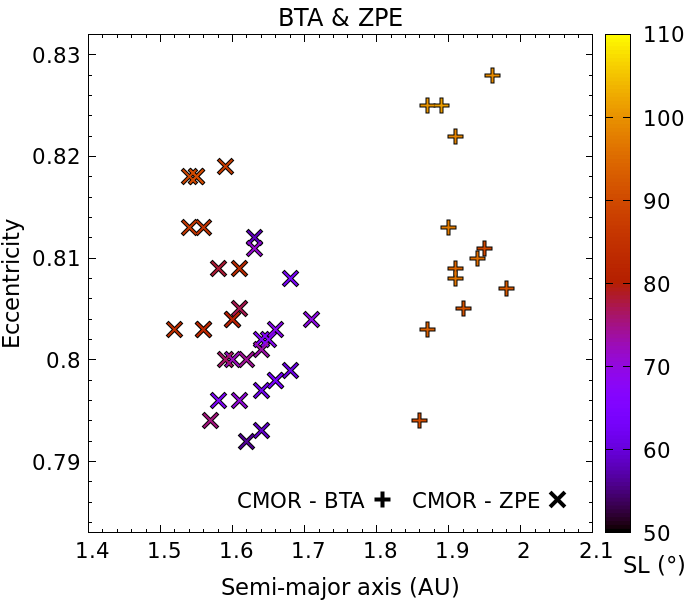}
	\caption{\label{fig:ae} Semi-major axis and eccentricity of the NTA (top), STA (middle), BTA and ZPE (bottom) as measured by CAMS, CMOR (wavelet analysis) or selected by \protect\cite{Svoren2011} from the IAU MDC database. }
\end{figure}

Figures \ref{fig:qW} and \ref{fig:ae} show correlations for all datasets in our analysis, as a function of solar longitude, of the perihelion distance ($q$), semi-major axis ($a$), eccentricity ($e$), longitude of the ascending node ($\Omega$) of the NTA, STA, BTA and ZPE. For the NTA and STA, in agreement with \citet{Steel1991}, we observe a general increase of the perihelion distance (Figure \ref{fig:qW}) and semi-major axis (Figure C3.1) with the longitude of the ascending node, reflecting the influence of orbital precession on the meteoroids' evolution. Though we do not see clear changes in slope or inflection in $q$ vs $\Omega$ as proposed by \cite{Steel1991}, we so observe two clusters of meteoroids around $\Omega=130-205\degree$ and  $\Omega=170-290\degree$ that may correspond to the NTA sub-groups around $\Omega=205\degree$ the authors reported (see solid line in Figure \ref{fig:qW}).
The daytime streams show different orbital correlations. We observe a slight decrease of $a$ as a function of $\Omega$ for the ZPE, but no clear dependence is visible for the BTA (see Figure C3.1). The meteoroids' $q$ also decrease with the solar longitude, in contrast with the NTA/STA  (cf. Figure \ref{fig:qW}). 

Unlike \cite{Steel1991}, we identified in Section \ref{sec:orbital_drift} a possible negative correlation of $e$ with SL for the NTA and STA, while a positive trend in $e$ as function of SL is observed for the BTA and ZPE. Optical measurements of the NTA and STA reveal a clear trend of increasing semi-major axis with eccentricity (see Figure \ref{fig:ae}). However, we do not observe this trend in any shower recorded by CMOR (wavelet analysis), which in fact shows a possible anti-correlation in $a$ vs $e$ for the STA. As detailed in Section \ref{sec:wavelet}, the wavelet approach identifies the core of a stream in a more robust way than the D-criteria method adopted in \cite{Steel1991}; we therefore interpret the negative ($a$,$e$) trend of the nighttime Taurids measured by the wavelet analysis as being real.

To investigate this difference between optical and radar data, we present in Figure C3.3. a comparison between the wavelet analysis and individual orbits measured by the CMOR radar. CMOR orbits robustly selected with the Convex Hull approach show generally good agreement with the optical measurements, and present a smaller dispersion than the CAMS and IAU MDC meteors. We note that for the NTA and STA, the radar orbits follow the positive trend in $a$-$e$ observed for optical data. Such dependence is also visible for the daytime BTA and ZPE (cf. Figure C3.3), while the wavelet analysis of these streams do not show any clear correlation. 

The wavelet approach describes the evolution over time of the regions with the highest concentration of meteoroids within a stream. The orbital correlations identified for these regions, defining the core of the meteor shower, may differ from those inferred from the broad STA stream. If the whole stream displays a positive trend in ($a$,$e$), meteoroids belonging to the core of the stream observed with CMOR present an anti-correlation of $a$ with $e$ (cf. Figure C3.3). These differences provide an additional constraint on the formation mechanism of smaller meteoroids compared to larger Taurids \citep[e.g.,][]{Steel1991}. 

We remark that such a negative trend could also be detected in optical data. Indeed, the fits of the STA orbital elements as function of the solar longitude presented in Table C1.4 for CAMS and IAU MDC meteors all suggest a positive evolution of $a$ and a negative variation of $e$ with time. Since these fits are highly sensitive to the presence of high-density regions in the orbital elements maps (cf. Figures C2.1.1 and C2.1.2), we propose that optical observations may also display a negative trend in ($a$,$e$) for the densest part of the STA stream. However, additional analysis of the CAMS and IAU MDC databases is necessary to confirm this hypothesis.

In contradiction to the interpretation of \cite{Steel1991} we find that the NTA has an anti-correlation with $i$ and SL in all datasets (cf. Figure C1.1) while the STA inclination distribution does not show any certain trend (see Figure C1.2). The STA do occur at higher $i$ than the NTA on average for any given SL, though considerable scatter is present. 

As shown in Appendix C1, the mean aphelion distance for all four streams lays between 3-4 AU over their respective main activity periods. Notably, many of the CAMS and some of the NTA/STA identified by \citet{Svoren2011} have aphelia beyond 4.5 AU. Such orbits would be very short-lived as they interact strongly with or cross Jupiter's orbit and either represent measurement errors or meteoroids on very young orbits.

\subsection{Age considerations} \label{sec:precession}

Taurid meteoroids ejected on an orbit similar to 2P/Encke are expected to show a strong dependence of the inclination $i$ with the perihelion argument $\omega$ (with both angles relative to Jupiter's orbital plane) during their evolution \citep{Asher1991,Steel1991}.  Maximum values of $i$ are expected when $\omega$ is close to 0 and 180$\degree$, while minima in $i$ should be attained near $\omega$=90 or 270$\degree$ \citep{Whipple1940}.

Performing meteor observations from Earth impose a significant constraint on the meteoroids' $\omega$ measured; indeed, the population of meteoroids intersecting the planet's orbit can not be expected to be representative of the full Taurid complex. This observational bias causes the inclinations of the meteoroids observed to be smaller compared with the complex as a whole. 
Despite this selection effect, the relation between $i$ and $\omega$ in the Jovian frame was expected to be visible when plotting the ($i$,$\omega$) distribution for each separate branch. However, the results of \cite{Steel1991} seem to indicate  that this $i-\omega$ dependence is not perceptible in meteor observations, due to conversion effects between the ecliptic and the Jovian frame \citep{Asher1991}. 

When comparing the observed $i-\omega$ distribution with the range of $i-\omega$ of all the Taurid meteoroids that could be observed on Earth (i.e., a representation of the sampling effect detailed above), \cite{Steel1991} identified some differences between the two distributions. In particular, he noticed that low inclinations, that should be measured for each branch, were detected in the NTA but not observed in the STA. Such differences, if real, could provide an additional constraint on the age and the formation of the four Taurid branches. 
In Appendix C3.4, we revisit the analysis of \cite{Steel1991} in order to analyze the $i-\omega$ evolution of each branch in the Jovian frame. Figure C3.4 compares the $i-\omega$ phase space expected for the Taurid meteoroid stream with the radar and optical observations. As noted in \cite{Asher1991}, after conversion into the Jovian frame, the extrema in $i$ at the expected $\omega$ values of $k\times90\degree,\mbox{ } k\in\{1,4\}$ are not observed anymore. 
In the top panels of the Figure, we observe that the observed NTA and STA fill most of the allowed area, with no obvious gaps or clusters in the $i-\omega$ distribution. With a higher number of meteor observations, we note that low-inclinations are reached by both the NTA and the STA, in contrast with \cite{Steel1991}'s result. 

In the bottom panels of Figure C3.4 we see that the daytime Taurids observed with CMOR fill most of the expected orbital range. The apparent NTA, BTA, ZPE cut-off at low inclinations (<1.5$\degree$) observed in CMOR data is likely caused by a selection effect, related to the choice of the parameters' limiting values considered during the Convex Hull extraction (cf. Table C2.2). Despite this effect, the fact that meteoroids in each branch have spread enough to fill a large portion of the allowed observational phase space indicate that the Taurid meteor showers are probably several millennia old.

\begin{table}
	\centering
	\begin{tabular}{ccccccc}
		Object & a & e & $\omega$ & $\Pi$ & $\Delta$T  & $\Delta$T ($50\degree$)\\
		& (AU) & & (ka) & (ka) & (ka) & (ka) \\
		\hline
		& & & & & & \\[-0.25cm]
		NTA & 1.88 & 0.79 & 15 & 80 & 3.3 & 11.1\\
		    & 2.18 & 0.83 & 8 & 55 & 2.3 & 7.6 \\
            & 2.29 & 0.83 & 6 & 45 & 1.9 & 6.3 \\
		\hline
		& & & & & &\\[-0.25cm]
		STA  & 1.41 & 0.77 & 31 & 147 & 14.7 & 20.4\\
		     & 1.82 & 0.81 & 16 & 90 & 9.0 & 12.5\\ 
		     & 2.09 & 0.85 & 9 & 68 & 6.8 & 9.4\\
		\hline
		& & & & & & \\[-0.25cm]
		BTA  & 1.56 & 0.77 & 25 & 122 & 7.8 & 16.9\\
		     & 1.79 & 0.79 & 17 & 90 & 5.8 & 12.5\\
		     & 2.07 & 0.83 & 10 & 67 & 4.3 & 9.3\\
		\hline
		& & & & &\\[-0.25cm]
        ZPE & 1.34 & 0.74 & 35 & 153 & 13.2 & 21.3\\
            & 1.43 & 0.78 & 30 & 146 & 12.6 & 20.3\\
            & 1.75 & 0.83 & 17 & 105 & 9.0 & 14.6\\
		\hline
		& & & & &\\[-0.25cm]
		2P/Encke & 2.2 & 0.85 & 7 & 55 & -- & 7.6 \\
	\end{tabular}
	\caption{\label{tab:prescale}The $a$, $e$ values framing the core activity of the Taurid showers (first and last rows) with associated precession timescales in thousands of years (ka). The middle rows refer to the average $a$, $e$ values summarized in Table \ref{tab:orbelts}. The core activity is extracted from the density plots of CMOR measurements given in C2.2 and represented by the black contour in Figure C3.3. $\Delta$T represents the minimum time required to cover the perihelion longitude ranges determined for each shower (or the average 50$\degree$ determined for the spring and autumn Taurids) with the secular period of revolution of $\Pi$ indicated in the fifth column.}
\end{table}

A crude estimate of the Taurids' age can be performed using the average value and dispersion of the streams' orbital elements. Restricting our discussion to the core of the stream, represented by the red-yellow areas in the density maps presented in Appendix C2.1 and C2.2, we find that the range in the longitude of perihelion ($\Pi$) for the NTA spans $150\degree<\Pi<165\degree$  while for the STA this is 117$\degree<\Pi<153\degree$, with the peak activity occurring with $\Pi$=160$\degree$ for the NTA and $\Pi$=140$\degree$ for the STA. For the daytime streams, the core duration spans 143$\degree<\Pi<166\degree$ for the BTA and 114$\degree<\Pi<145\degree$ for the ZPE. The peak activity occurs with $\Pi$=132$\degree$ for the ZPE and $\Pi$=160$\degree$ for the BTA. The expected correspondence between the $\Pi$ values for the twins BTA/NTA and ZPE/STA is clear. In total, the daytime and nighttime streams cover a range of $\Delta\Pi= 48\degree$ and $\Delta\Pi=52\degree$ respectively. 

Table \ref{tab:prescale} presents the estimated secular periods of revolution of $\omega$ and $\Pi$ (in thousands of years) corresponding to the core's $a$ and $e$ value. This provide some estimate for timescales needed for branch formation, following \citet{Asher1991}. For each shower, we defined the core as the highest concentration of orbits measured by CMOR (i.e. yellow-red areas in maps of Appendix C2.2) represented by the black contours in the $a$-$e$ plots of Figure C3.3. The limiting semi-major axis and eccentricity ranges of the core are represented by the first and third rows in the table, while the middle row indicate the average $a$ and $e$ values determined in Table \ref{tab:orbelts}. The last columns of Table \ref{tab:prescale} indicate the time required for an orbit of given $a$, $e$ value to cover the range of perihelion longitudes observed for each stream, that is 15$\degree$ for the NTA, 36$\degree$ for the STA, 23$\degree$ for the BTA and 31$\degree$ for the ZPE, or the average of 50$\degree$ determined for the spring and autumn Taurids.  

We observe that the precession periods for each stream follow a similar pattern, with particles of large semi-major axis and eccentricity showing the fastest precession cycles. Average NTA orbits present the smallest secular periods of revolution, while the ZPE display the longest precession timescales. When looking at the minimum time required to explain the observed range in perihelion longitudes ($\Delta$T), we see that the core of the NTA and BTA can be formed over shorter timescales (2 to 8 ka) than the STA and ZPE (7 to 15 ka). An object with a typical Taurid orbit, that is with $a$-$e$ values matching the cores' limits of any of the four Taurid showers, would have to precess during 6 to 20 ka to explain the perihelion range of about 50$\degree$ observed for the NTA/STA and the BTA/ZPE. These estimates are in reasonable agreement with the results of \cite{Asher1991} and \citet{Babadzhanov1990}, who determined timescales of <10 ka to up to 18 ka for the branches formation (cf. Section \ref{sec:age}). 

\section{Taurid resonant swarm} \label{sec:resonance}

 \subsection{Description}
 
\cite{Asher1993b} postulated the existence of a dense swarm of meteoroids within the Taurid stream, resulting from material trapped into the 7:2 MMR with Jupiter. By protecting the meteoroids from close encounters with Jupiter when they approach aphelion, the resonance prevents the orbital dispersion of dust along the orbit. As a consequence,  meteoroids tend to cluster in mean anomaly, creating a structure they called the Taurid resonant swarm (hereafter termed the Taurid Swarm Complex or TSC). Enhanced meteor activity is expected to occur at each return of the swarm in the Earth's vicinity, roughly every 3 to 7 years. The TSC is expected to be particularly rich in large meteoroids, small particles being removed faster from the resonance due to radiative forces \citep{Asher1991}. We therefore expect to observe a higher proportion of fireballs during a swarm year than during a regular Taurid apparition (cf. Section \ref{sec:fireballs}).

The existence of a swarm of meteoroids trapped in the 7:2 MMR was suggested as an explanation for several unusual events in recent meteor history. These include strong fireball activity reported over two weeks in November 1951, the 10-day bombardment of the Moon by meteoroids in June 1975 and possibly even the Tunguska event \citep{VD1955,Dorman1978,Asher1998}. Such events allowed \cite{Asher1993b} to define the location of the resonance centre along the orbit, and to predict dates of enhanced Taurid activity due to the 7:2 MMR with Jupiter\footnote{\url{https://www.cantab.net/users/davidasher/taurid/swarmyears.html}, accessed in October 2021}. Further observations have largely confirmed the suggested years of apparition of the TSC predicted by Asher.

For example, observations collected by the Nippon Meteor Society over six decades revealed that Taurid fireballs were more abundant in 1988, 1991 and 1995. An inspection of older data indicated that notable Taurid activity was reported in  1934, 1937, 1954, 1964, 1971, 1978, and 1988, all corresponding to years of predicted TSC returns \citep{Asher1998}. \cite{Beech2004} analyzed Taurid fireball activity from 6 different surveys between 1962 and 2002, combining visual observations, cameras records, photographic surveys and space-based sensors. The authors found the observed peaks in fireball rates to be consistent with the years of predicted swarm returns.  

Analyzing the IMO VMDB between 1985 and 2005, \cite{Dubietis2007} confirmed enhanced Taurid activity in 1988, 1991 and 1995, as well as in 1998 and 2005 - these also match years of predicted encounters with the TSC. Increased fireball rates were widely reported \citep{Johannink2006,Miskotte2006,Dubietis2006,Olech2017} in 2005 while \cite{Dubietis2006} noted enhanced activity in the number of visual Taurids (not just fireballs) observed that year.

The centre of the resonant swarm was predicted to be -30$\degree$ and 35$\degree$ in mean anomaly from the Earth in autumn 2008 and 2012, respectively. Although these distances correspond to the limits of the proposed extent of the TSC in \cite{Asher1993b}'s model, an outburst with a tight radiant concentration in the STA was reported in video records by SonotaCo in 2008 \citep{Kanamori2009}. To our knowledge, no special enhancement of the Taurid activity was detected in 2012. However, we note that few visual observations were reported around the predicted time of the TSC return that year (cf. Figures A.2.2 and A.2.5), because of unfavourable moonlight conditions (full moon on October 30, 2012).

In 2015, a favorable return of the TSC (center of the TSC only -7$\degree$ from Earth) resulted in significant fireball activity in autumn \citep{Olech2017, Spurny2017, Shiba2016, Devillepoix2021}. For the first time, highly accurate meteor orbits \citep{Spurny2017} allowed the unequivocal identification of a branch of STA meteoroids trapped in the 7:2 MMR with Jupiter through direct measurement of the semi-major axis of associated fireballs \citep{Spurny2017,Devillepoix2021}.

 We note that enhanced rates in 1991 reported by several authors \citep[e.g.,][]{Asher1998,Beech2004,Johannink2006,Dubietis2007}, actually relate to a "near miss" resonance year, suggesting that the centre of the resonance may be as far as 30$\degree$ in mean anomaly from the Earth and still produce significant fireball activity. Despite the original expectations of \cite{Asher1998}, that the resonant swarm is the source of northern rather than southern Taurid meteors, the TSC seem to originate exclusively from the STA \citep{Johannink2006,Olech2017,Shiba2016,Spurny2017,Devillepoix2021}.
 
 No detection of enhanced Taurid meteor activity has been reported so far for the past returns of the TSC in May and June (expected in 2002, 2009, 2012, 2019). However, this apparent negative observation is less clear as only a large enhancement in fireball rates  would be detectable for these daylight streams. Following the call for observations of \cite{Clark2019}, several new asteroids residing in the 7:2 MMR with Jupiter were discovered during the last return of the TSC in June 2019 \citep{Egal2021}. Future swarm years are expected to occur in 2022, 2025 and 2032 for the autumn showers and 2026 and 2029 for the daytime Taurids. 
 
 \subsection{No TSC in CMOR data}
  
  Past searches focussing on Taurid orbital elements measured by CMOR have failed to identify the swarm of meteoroids trapped into the 7:2 Jovian resonance \citep{Soja2011,Dewsnap2021}. In this section, we examine the ZHR$_v$ profiles presented in Appendix A2 to examine possible variations in the Taurid level of activity recorded by CMOR, in particular during swarm years. 
  
  To investigate the correlation between the TSC returns and the meteor rates observed by the radar, we present in Figure A3.1 the maximum ZHR$_v$ measured by the 29 and 38 MHz systems for each apparition of the showers between 2002 and 2021. Gaps in the figure are caused by years for which an incomplete activity profile was recorded (see for example the NTA recorded by the 29 MHz antennas). Note that the radar echoes detected by CMOR used to measure ZHR$_v$ are strongly filtered to only include underdense echoes. Thus no Taurids brighter than magnitude +5 are included in these measurements, which are appropriate to a Taurid mass range from 10$^{-7}$ to 10$^{-5}$ kg.
  
  To identify the maximum, we focused our search on the solar longitude intervals [210$\degree$,250$\degree$], [180$\degree$,245$\degree$] and [85$\degree$,105$\degree$] for the NTA, STA and BTA respectively. In the case of the ZPE, we retained the maximum ZHR$_v$ measured between 73$\degree$ and 95$\degree$ SL, in order to avoid the strong Arietid contamination around 70-72$\degree$ SL (see Section \ref{sec:average_activity}). The returns of the TSC since the construction of CMOR, i.e. in 2005 and 2015 (and possibly 2008 and 2012) for the nighttime showers and 2002, 2009, 2012 and 2019 for the daytime showers, are indicated by vertical blue arrows in Figure A3.1. The average activity level of the showers measured by each system is shown with a horizontal red line in the figure.
  
  Since 2002, the maximum activity of the Taurids recorded by CMOR is scattered around the average mean rates determined in Section \ref{sec:average_activity}, with no significant increase of the shower's level during the return of the TSC. In line with previous work, we detect no enhanced NTA activity in 2005, 2008, 2012 or 2015; however, we also measure no increased STA rates for those years either. In Figure A3.1,  we observe a peak in STA activity measured by the CMOR 38 MHz system in 2011. However, this peak is absent from the 29 MHz data. After investigation, we found that the antennas of the 38 MHz were rotated by 90$\degree$ between October 3 and October 11 2011, exactly when the peak occurred,  artificially increasing the activity measured. The enhanced STA activity in 2011 measured by the CMOR 38 MHz system is therefore an instrumental artifact. 
  
  As is the case with the nighttime showers, we observe no unusual BTA or ZPE activity during the TSC returns in the spring with CMOR. Enhanced meteor rates were expected to be observed in 2002, 2009, 2019, and to a lesser degree in 2012. In June 2019, in particular, the Earth approached the centre of the TSC within 5$\degree$ in mean anomaly. However, the showers recorded by CMOR that year showed low intensity levels. In Figure A3.1, we see scatter in the maximum ZHR$_v$ measured for the ZPE in the 29 MHz and 38 MHz systems of about 20\% relative to the long-term mean, reflecting the variable activity of the shower since 2002 (cf. profiles in A2). However, none of the maximum ZHR$_v$ peaks measured from the 38 MHz system correspond to the return of a resonant year. In contrast, the apparent enhanced activity of the ZPE observed in 2008 and 2013 by this system does not correspond to any predicted return of the resonant swarm. 
  
  We note that the highest BTA rates measured by CMOR (on both systems) occurred in 2002, when the centre of the resonant swarm was expected to be at about 11$\degree$ in mean anomaly from the Earth. However, the rates measured by the 38 MHz system, if significant, do not differ much from later estimates of the shower's maximum activity level (for example in 2013). In addition, enhanced meteor rates in 2002 were recorded for all the Taurid showers by the 29 MHz system, suggesting the possibility that these are due to calibration uncertainties during the first year of activity of the CMOR radar when system parameters were being frequently changed (see Section \ref{sec:general_considerations}).  We therefore find no conclusive evidence that the TSC induces visible variations of the Taurids activity in CMOR data and suggest this represents a true absence of smaller (sub-mm) particles in the TSC. 
 
 \section{Discussion} \label{sec:discussion}
 
 \subsection{The Taurid Swarm at different sizes}
 
 The semi-major axis extent of the 7:2 MMR at an eccentricity of 0.85-0.87 ranges from about 2.23 to 2.28 AU \citep{Asher1991,Egal2021}. The distribution of the orbital elements of the NTA and STA largely spans the full extent of the resonance (see Figure C3.1), making it possible in principle to observe the TSC returns with different detection methods measuring particles at different sizes. The effects of the resonant swarm on the STA activity have been identified in visual observations \citep[e.g.,][]{Asher1998,Dubietis2007}, photographic and video measurements \citep[e.g.,][]{Spurny2017,Devillepoix2021}, but have yet to be reported in radar data. 
 
While the direct identification of resonant meteoroids in CMOR data may be hampered by the accuracy of the radar orbits \citep{Soja2011}, we note that in contrast with other devices, we do not observe any increase of the STA activity during the returns of the TSC. Typical masses detected by CMOR lie around 10$^{-7}$ kg \citep{Brown2010}, 
which are much smaller than the cm-sized meteoroids producing the fireballs observed during the swarm years. In addition, overdense echoes were excluded from our analysis of CMOR's data, preventing the detection of Taurid fireballs \citep{Dewsnap2021}. This may support the hypothesis that radiative forces sweep out small particles from the resonance faster than large meteoroids, that tend to stay on similar orbits for long periods of time \citep{Asher1998}.  

\subsection{The relative strengths of the Taurid branches}

The meteor observations we have summarized here reveal that the four meteoroid branches detected on Earth differ in strength. In Section \ref{sec:average_activity}, we determined that the STA are on average more active than the NTA, particularly in radar measurements. The number of STA also exceeds the number of NTA meteors selected by \cite{Svoren2011} in the IAU MDC database (143 STA for 84 NTA) in CAMS data ($\sim$2000 STA for $\sim$1800 NTA in the catalogue 3.0) or in CMOR measurements (cf. Section \ref{sec:convex_hull}). For the daytime showers, we confirm in Section \ref{sec:average_activity} that the ZPE intensity is higher than the BTA, as reported by \cite{Kronk1988} and \cite{Dewsnap2021}.
 
 The northern meteoroid branch is therefore predominant during the daytime showers, while the southern branch is dominant during the nighttime intersection with Earth. Since the ZPE is the twin shower of the STA, and the BTA the twin of the BTA, such an observation is not surprising. The strength differences between the BTA/NTA and ZPE/STA branches may point towards different ejection epochs or precession rates of the meteoroids belonging to each shower (cf. Table \ref{tab:prescale}). This may indicate that the evolution of the Taurid meteoroid complex is not advanced enough to have meteoroids spread equally across all branches.
 
 \subsection{The age of the Taurids} \label{sec:age}
 
 The age of the TSC and the Taurid meteoroid complex is still subject to debate. The existence of several branches in the Taurid stream has often been interpreted as a sign of old age. It takes about 5000 to 6000 years for 2P/Encke's orbit to cover one full cycle in perihelion argument \citep{Egal2021}, depending on semi-major axis.  Assuming 2P/Encke (or an object on a similar orbit) is the unique parent body of the four major Taurid showers, a time of at least one precession period is required to create a meteor shower at the ascending and descending node of the stream, with a northern and a southern branch. Several models suggest that several rotations of the apsidal lines are necessary to create the NTA, STA, BTA and ZPE observed today, leading to higher age estimates of 5-14 ka \citep{Whipple1940,Whipple1952}, 4.7-18 ka \citep{Babadzhanov1990}, 20-30 ka \citep{Steel1996}, or 10-100 ka \citep{Jones1986}. 
 
 In Table \ref{tab:prescale}, we observe that the precession timescale in $\omega$ associated with the core orbital elements of the stream suggest an age between 6 and 30 ka to allow time for the twin showers to appear. Depending on the semi-major axis and eccentricity of the parent body orbit, a minimum precession duration of 6 to 20 ka is necessary to reproduce the average $\Pi$ dispersion of 50$\degree$ observed for the NTA/STA or the BTA/ZPE branch considered together (cf. Section \ref{sec:precession}). If the Taurid showers are created from an unique parent body, the stream complex is probably older than 6 ka. 
 
 The dispersion in $\Pi$ of the core of each Taurid shower also suggest that the present NTA/BTA branch may have formed over a shorter timescale than the STA/ZPE branch. If the difference in strength between the branches reflects different ejection epochs, this may indicate that the NTA and the BTA are younger than the STA and ZPE.

 \subsection{Multiple parent bodies hypothesis}
 
 Another possible explanation for the different intensities of the major Taurid showers is the presence of multiple parent bodies. \citet{Asher1991} suggested multiple fragmenting objects of a larger precursor cometary nucleus could explain the observed features of the TC. More recently, the different characteristics of the NTA and STA led \cite{Jenniskens2006} to propose that the Taurids are younger than the age estimates mentioned above having been created directly from other parent bodies not yet discovered in the stream. Subsequently, several associations between the Taurid showers, the TSC and asteroids of the Taurid complex have also been proposed \citep[e.g.,][]{Porubcan2004,Porubcan2006,Babadzhanov2008,Jopek2011,Olech2016,Spurny2017,Devillepoix2021}. 
 
 Such associations support the original hypothesis of \cite{Clube1984} that the Taurid stream was formed as a result of successive fragmentations of a large parent body a few thousand years ago. However, one must proceed with caution when searching for possible linkages among the different objects of the Taurid complex, given the multiple dynamical pathways which transit the Taurid orbital phase space \citep[for more details, see][]{Egal2021}.
 
 We also do not find clear evidence for multiple sub-streams from radar, the VMN or visual observations based on activity. The activity profiles of the NTA and STA are noisy and show significant variations from year to year which we suggest is most likely a consequence of the relatively weak showers embedded within the anti-helion sporadic source. We do not see any of the strong local maxima in relative activity as reported by \citet{Jenniskens2016} in our average profiles for any of the Taurid branches (cf. Figure \ref{fig:average_NSTA}). The appearance of different radiant areas during the time of the NTA/STA, which \citet{Jenniskens2016} also reported as further evidence for sub-streams, we cannot determine from our wavelet radar measurements as the analysis localizes the strongest maxima within our search window and links maxima together. 
 
\section{Conclusions}

We have attempted to synthesize available observational measurements of the four major Taurid showers, namely the NTA, STA, BTA and ZPE. We collected, processed and compared decades of meteor measurements gathered by visual observers (IMO VMDB), video and photographic cameras (IMO VMN, CAMS, IAU MDC) and the CMOR radar. When possible, results were compared with previous analyses published in the literature.

We computed the average activity profile of each shower following the methodology described in \cite{Egal2020b} from the VMN, VMDB and CMOR 38 MHz and 29 MHz systems (since 2011, 1989 and 2002 respectively). Individual activity profiles obtained for each system since 2002 are provided as a supplementary Appendix. In addition, we measured the radiant drift and orbital elements variations of the Taurids from \cite{Svoren2011}'s selection of photographic meteors, the CAMS database (v3.0) coupled with the analysis of \cite{Jenniskens2016} and CMOR measurements. 

The main characteristics of the Taurid showers as derived from our analysis are: 

\begin{enumerate}
    \item The NTA activity level is generally low, with a maximum annual ZHR$_v$ around 5-6 meteors per hour. The shower's total duration period ranges between 197$\degree$ and 255$\degree$ SL, but most of the activity is confined to an interval of 210-250$\degree$ in SL. The NTA does not display a clear maximum, with the highest meteor rates  forming a plateau between 220$\degree$ and 232$\degree$ SL.
    \item STA meteors are generally observed between 170$\degree$ and 245$\degree$ SL, and the shower is slightly more active than the NTA. 
    The STA displays an early peak of activity around  197-198$\degree$, that is twice as strong in radar records than in optical data (reaching 10-11 meteors per hour). A second peak is observed around 220$\degree$ SL, reaching an average level of 5-6 meteors per hour across all detection methods. The STA are noticeably enriched in sub-mm meteoroids at the beginning of their activity period. 
    \item BTA meteors are visible on radar measurements between  84$\degree$ and 106$\degree$ SL. The shower peaks between 91 and 95$\degree$ SL, with an average maximum activity of about 9 meteors per hour.  
    \item The ZPE are detectable between 55$\degree$ and 94$\degree$ SL, and present a broad maximum close to 12-14 meteors per hour between 76.5 and 81.5$\degree$ SL. ZPE meteors are contaminated by the nearby and much stronger daytime Arietids around SL 69-73$\degree$. 
    \item For each shower, the location and magnitude of peak activity vary from apparition to apparition, indicating a filamentary structure of the stream and/or the effects of small number statistics.
    \item Available spectral and physical proxies for Taurid meteoroids suggests material with cometary affinities and an anti-correlation of strength with particle size. 
    \item The mass distribution index of NTA/STA at their time of maximum is estimated to be 1.9. Radar measurements of the mass index for the ZPE yield 1.81 $\pm$ 0.05 and 1.87 $\pm$ 0.05 for the BTA.
    \item The return of the Taurid Swarm Complex (TSC), known to produce enhanced meteor activity and particularly unusual fireballs rates, is only detected by visual/optical devices in the STA shower. No resonant meteoroids have been identified in the NTA, ZPE or BTA. 
    \item We find no trace of the TSC existence in CMOR radar data either, indicating that smaller meteoroids may be removed from the 7:2 MMR much faster than fireball-producing meteoroids.
    \item We see a significant variation of the streams' orbital elements with solar longitude for the NTA and the STA, while variations of the BTA and ZPE mean orbits estimated from the wavelet-based analysis of CMOR data are more discrete. Such trends in orbital elements can help modellers in discriminating between different formation scenarios of the Taurid meteoroid complex.
    \item Optical and radar measurements show a positive correlation between ($a$,$e$) for the NTA and STA meteors, but an opposite trend for the core of the stream in CMOR data. We suggest this reflects a difference in age or evolution between the core of these streams and other meteoroids of the complex.
    \item The precession timescales associated with the core of the Taurid streams in optical and radar data indicate a minimum time of 6 ka required to create the four Taurid branches from an unique parent body. Published estimates of the TMC complex range between 5 to 100 thousand years.
    \item The northern meteoroid branch is predominant during the daytime showers (ZPE), while the southern branch is more active during the nighttime showers (STA). The dispersion of perihelion longitudes observed for the core of each shower may indicate that the NTA/BTA branch can be formed over shorter timescale than the STA/ZPE branch. 
\end{enumerate}

Disparities between the NTA and the STA (or between the ZPE and the BTA) provide age constraints for modellers presuming all originate from the same parent body.  However, it has been increasingly suggested that different parent bodies may be associated with minor Taurid showers, as well as with the resonant meteoroid swarm (cf. Section \ref{sec:discussion}). Identifying the number of parent bodies of the Taurid stream and their characteristics is essential to understand the formation and evolution of this broader complex. This work aims to be an observational starting point providing an overview of the NTA, STA, BTA and ZPE to serve as the basis of future models of the Taurid Meteoroid Complex. 

\section*{Acknowledgements}

Funding for this work was provided in part through NASA co-operative agreement 80NSSC21M0073. This work was funded in part by the Natural Sciences and Engineering Research Council of Canada Discovery Grants program (Grants no. RGPIN-2016-04433 \& RGPIN-2018-05659) and the Canada Research Chairs program. We would like to thank Prof. Margaret Campbell-Brown and Logan Dewsnap for their help with the processing of CMOR fluxes. We are thankful to the referee for their careful review that helped improve an earlier version of the manuscript.

\section*{Data availability statement}
 	
The data underlying this article will be shared on reasonable request to the corresponding author.




\bibliographystyle{mnras}
\bibliography{references} 

\begin{thebibliography}{}
\makeatletter
\relax
\def\mn@urlcharsother{\let\do\@makeother \do\$\do\&\do\#\do\^\do\_\do\%\do\~}
\def\mn@doi{\begingroup\mn@urlcharsother \@ifnextchar [ {\mn@doi@}
  {\mn@doi@[]}}
\def\mn@doi@[#1]#2{\def\@tempa{#1}\ifx\@tempa\@empty \href
  {http://dx.doi.org/#2} {doi:#2}\else \href {http://dx.doi.org/#2} {#1}\fi
  \endgroup}
\def\mn@eprint#1#2{\mn@eprint@#1:#2::\@nil}
\def\mn@eprint@arXiv#1{\href {http://arxiv.org/abs/#1} {{\tt arXiv:#1}}}
\def\mn@eprint@dblp#1{\href {http://dblp.uni-trier.de/rec/bibtex/#1.xml}
  {dblp:#1}}
\def\mn@eprint@#1:#2:#3:#4\@nil{\def\@tempa {#1}\def\@tempb {#2}\def\@tempc
  {#3}\ifx \@tempc \@empty \let \@tempc \@tempb \let \@tempb \@tempa \fi \ifx
  \@tempb \@empty \def\@tempb {arXiv}\fi \@ifundefined
  {mn@eprint@\@tempb}{\@tempb:\@tempc}{\expandafter \expandafter \csname
  mn@eprint@\@tempb\endcsname \expandafter{\@tempc}}}

\bibitem[\protect\citeauthoryear{{Ahn}}{{Ahn}}{2003}]{Ahn2003}
{Ahn} S.-H.,  2003, \mnras, \href
  {http://adsabs.harvard.edu/abs/2003MNRAS.343.1095A} {343, 1095}

\bibitem[\protect\citeauthoryear{{Arlt}}{{Arlt}}{2000}]{Arlt2000}
{Arlt} R.,  2000, in {Arlt} R.,  ed., Proceedings of the International Meteor
  Conference, 18th IMC, Frasso Sabino, Italy, 1999. pp 112--120

\bibitem[\protect\citeauthoryear{{Asher}}{{Asher}}{1991}]{Asher1991}
{Asher} D.~J.,  1991, PhD thesis, Oxford Univ. (England).

\bibitem[\protect\citeauthoryear{Asher \& Clube}{Asher \&
  Clube}{1993}]{Asher1993b}
Asher D.~J.,  Clube S.,  1993, Quarterly Journal of the Royal Astronomical
  Society, 34, 481

\bibitem[\protect\citeauthoryear{{Asher} \& {Izumi}}{{Asher} \&
  {Izumi}}{1998}]{Asher1998}
{Asher} D.~J.,  {Izumi} K.,  1998, \mn@doi [\mnras]
  {10.1046/j.1365-8711.1998.01395.x}, \href
  {https://ui.adsabs.harvard.edu/abs/1998MNRAS.297...23A} {297, 23}

\bibitem[\protect\citeauthoryear{{Aspinall} \& {Hawkins}}{{Aspinall} \&
  {Hawkins}}{1951}]{Aspinall1951}
{Aspinall} A.,  {Hawkins} G.~S.,  1951, \mn@doi [\mnras]
  {10.1093/mnras/111.1.18}, \href
  {https://ui.adsabs.harvard.edu/abs/1951MNRAS.111...18A} {111, 18}

\bibitem[\protect\citeauthoryear{{Babadzhanov}}{{Babadzhanov}}{2001}]{Babadzhanov2001}
{Babadzhanov} P.~B.,  2001, \mn@doi [\aap] {10.1051/0004-6361:20010583}, \href
  {https://ui.adsabs.harvard.edu/abs/2001A%26A...373..329B} {373, 329}

\bibitem[\protect\citeauthoryear{{Babadzhanov} \& {Kokhirova}}{{Babadzhanov} \&
  {Kokhirova}}{2009}]{Babadzhanov2009}
{Babadzhanov} P.~B.,  {Kokhirova} G.~I.,  2009, \mn@doi [\aap]
  {10.1051/0004-6361:200810460}, \href
  {https://ui.adsabs.harvard.edu/abs/2009A&A...495..353B} {495, 353}

\bibitem[\protect\citeauthoryear{{Babadzhanov}, {Obrubov}  \&
  {Makhmudov}}{{Babadzhanov} et~al.}{1990}]{Babadzhanov1990}
{Babadzhanov} P.~B.,  {Obrubov} Y.~V.,   {Makhmudov} N.,  1990, Solar System
  Research, \href {https://ui.adsabs.harvard.edu/abs/1990SoSyR..24...12B} {24,
  12}

\bibitem[\protect\citeauthoryear{{Babadzhanov}, {Williams}  \&
  {Kokhirova}}{{Babadzhanov} et~al.}{2008}]{Babadzhanov2008}
{Babadzhanov} P.~B.,  {Williams} I.~P.,   {Kokhirova} G.~I.,  2008, \mn@doi
  [\mnras] {10.1111/j.1365-2966.2008.13096.x}, \href
  {https://ui.adsabs.harvard.edu/abs/2008MNRAS.386.1436B} {386, 1436}

\bibitem[\protect\citeauthoryear{Baggaley}{Baggaley}{2002}]{Baggaley2002}
Baggaley W.,  2002, in {Murad, E. and Williams} I.,  ed., , Meteors in the
  Earth's Atmosphere.
Cambridge Univ Press, Cambridge, U.K., Chapt. Radar Obse, pp 123--147

\bibitem[\protect\citeauthoryear{{Beech}, {Hargrove}  \& {Brown}}{{Beech}
  et~al.}{2004}]{Beech2004}
{Beech} M.,  {Hargrove} M.,   {Brown} P.,  2004, The Observatory, \href
  {https://ui.adsabs.harvard.edu/abs/2004Obs...124..277B} {124, 277}

\bibitem[\protect\citeauthoryear{{Bellot Rubio}, {Mart{\'\i}nez Gonz{\'a}lez},
  {Ruiz Herrera}, {Licandro}, {Mart{\'\i}nez-Delgado}, {Rodr{\'\i}guez-Gil}  \&
  {Serra-Ricart}}{{Bellot Rubio} et~al.}{2002}]{BR2002}
{Bellot Rubio} L.~R.,  {Mart{\'\i}nez Gonz{\'a}lez} M.~J.,  {Ruiz Herrera} L.,
  {Licandro} J.,  {Mart{\'\i}nez-Delgado} D.,  {Rodr{\'\i}guez-Gil} P.,
  {Serra-Ricart} M.,  2002, \mn@doi [\aap] {10.1051/0004-6361:20020672}, \href
  {https://ui.adsabs.harvard.edu/abs/2002A&A...389..680B} {389, 680}

\bibitem[\protect\citeauthoryear{{Bone}}{{Bone}}{1991}]{Bone1991}
{Bone} N.~M.,  1991, Journal of the British Astronomical Association, \href
  {https://ui.adsabs.harvard.edu/abs/1991JBAA..101..145B} {101, 145}

\bibitem[\protect\citeauthoryear{{Borovi{\v{c}}ka} \&
  {Spurn{\'y}}}{{Borovi{\v{c}}ka} \& {Spurn{\'y}}}{2020}]{Borovicka2020}
{Borovi{\v{c}}ka} J.,  {Spurn{\'y}} P.,  2020, \mn@doi [\planss]
  {10.1016/j.pss.2020.104849}, \href
  {https://ui-adsabs-harvard-edu.ezproxy.obspm.fr/abs/2020P&SS..18204849B}
  {182, 104849}

\bibitem[\protect\citeauthoryear{Brown, Jones, Weryk  \& Campbell-Brown}{Brown
  et~al.}{2005}]{Brown2005}
Brown P.~G.,  Jones J.,  Weryk R.,   Campbell-Brown M.,  2005, \mn@doi [Earth,
  Moon and Planets] {10.1007/s11038-005-5041-1}, pp 617--626

\bibitem[\protect\citeauthoryear{{Brown}, {Weryk}, {Wong}  \& {Jones}}{{Brown}
  et~al.}{2008}]{Brown2008}
{Brown} P.,  {Weryk} R.~J.,  {Wong} D.~K.,   {Jones} J.,  2008, Earth Moon and
  Planets, \href
  {https://ui-adsabs-harvard-edu.ezproxy.obspm.fr/abs/2008EM&P..102..209B}
  {102, 209}

\bibitem[\protect\citeauthoryear{{Brown}, {Wong}, {Weryk}  \&
  {Wiegert}}{{Brown} et~al.}{2010}]{Brown2010}
{Brown} P.,  {Wong} D.~K.,  {Weryk} R.~J.,   {Wiegert} P.,  2010, \icarus,
  \href
  {https://ui-adsabs-harvard-edu.ezproxy.obspm.fr/abs/2010Icar..207...66B}
  {207, 66}

\bibitem[\protect\citeauthoryear{{Brown}, {Marchenko}, {Moser}, {Weryk}  \&
  {Cooke}}{{Brown} et~al.}{2013}]{Brown2013}
{Brown} P.,  {Marchenko} V.,  {Moser} D.~E.,  {Weryk} R.,   {Cooke} W.,  2013,
  Meteoritics \& Planetary Science, \href
  {https://ui.adsabs.harvard.edu/abs/2013M&PS...48..270B} {48, 270}

\bibitem[\protect\citeauthoryear{{Bu{\v{c}}ek} \&
  {Porub{\v{c}}an}}{{Bu{\v{c}}ek} \& {Porub{\v{c}}an}}{2014}]{Bucek2014}
{Bu{\v{c}}ek} M.,  {Porub{\v{c}}an} V.,  2014, in {Jopek} T.~J.,  {Rietmeijer}
  F.~J.~M.,  {Watanabe} J.,   {Williams} I.~P.,  eds, Meteoroids 2013. p.~193

\bibitem[\protect\citeauthoryear{{Campbell-Brown} \& {Brown}}{{Campbell-Brown}
  \& {Brown}}{2015}]{CB2015}
{Campbell-Brown} M.,  {Brown} P.~G.,  2015, \mnras, \href
  {https://ui.adsabs.harvard.edu/abs/2015MNRAS.446.3669C} {446, 3669}

\bibitem[\protect\citeauthoryear{{Clark}, {Wiegert}  \& {Brown}}{{Clark}
  et~al.}{2019}]{Clark2019}
{Clark} D.~L.,  {Wiegert} P.,   {Brown} P.~G.,  2019, \mn@doi [\mnras]
  {10.1093/mnrasl/slz076}, \href
  {https://ui-adsabs-harvard-edu.ezproxy.obspm.fr/abs/2019MNRAS.487L..35C}
  {487, L35}

\bibitem[\protect\citeauthoryear{Clube \& Napier}{Clube \&
  Napier}{1984}]{Clube1984}
Clube S.,  Napier W.,  1984, Monthly Notices of the Royal Astronomical Society,
  211, 953

\bibitem[\protect\citeauthoryear{{Cook}}{{Cook}}{1973}]{Cook1973}
{Cook} A.~F.,  1973, {A Working List of Meteor Streams}.
p.~183

\bibitem[\protect\citeauthoryear{{Denning}}{{Denning}}{1928}]{Denning1928}
{Denning} W.~F.,  1928, Journal of the British Astronomical Association, 38,
  302

\bibitem[\protect\citeauthoryear{{Devillepoix} et~al.,}{{Devillepoix}
  et~al.}{2021}]{Devillepoix2021}
{Devillepoix} H. A.~R.,  et~al., 2021, arXiv e-prints, \href
  {https://ui.adsabs.harvard.edu/abs/2021arXiv210808450D} {p. arXiv:2108.08450}

\bibitem[\protect\citeauthoryear{{Dewsnap} \& {Campbell-Brown}}{{Dewsnap} \&
  {Campbell-Brown}}{2021}]{Dewsnap2021}
{Dewsnap} R.~L.,  {Campbell-Brown} M.,  2021, \mn@doi [\mnras]
  {10.1093/mnras/stab2351}, \href
  {https://ui-adsabs-harvard-edu.ezproxy.obspm.fr/abs/2021MNRAS.507.4521D}
  {507, 4521}

\bibitem[\protect\citeauthoryear{{Dorman}, {Evans}, {Nakamura}  \&
  {Latham}}{{Dorman} et~al.}{1978}]{Dorman1978}
{Dorman} J.,  {Evans} S.,  {Nakamura} Y.,   {Latham} G.,  1978, Lunar and
  Planetary Science Conference Proceedings, \href
  {https://ui.adsabs.harvard.edu/abs/1978LPSC....9.3615D} {3, 3615}

\bibitem[\protect\citeauthoryear{{Dubietis} \& {Arlt}}{{Dubietis} \&
  {Arlt}}{2006}]{Dubietis2006}
{Dubietis} A.,  {Arlt} R.,  2006, WGN, Journal of the International Meteor
  Organization, \href {https://ui.adsabs.harvard.edu/abs/2006JIMO...34....3D}
  {34, 3}

\bibitem[\protect\citeauthoryear{{Dubietis} \& {Arlt}}{{Dubietis} \&
  {Arlt}}{2007}]{Dubietis2007}
{Dubietis} A.,  {Arlt} R.,  2007, \mn@doi [\mnras]
  {10.1111/j.1365-2966.2007.11488.x}, \href
  {https://ui.adsabs.harvard.edu/abs/2007MNRAS.376..890D} {376, 890}

\bibitem[\protect\citeauthoryear{{Egal}, {Brown}, {Rendtel}, {Campbell-Brown}
  \& {Wiegert}}{{Egal} et~al.}{2020}]{Egal2020b}
{Egal} A.,  {Brown} P.~G.,  {Rendtel} J.,  {Campbell-Brown} M.,   {Wiegert} P.,
   2020, \mn@doi [\aap] {10.1051/0004-6361/202038115}, \href
  {https://ui-adsabs-harvard-edu.ezproxy.obspm.fr/abs/2020A&A...640A..58E}
  {640, A58}

\bibitem[\protect\citeauthoryear{{Egal}, {Wiegert}, {Brown}, {Spurn{\'y}},
  {Borovi{\v{c}}ka}  \& {Valsecchi}}{{Egal} et~al.}{2021}]{Egal2021}
{Egal} A.,  {Wiegert} P.,  {Brown} P.~G.,  {Spurn{\'y}} P.,  {Borovi{\v{c}}ka}
  J.,   {Valsecchi} G.~B.,  2021, \mn@doi [\mnras] {10.1093/mnras/stab2237},
  \href
  {https://ui-adsabs-harvard-edu.ezproxy.obspm.fr/abs/2021MNRAS.507.2568E}
  {507, 2568}

\bibitem[\protect\citeauthoryear{Froncisz, Brown  \& Weryk}{Froncisz
  et~al.}{2020}]{Froncisz2020}
Froncisz M.,  Brown P.,   Weryk R.,  2020, \mn@doi [Planetary and Space
  Science] {10.1016/j.pss.2020.104980}, 190, 104980

\bibitem[\protect\citeauthoryear{{Gartrell} \& {Elford}}{{Gartrell} \&
  {Elford}}{1975}]{Gartrell1975}
{Gartrell} G.,  {Elford} W.~G.,  1975, \mn@doi [Australian Journal of Physics]
  {10.1071/PH750591}, \href
  {https://ui.adsabs.harvard.edu/abs/1975AuJPh..28..591G} {28, 591}

\bibitem[\protect\citeauthoryear{{Hindley}}{{Hindley}}{1972}]{Hindley1972}
{Hindley} K.~B.,  1972, Journal of the British Astronomical Association, \href
  {https://ui.adsabs.harvard.edu/abs/1972JBAA...82..287H} {82, 287}

\bibitem[\protect\citeauthoryear{{Jenniskens}}{{Jenniskens}}{1994}]{Jenniskens1994}
{Jenniskens} P.,  1994, \aap, \href
  {https://ui.adsabs.harvard.edu/abs/1994A&A...287..990J} {287, 990}

\bibitem[\protect\citeauthoryear{{Jenniskens}}{{Jenniskens}}{2006}]{Jenniskens2006}
{Jenniskens} P.,  2006, {Meteor Showers and their Parent Comets}

\bibitem[\protect\citeauthoryear{{Jenniskens} et~al.,}{{Jenniskens}
  et~al.}{2016}]{Jenniskens2016}
{Jenniskens} P.,  et~al., 2016, \mn@doi [\icarus]
  {10.1016/j.icarus.2015.09.013}, \href
  {https://ui-adsabs-harvard-edu.ezproxy.obspm.fr/abs/2016Icar..266..331J}
  {266, 331}

\bibitem[\protect\citeauthoryear{{Johannink} \& {Miskotte}}{{Johannink} \&
  {Miskotte}}{2006}]{Johannink2006}
{Johannink} C.,  {Miskotte} K.,  2006, WGN, Journal of the International Meteor
  Organization, \href {https://ui.adsabs.harvard.edu/abs/2006JIMO...34....7J}
  {34, 7}

\bibitem[\protect\citeauthoryear{Jones}{Jones}{1986}]{Jones1986}
Jones J.,  1986, \mn@doi [Monthly Notices of the Royal Astronomical Society]
  {10.1093/mnras/221.2.257}, 221, 257

\bibitem[\protect\citeauthoryear{{Jopek}}{{Jopek}}{2011}]{Jopek2011}
{Jopek} T.~J.,  2011, \memsai, \href
  {https://ui.adsabs.harvard.edu/abs/2011MmSAI..82..310J} {82, 310}

\bibitem[\protect\citeauthoryear{Kanamori}{Kanamori}{2012}]{Kanamori2009}
Kanamori T.~S.,  2012, WGN, Journal of the International Meteor Organization,
  37, 55

\bibitem[\protect\citeauthoryear{{Ka{\v{n}}uchov{\'a}} \&
  {Svore{\v{n}}}}{{Ka{\v{n}}uchov{\'a}} \&
  {Svore{\v{n}}}}{2012}]{Kanuchova2012}
{Ka{\v{n}}uchov{\'a}} Z.,  {Svore{\v{n}}} J.,  2012, Contributions of the
  Astronomical Observatory Skalnate Pleso, \href
  {https://ui.adsabs.harvard.edu/abs/2012CoSka..42..115K} {42, 115}

\bibitem[\protect\citeauthoryear{{Ka{\v{n}}uchov{\'a}} \&
  {Svore{\v{n}}}}{{Ka{\v{n}}uchov{\'a}} \&
  {Svore{\v{n}}}}{2014}]{Kanuchova2014}
{Ka{\v{n}}uchov{\'a}} Z.,  {Svore{\v{n}}} J.,  2014, Contributions of the
  Astronomical Observatory Skalnate Pleso, \href
  {https://ui.adsabs.harvard.edu/abs/2014CoSka..44..109K} {44, 109}

\bibitem[\protect\citeauthoryear{{Konovalova}}{{Konovalova}}{2003}]{Konovalova2003}
{Konovalova} N.~A.,  2003, \mn@doi [\aap] {10.1051/0004-6361:20030521}, \href
  {https://ui.adsabs.harvard.edu/abs/2003A&A...404.1145K} {404, 1145}

\bibitem[\protect\citeauthoryear{{Kronk}}{{Kronk}}{1988}]{Kronk1988}
{Kronk} G.~W.,  1988, {Meteor showers. A descriptive catalog}.
Enslow Publishers, Hillside, NJ

\bibitem[\protect\citeauthoryear{{Lovell}}{{Lovell}}{1954}]{Lovell1954}
{Lovell} A. C.~B.,  1954, {Meteor astronomy.}

\bibitem[\protect\citeauthoryear{{Madiedo}, {Trigo-Rodr{\'\i}guez}, {Williams},
  {Ortiz}  \& {Cabrera}}{{Madiedo} et~al.}{2013}]{Madiedo2013}
{Madiedo} J.~M.,  {Trigo-Rodr{\'\i}guez} J.~M.,  {Williams} I.~P.,  {Ortiz}
  J.~L.,   {Cabrera} J.,  2013, \mn@doi [\mnras] {10.1093/mnras/stt342}, \href
  {https://ui-adsabs-harvard-edu.ezproxy.obspm.fr/abs/2013MNRAS.431.2464M}
  {431, 2464}

\bibitem[\protect\citeauthoryear{{Madiedo}, {Ortiz}, {Trigo-Rodr{\'\i}guez},
  {Dergham}, {Castro-Tirado}, {Cabrera-Ca{\~n}o}  \& {Pujols}}{{Madiedo}
  et~al.}{2014}]{Madiedo2014}
{Madiedo} J.~M.,  {Ortiz} J.~L.,  {Trigo-Rodr{\'\i}guez} J.~M.,  {Dergham} J.,
  {Castro-Tirado} A.~J.,  {Cabrera-Ca{\~n}o} J.,   {Pujols} P.,  2014, \mn@doi
  [\icarus] {10.1016/j.icarus.2013.12.025}, \href
  {https://ui.adsabs.harvard.edu/abs/2014Icar..231..356M} {231, 356}

\bibitem[\protect\citeauthoryear{{Matlovi{\v{c}}}, {T{\'o}th}, {Rudawska}  \&
  {Korno{\v{s}}}}{{Matlovi{\v{c}}} et~al.}{2017}]{Matlovic2017}
{Matlovi{\v{c}}} P.,  {T{\'o}th} J.,  {Rudawska} R.,   {Korno{\v{s}}} L.,
  2017, \mn@doi [\planss] {10.1016/j.pss.2017.02.007}, \href
  {https://ui.adsabs.harvard.edu/abs/2017P&SS..143..104M} {143, 104}

\bibitem[\protect\citeauthoryear{McKinley}{McKinley}{1961}]{McKinley1961}
McKinley D. W.~R.,  1961, New York, McGraw-Hill, 1961.

\bibitem[\protect\citeauthoryear{{Miskotte} \& {Johannink}}{{Miskotte} \&
  {Johannink}}{2006}]{Miskotte2006}
{Miskotte} K.,  {Johannink} C.,  2006, WGN, Journal of the International Meteor
  Organization, \href {https://ui.adsabs.harvard.edu/abs/2006JIMO...34...11M}
  {34, 11}

\bibitem[\protect\citeauthoryear{{Molau} \& {Kac}}{{Molau} \&
  {Kac}}{2009}]{Molau2009}
{Molau} S.,  {Kac} J.,  2009, WGN, Journal of the International Meteor
  Organization, \href {http://adsabs.harvard.edu/abs/2009JIMO...37..188M} {37,
  188}

\bibitem[\protect\citeauthoryear{{Olech} et~al.,}{{Olech}
  et~al.}{2016}]{Olech2016}
{Olech} A.,  et~al., 2016, \mn@doi [\mnras] {10.1093/mnras/stw1261}, \href
  {https://ui.adsabs.harvard.edu/abs/2016MNRAS.461..674O} {461, 674}

\bibitem[\protect\citeauthoryear{{Olech} et~al.,}{{Olech}
  et~al.}{2017}]{Olech2017}
{Olech} A.,  et~al., 2017, \mn@doi [\mnras] {10.1093/mnras/stx716}, \href
  {https://ui.adsabs.harvard.edu/abs/2017MNRAS.469.2077O} {469, 2077}

\bibitem[\protect\citeauthoryear{{Olsson-Steel}}{{Olsson-Steel}}{1988}]{OS1988}
{Olsson-Steel} D.,  1988, \mn@doi [\icarus] {10.1016/0019-1035(88)90127-3},
  \href {https://ui.adsabs.harvard.edu/abs/1988Icar...75...64O} {75, 64}

\bibitem[\protect\citeauthoryear{{Pecina}, {Porub\v{c}an}, {Pecinov{\'a}}  \&
  {Toth}}{{Pecina} et~al.}{2004}]{Pecina2004}
{Pecina} P.,  {Porub\v{c}an} V.,  {Pecinov{\'a}} D.,   {Toth} J.,  2004,
  \mn@doi [Earth Moon and Planets] {10.1007/s11038-005-4504-8}, \href
  {https://ui.adsabs.harvard.edu/abs/2004EM&P...95..681P} {95, 681}

\bibitem[\protect\citeauthoryear{{Porubcan} \& {Ocenas}}{{Porubcan} \&
  {Ocenas}}{1992}]{Porubcan1992}
{Porubcan} V.,  {Ocenas} D.,  1992, in Proceedings of the International Meteor
  Conference, 10th IMC, Potsdam, Germany, 1991. p.~7

\bibitem[\protect\citeauthoryear{{Porub{\v c}an}, {Williams}  \& {Korno{\v
  s}}}{{Porub{\v c}an} et~al.}{2004}]{Porubcan2004}
{Porub{\v c}an} V.,  {Williams} I.~P.,   {Korno{\v s}} L.,  2004, \mn@doi
  [Earth Moon and Planets] {10.1007/s11038-005-2243-5}, \href
  {https://ui.adsabs.harvard.edu/abs/2004EM%26P...95..697P} {95, 697}

\bibitem[\protect\citeauthoryear{{Porub{\v{c}}an} \&
  {Korno{\v{s}}}}{{Porub{\v{c}}an} \& {Korno{\v{s}}}}{2002}]{Porubcan2002}
{Porub{\v{c}}an} V.,  {Korno{\v{s}}} L.,  2002, in {Warmbein} B.,  ed.,  ESA
  Special Publication Vol. 500, Asteroids, Comets, and Meteors: ACM 2002. pp
  177--180

\bibitem[\protect\citeauthoryear{{Porub{\v{c}}an}, {Korno{\v{s}}}  \&
  {Williams}}{{Porub{\v{c}}an} et~al.}{2006}]{Porubcan2006}
{Porub{\v{c}}an} V.,  {Korno{\v{s}}} L.,   {Williams} I.~P.,  2006,
  Contributions of the Astronomical Observatory Skalnate Pleso, \href
  {https://ui.adsabs.harvard.edu/abs/2006CoSka..36..103P} {36, 103}

\bibitem[\protect\citeauthoryear{{Porub{\v{c}}an}, {Zigo}, {Pecina},
  {Pecinov{\'a}}, {Cevolani}, {Pupillo}  \& {Rozboril}}{{Porub{\v{c}}an}
  et~al.}{2007}]{Porubcan2007}
{Porub{\v{c}}an} V.,  {Zigo} P.,  {Pecina} P.,  {Pecinov{\'a}} D.,  {Cevolani}
  G.,  {Pupillo} G.,   {Rozboril} J.,  2007, Contributions of the Astronomical
  Observatory Skalnate Pleso, \href
  {https://ui.adsabs.harvard.edu/abs/2007CoSka..37...31P} {37, 31}

\bibitem[\protect\citeauthoryear{Rendtel}{Rendtel}{2014}]{rendtel2014}
Rendtel J.,  2014, {Meteor Shower Workbook 2014}.
International Meteor Organization

\bibitem[\protect\citeauthoryear{Rendtel \& Arlt}{Rendtel \&
  Arlt}{2017}]{Rendtel2017}
Rendtel J.,  Arlt R.,  2017, {Handbook For Meteor Observers}, 2014 edn.
International Meteor Organization

\bibitem[\protect\citeauthoryear{Rendtel, Arlt  \& McBeath}{Rendtel
  et~al.}{1995}]{Rendtel1995}
Rendtel J.,  Arlt R.,   McBeath A.,  1995, {Handbook for visual meteor
  observers}.
International Meteor Organization

\bibitem[\protect\citeauthoryear{{Roggemans}}{{Roggemans}}{1989}]{Roggemans1989}
{Roggemans} P.,  1989, WGN, Journal of the International Meteor Organization,
  \href {https://ui.adsabs.harvard.edu/abs/1989JIMO...17..104R} {17, 104}

\bibitem[\protect\citeauthoryear{{Sekanina}}{{Sekanina}}{1973}]{Sekanina1973}
{Sekanina} Z.,  1973, \mn@doi [\icarus] {10.1016/0019-1035(73)90210-8}, \href
  {https://ui-adsabs-harvard-edu.ezproxy.obspm.fr/abs/1973Icar...18..253S} {18,
  253}

\bibitem[\protect\citeauthoryear{{Sekanina}}{{Sekanina}}{1976}]{Sekanina1976}
{Sekanina} Z.,  1976, \mn@doi [\icarus] {10.1016/0019-1035(76)90009-9}, \href
  {https://ui-adsabs-harvard-edu.ezproxy.obspm.fr/abs/1976Icar...27..265S} {27,
  265}

\bibitem[\protect\citeauthoryear{{Shiba}}{{Shiba}}{2016}]{Shiba2016}
{Shiba} Y.,  2016, WGN, Journal of the International Meteor Organization, \href
  {https://ui.adsabs.harvard.edu/abs/2016JIMO...44...78S} {44, 78}

\bibitem[\protect\citeauthoryear{{Soja}, {Baggaley}, {Brown}  \&
  {Hamilton}}{{Soja} et~al.}{2011}]{Soja2011}
{Soja} R.~H.,  {Baggaley} W.~J.,  {Brown} P.,   {Hamilton} D.~P.,  2011,
  \mn@doi [\mnras] {10.1111/j.1365-2966.2011.18442.x}, \href
  {https://ui-adsabs-harvard-edu.ezproxy.obspm.fr/abs/2011MNRAS.414.1059S}
  {414, 1059}

\bibitem[\protect\citeauthoryear{{SonotaCo}}{{SonotaCo}}{2009}]{SonotaCo2009}
{SonotaCo} 2009, WGN, Journal of the International Meteor Organization, \href
  {https://ui.adsabs.harvard.edu/abs/2009JIMO...37...55S} {37, 55}

\bibitem[\protect\citeauthoryear{{Southworth} \& {Hawkins}}{{Southworth} \&
  {Hawkins}}{1963}]{Southworth1963}
{Southworth} R.~B.,  {Hawkins} G.~S.,  1963, Smithsonian Contributions to
  Astrophysics, \href
  {https://ui-adsabs-harvard-edu.ezproxy.obspm.fr/abs/1963SCoA....7..261S} {7,
  261}

\bibitem[\protect\citeauthoryear{{Spurn{\'y}}, {Borovi{\v c}ka}, {Mucke}  \&
  {Svore{\v n}}}{{Spurn{\'y}} et~al.}{2017}]{Spurny2017}
{Spurn{\'y}} P.,  {Borovi{\v c}ka} J.,  {Mucke} H.,   {Svore{\v n}} J.,  2017,
  \mn@doi [\aap] {10.1051/0004-6361/201730787}, \href
  {https://ui.adsabs.harvard.edu/abs/2017A%26A...605A..68S} {605, A68}

\bibitem[\protect\citeauthoryear{{Steel} \& {Asher}}{{Steel} \&
  {Asher}}{1996}]{Steel1996}
{Steel} D.~I.,  {Asher} D.~J.,  1996, \mn@doi [\mnras]
  {10.1093/mnras/281.3.937}, \href
  {https://ui.adsabs.harvard.edu/abs/1996MNRAS.281..937S} {281, 937}

\bibitem[\protect\citeauthoryear{{Steel}, {Asher}  \& {Clube}}{{Steel}
  et~al.}{1991}]{Steel1991}
{Steel} D.~I.,  {Asher} D.~J.,   {Clube} S.~V.~M.,  1991, \mn@doi [\mnras]
  {10.1093/mnras/251.4.632}, \href
  {https://ui.adsabs.harvard.edu/abs/1991MNRAS.251..632S} {251, 632}

\bibitem[\protect\citeauthoryear{Stohl}{Stohl}{1986}]{Stohl1986}
Stohl J.,  1986, in Asteroids, Comets, Meteors II. pp 565--574, \url
  {http://adsabs.harvard.edu/full/1986acm..proc..565S}

\bibitem[\protect\citeauthoryear{{Stohl} \& {Porubcan}}{{Stohl} \&
  {Porubcan}}{1990}]{Stohl1990}
{Stohl} J.,  {Porubcan} V.,  1990, in {Lagerkvist} C.~I.,  {Rickman} H.,
  {Lindblad} B.~A.,  eds, Asteroids, Comets, Meteors III. p.~571

\bibitem[\protect\citeauthoryear{{Svore{\v{n}}}, {Kri{\v{s}}andov{\'a}}  \&
  {Ka{\v{n}}uchov{\'a}}}{{Svore{\v{n}}} et~al.}{2011}]{Svoren2011}
{Svore{\v{n}}} J.,  {Kri{\v{s}}andov{\'a}} Z.,   {Ka{\v{n}}uchov{\'a}} Z.,
  2011, Contributions of the Astronomical Observatory Skalnate Pleso, \href
  {https://ui.adsabs.harvard.edu/abs/2011CoSka..41...23S} {41, 23}

\bibitem[\protect\citeauthoryear{{Tomko} \& {Neslu{\v{s}}an}}{{Tomko} \&
  {Neslu{\v{s}}an}}{2019}]{Tomko2019}
{Tomko} D.,  {Neslu{\v{s}}an} L.,  2019, \mn@doi [\aap]
  {10.1051/0004-6361/201833868}, \href
  {https://ui.adsabs.harvard.edu/abs/2019A&A...623A..13T} {623, A13}

\bibitem[\protect\citeauthoryear{{Triglav-{\v{C}}ekada} \&
  {Arlt}}{{Triglav-{\v{C}}ekada} \& {Arlt}}{2005}]{TC2005}
{Triglav-{\v{C}}ekada} M.,  {Arlt} R.,  2005, WGN, Journal of the International
  Meteor Organization, \href
  {https://ui.adsabs.harvard.edu/abs/2005JIMO...33...41T} {33, 41}

\bibitem[\protect\citeauthoryear{{Verniani}}{{Verniani}}{1967}]{Verniani1967}
{Verniani} F.,  1967, {Meteor masses and luminosity}.
 Vol. 10, 181-195, Smithsonian Contributions to Astrophysics

\bibitem[\protect\citeauthoryear{Verniani}{Verniani}{1973}]{Verniani1973}
Verniani F.,  1973, Journal of Geophysical Research, 78, 8429

\bibitem[\protect\citeauthoryear{{Whipple}}{{Whipple}}{1940}]{Whipple1940}
{Whipple} F.~L.,  1940, Proceedings of the American Philosophical Society,
  \href
  {https://ui-adsabs-harvard-edu.ezproxy.obspm.fr/abs/1940PAPhS..83..711W} {83,
  711}

\bibitem[\protect\citeauthoryear{{Whipple} \& {El-Din Hamid}}{{Whipple} \&
  {El-Din Hamid}}{1952}]{Whipple1952}
{Whipple} F.~L.,  {El-Din Hamid} S.,  1952, Helwan Institute of Astronomy and
  Geophysics Bulletins, \href
  {https://ui-adsabs-harvard-edu.ezproxy.obspm.fr/abs/1952HelOB..41....3W} {41,
  3}

\bibitem[\protect\citeauthoryear{{van Diggelen}}{{van Diggelen}}{1955}]{VD1955}
{van Diggelen} J.,  1955, Ciel et Terre, \href
  {https://ui.adsabs.harvard.edu/abs/1955C&T....71..180V} {71, 180}

\makeatother
\end{thebibliography}


\begin{thebibliography}{}
\makeatletter
\relax
\def\mn@urlcharsother{\let\do\@makeother \do\$\do\&\do\#\do\^\do\_\do\%\do\~}
\def\mn@doi{\begingroup\mn@urlcharsother \@ifnextchar [ {\mn@doi@}
  {\mn@doi@[]}}
\def\mn@doi@[#1]#2{\def\@tempa{#1}\ifx\@tempa\@empty \href
  {http://dx.doi.org/#2} {doi:#2}\else \href {http://dx.doi.org/#2} {#1}\fi
  \endgroup}
\def\mn@eprint#1#2{\mn@eprint@#1:#2::\@nil}
\def\mn@eprint@arXiv#1{\href {http://arxiv.org/abs/#1} {{\tt arXiv:#1}}}
\def\mn@eprint@dblp#1{\href {http://dblp.uni-trier.de/rec/bibtex/#1.xml}
  {dblp:#1}}
\def\mn@eprint@#1:#2:#3:#4\@nil{\def\@tempa {#1}\def\@tempb {#2}\def\@tempc
  {#3}\ifx \@tempc \@empty \let \@tempc \@tempb \let \@tempb \@tempa \fi \ifx
  \@tempb \@empty \def\@tempb {arXiv}\fi \@ifundefined
  {mn@eprint@\@tempb}{\@tempb:\@tempc}{\expandafter \expandafter \csname
  mn@eprint@\@tempb\endcsname \expandafter{\@tempc}}}

\bibitem[\protect\citeauthoryear{{Arlt}}{{Arlt}}{2000}]{Arlt2000}
{Arlt} R.,  2000, in {Arlt} R.,  ed., Proceedings of the International Meteor
  Conference, 18th IMC, Frasso Sabino, Italy, 1999. pp 112--120

\bibitem[\protect\citeauthoryear{Asher \& Clube}{Asher \&
  Clube}{1993}]{Asher1993b}
Asher D.~J.,  Clube S.,  1993, Quarterly Journal of the Royal Astronomical
  Society, 34, 481

\bibitem[\protect\citeauthoryear{Brown, Jones, Weryk  \& Campbell-Brown}{Brown
  et~al.}{2005}]{Brown2005}
Brown P.~G.,  Jones J.,  Weryk R.,   Campbell-Brown M.,  2005, \mn@doi [Earth,
  Moon and Planets] {10.1007/s11038-005-5041-1}, pp 617--626

\bibitem[\protect\citeauthoryear{{Devillepoix} et~al.,}{{Devillepoix}
  et~al.}{2021}]{Devillepoix2021}
{Devillepoix} H. A.~R.,  et~al., 2021, arXiv e-prints, \href
  {https://ui.adsabs.harvard.edu/abs/2021arXiv210808450D} {p. arXiv:2108.08450}

\bibitem[\protect\citeauthoryear{{Dubietis} \& {Arlt}}{{Dubietis} \&
  {Arlt}}{2007}]{Dubietis2007}
{Dubietis} A.,  {Arlt} R.,  2007, \mn@doi [\mnras]
  {10.1111/j.1365-2966.2007.11488.x}, \href
  {https://ui.adsabs.harvard.edu/abs/2007MNRAS.376..890D} {376, 890}

\bibitem[\protect\citeauthoryear{Froncisz, Brown  \& Weryk}{Froncisz
  et~al.}{2020}]{Froncisz2020}
Froncisz M.,  Brown P.,   Weryk R.,  2020, \mn@doi [Planetary and Space
  Science] {10.1016/j.pss.2020.104980}, 190, 104980

\bibitem[\protect\citeauthoryear{{Jenniskens} et~al.,}{{Jenniskens}
  et~al.}{2016}]{Jenniskens2016}
{Jenniskens} P.,  et~al., 2016, \mn@doi [\icarus]
  {10.1016/j.icarus.2015.09.013}, \href
  {https://ui-adsabs-harvard-edu.ezproxy.obspm.fr/abs/2016Icar..266..331J}
  {266, 331}

\bibitem[\protect\citeauthoryear{{Johannink} \& {Miskotte}}{{Johannink} \&
  {Miskotte}}{2006}]{Johannink2006}
{Johannink} C.,  {Miskotte} K.,  2006, WGN, Journal of the International Meteor
  Organization, \href {https://ui.adsabs.harvard.edu/abs/2006JIMO...34....7J}
  {34, 7}

\bibitem[\protect\citeauthoryear{{Miskotte} \& {Johannink}}{{Miskotte} \&
  {Johannink}}{2006}]{Miskotte2006}
{Miskotte} K.,  {Johannink} C.,  2006, WGN, Journal of the International Meteor
  Organization, \href {https://ui.adsabs.harvard.edu/abs/2006JIMO...34...11M}
  {34, 11}

\bibitem[\protect\citeauthoryear{{Porub{\v{c}}an} \&
  {Korno{\v{s}}}}{{Porub{\v{c}}an} \& {Korno{\v{s}}}}{2002}]{Porubcan2002}
{Porub{\v{c}}an} V.,  {Korno{\v{s}}} L.,  2002, in {Warmbein} B.,  ed.,  ESA
  Special Publication Vol. 500, Asteroids, Comets, and Meteors: ACM 2002. pp
  177--180

\bibitem[\protect\citeauthoryear{{Roggemans}}{{Roggemans}}{1989}]{Roggemans1989}
{Roggemans} P.,  1989, WGN, Journal of the International Meteor Organization,
  \href {https://ui.adsabs.harvard.edu/abs/1989JIMO...17..104R} {17, 104}

\bibitem[\protect\citeauthoryear{{Steel}, {Asher}  \& {Clube}}{{Steel}
  et~al.}{1991}]{Steel1991}
{Steel} D.~I.,  {Asher} D.~J.,   {Clube} S.~V.~M.,  1991, \mn@doi [\mnras]
  {10.1093/mnras/251.4.632}, \href
  {https://ui.adsabs.harvard.edu/abs/1991MNRAS.251..632S} {251, 632}

\bibitem[\protect\citeauthoryear{{Svore{\v{n}}}, {Kri{\v{s}}andov{\'a}}  \&
  {Ka{\v{n}}uchov{\'a}}}{{Svore{\v{n}}} et~al.}{2011}]{Svoren2011}
{Svore{\v{n}}} J.,  {Kri{\v{s}}andov{\'a}} Z.,   {Ka{\v{n}}uchov{\'a}} Z.,
  2011, Contributions of the Astronomical Observatory Skalnate Pleso, \href
  {https://ui.adsabs.harvard.edu/abs/2011CoSka..41...23S} {41, 23}

\makeatother
\end{thebibliography}



\appendix

\bsp	
\label{lastpage}
\end{document}


\begin{center}
  \Large \textbf{SUPPLEMENTARY APPENDICES}\\[1.0cm]
  \end{center}

\setcounter{topnumber}{4}
\setcounter{totalnumber}{4}

\newcommand{\fignumprefix}{}
\renewcommand{\thefigure}{\fignumprefix\arabic{figure}}
\renewcommand{\theHfigure}{figure.\thefigure}
\newcommand{\setfignumprefix}[1]{%
	\renewcommand{\fignumprefix}{#1}
	\setcounter{figure}{0}
}

\newcommand{\tabnumprefix}{}
\renewcommand{\thetable}{\tabnumprefix\arabic{table}}
\renewcommand{\theHtable}{table.\thetable}
\newcommand{\settabnumprefix}[2]{%
	\renewcommand{\tabnumprefix}{#1}
	\setcounter{table}{#2}
}

\textbf{APPENDIX A: ACTIVITY PROFILES}\\

The following graphs provide activity profiles for different annual returns for the STA, NTA, ZPE and BTA. These are compiled from various literature sources (provided in the captions to each figure subsets) and from the present analysis using visual, video and radar datasets. \\

 \textbf{A1 Resonant years}\\
 
\begin{figure}[!ht]
    \centering
 	\includegraphics[width=0.45\textwidth]{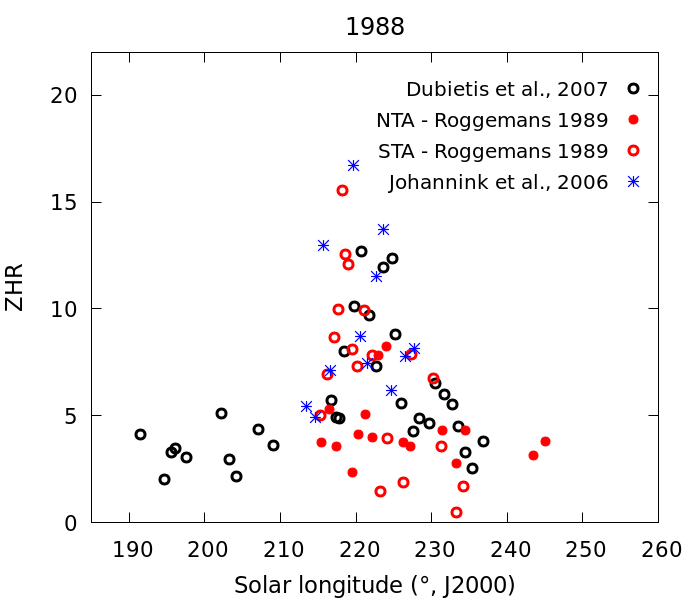}
 	\includegraphics[width=0.45\textwidth]{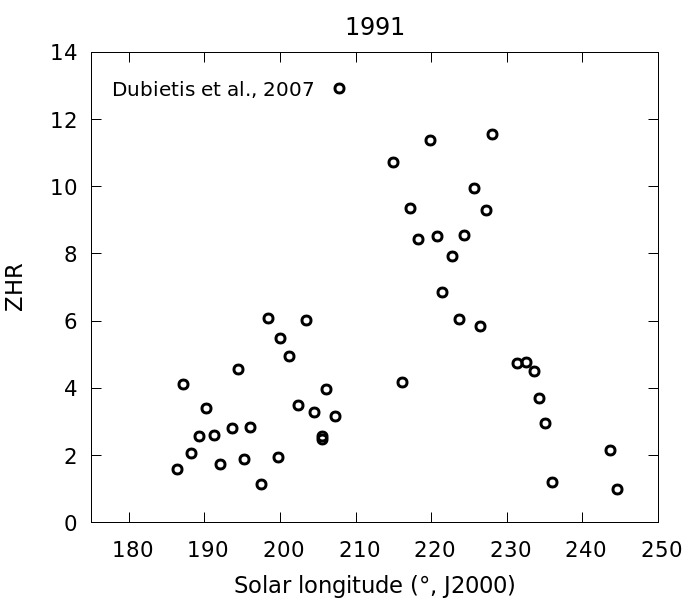}\\
 	\includegraphics[width=0.45\textwidth]{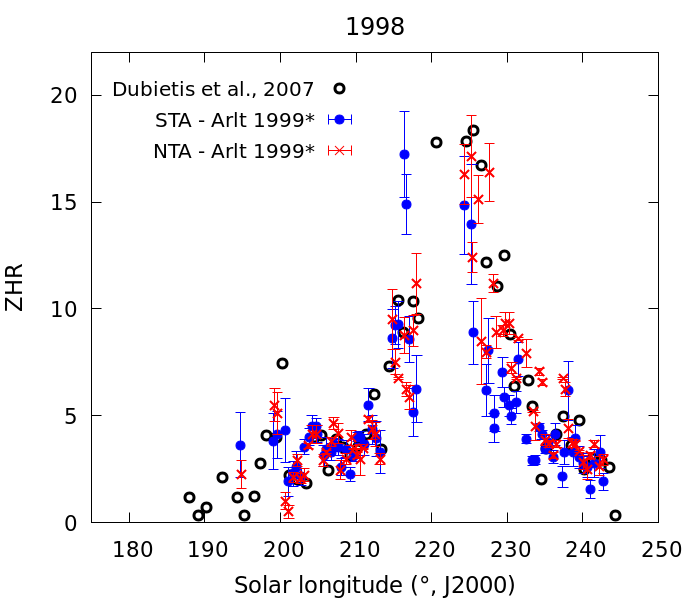}
 	\includegraphics[width=0.45\textwidth]{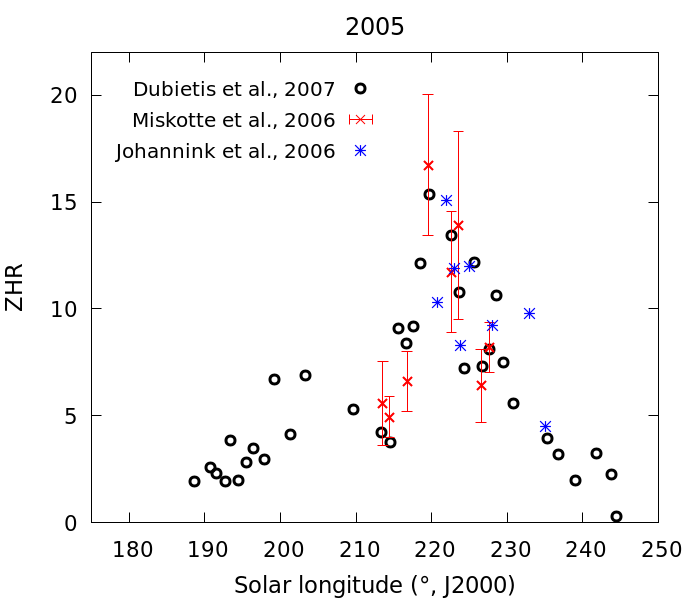}
 	\caption{Published activity profiles computed from visual observations in 1988, 1991, 1998 and 2005. The original activity profiles of Arlt 1999 have been rescaled to match Dubietis \& Arlt 2007.} \label{fig:visual_resonant_years}
 \end{figure}

\begin{figure}[!ht]
 	\includegraphics[width=0.49\textwidth]{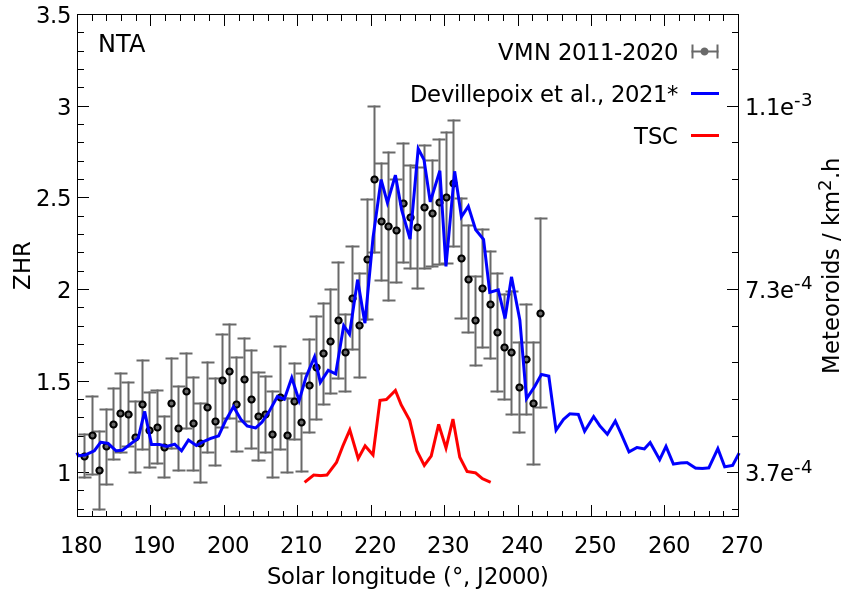}
 	\includegraphics[width=0.49\textwidth]{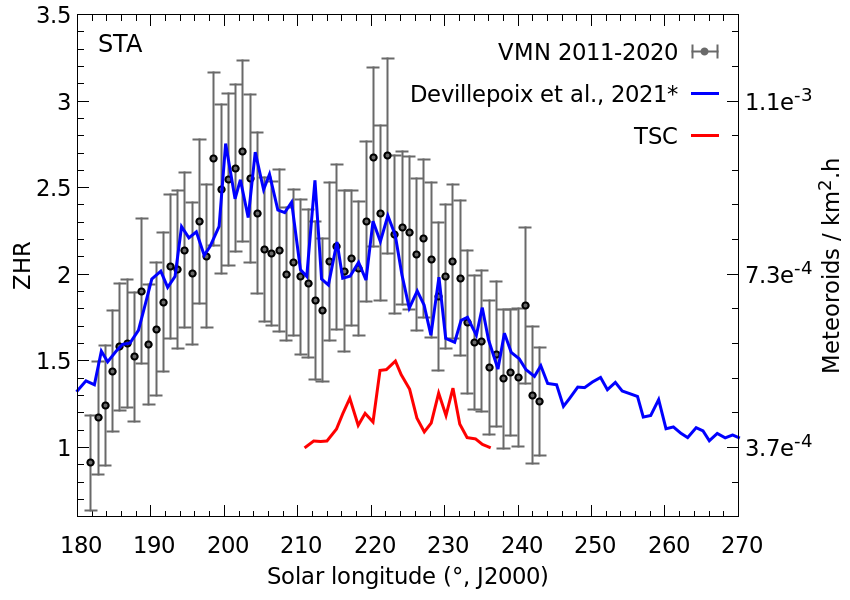}
 	\caption{\label{fig:video_profiles} Activity profiles from video (VMN) and photographic (DFN/CAMS) camera networks. \cite{Devillepoix2021} profiles based on CAMS and DFN records in 2015 were both obtained by converting the meteor rates to a ZHR estimate (with r=2.3) and adding 1 to the ZHR to match the VMN rates. The red curve corresponds to the 2015 return of the Taurid resonant swarm.  }
 \end{figure}

 Figures \ref{fig:visual_resonant_years} and \ref{fig:video_profiles} present the activity profiles of the autumn Taurid showers (NTA and STA) at each return of the resonant swarm years predicted by \cite{Asher1993b} since 1988. The plots compile the observations published in \cite{Roggemans1989}, \cite{Arlt2000}, \cite{Miskotte2006}, \cite{Johannink2006}, \cite{Dubietis2007}, and \cite{Devillepoix2021}, as well as measurements provided by the IMO VMN (see main text for details).\\
 
 \setfignumprefix{A1.}

\newpage
\textbf{A2 Activity profiles 2002 - 2021} \\

This section summarizes the annual activity profile of each apparition of the NTA, STA, BTA and ZPE since 2002 as observed with CMOR 29 MHz and 38 MHz systems (grey and black curves respectively). When available, the profiles are compared with the observations provided by the IMO VMDB (blue dots) and the IMO VMN (red dots). The standard IMO VMDB profiles were obtained assuming a constant population index of 2.3 over the STA and NTA activity period (see discussion in Section 2.1).\\

\textbf{NTA} \\

\setfignumprefix{A2.}
\begin{figure}[!ht]
 	\centering
 	\includegraphics[width=.45\textwidth]{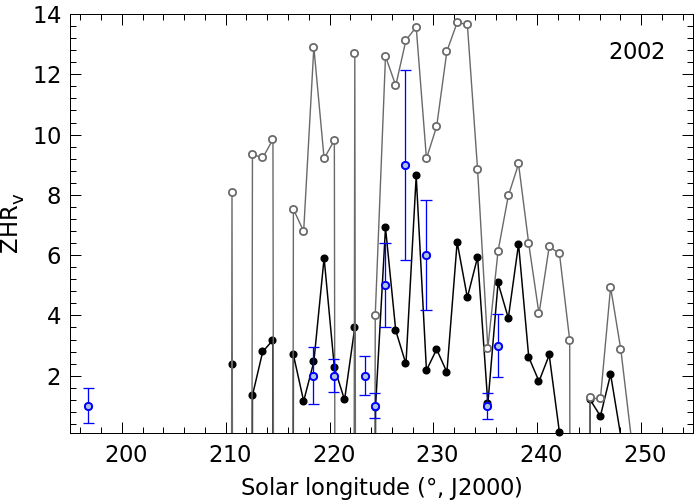}
 	\includegraphics[width=.45\textwidth]{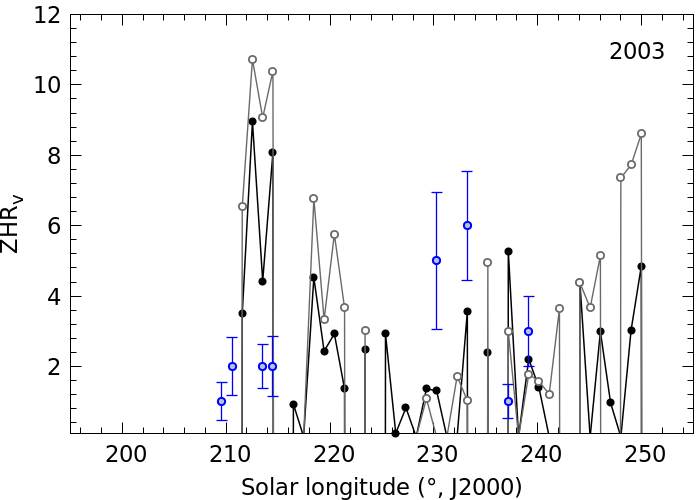}\\
 	\includegraphics[width=.45\textwidth]{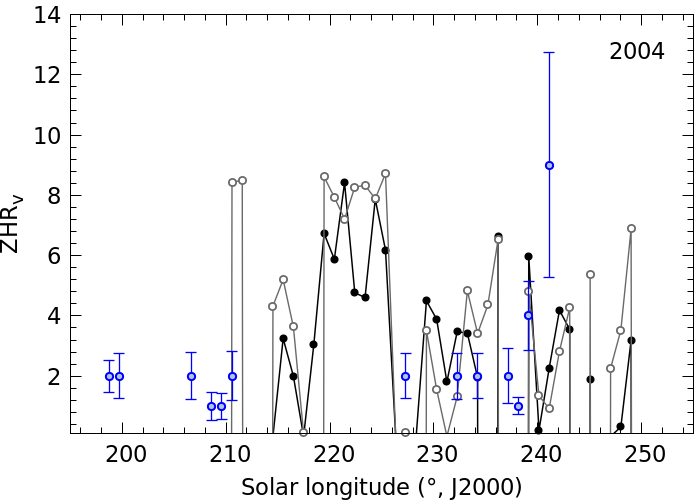}
 	\includegraphics[width=.45\textwidth]{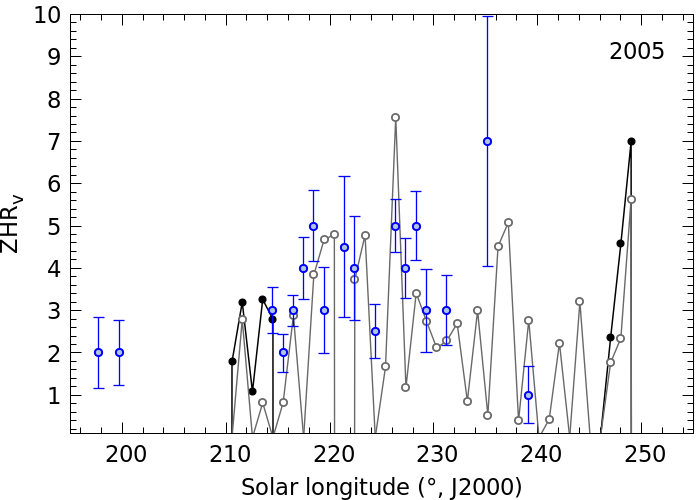}
 	\caption{ Activity profiles of the NTA between 2002 and 2005. Blue dots: IMO VMDB, black line: CMOR 38 MHz, grey line: CMOR 29 MHz.}
 \end{figure}
 
\begin{figure}[!ht]
 	\centering
 	\includegraphics[width=.45\textwidth]{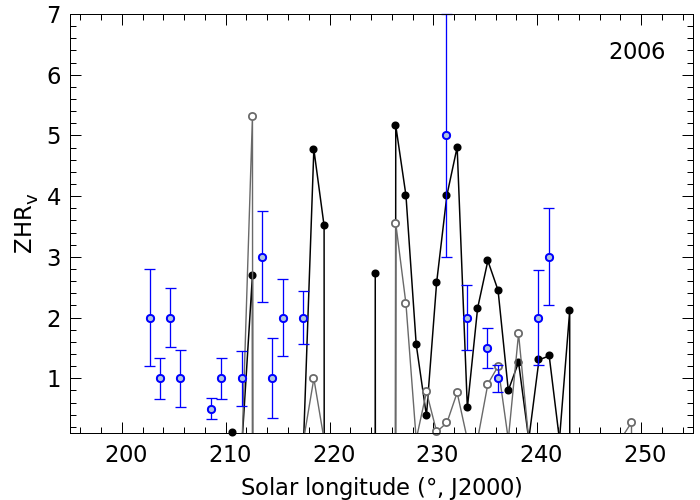}
 	\includegraphics[width=.45\textwidth]{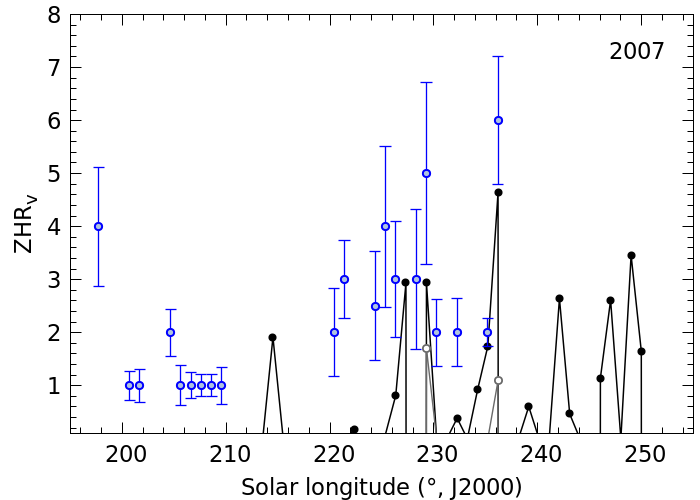}\\
 	\includegraphics[width=.45\textwidth]{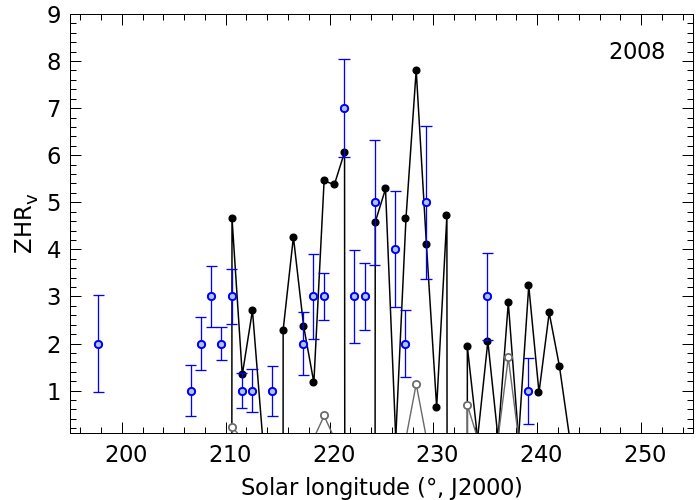}
 	\includegraphics[width=.45\textwidth]{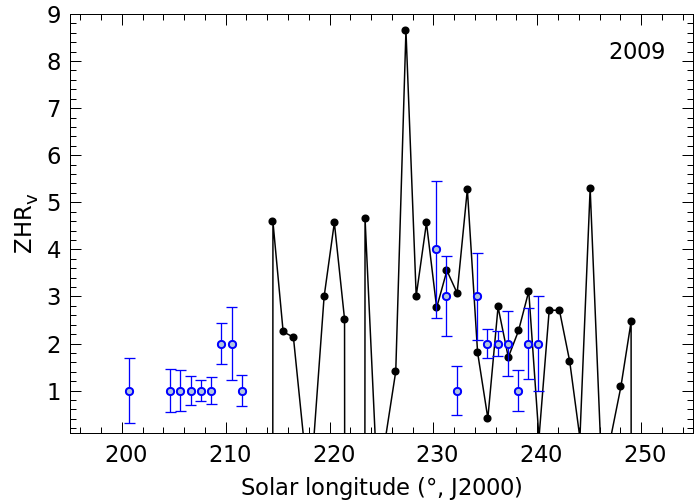}\\
	\includegraphics[width=.45\textwidth]{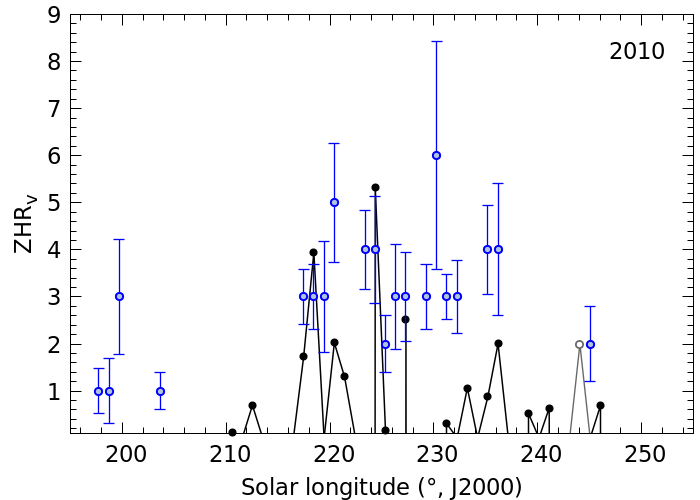}
	\includegraphics[width=.45\textwidth]{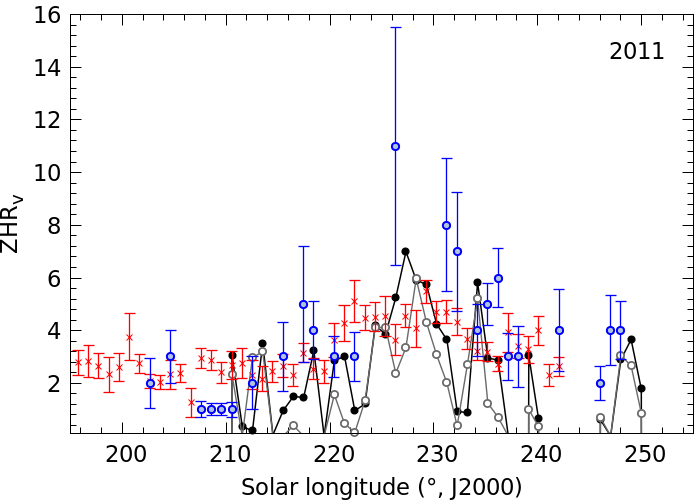}
	\includegraphics[width=.45\textwidth]{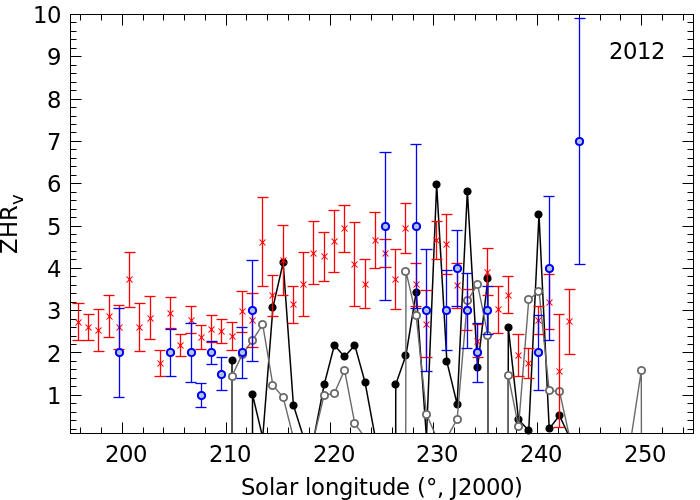}
	\includegraphics[width=.45\textwidth]{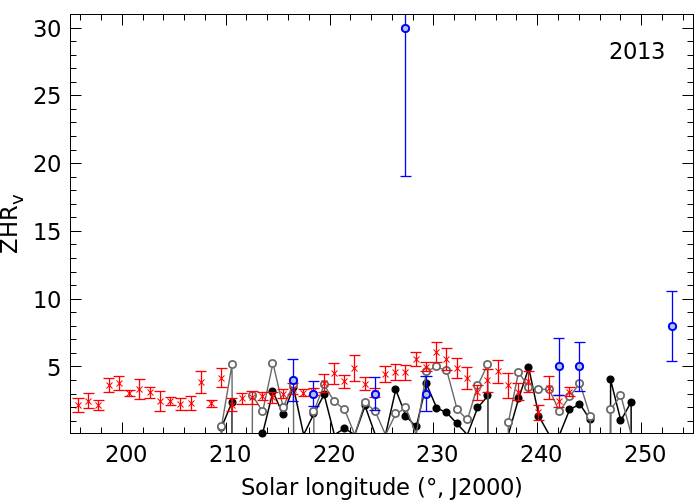}
 	\caption{ Activity profiles of the NTA between 2006 and 2013. Blue dots: IMO VMDB, black line: CMOR 38 MHz, grey line: CMOR 29 MHz, red dots: IMO VMN.}
 \end{figure}
 
\newpage
\begin{figure}[!ht]
	\centering
	\includegraphics[width=.45\textwidth]{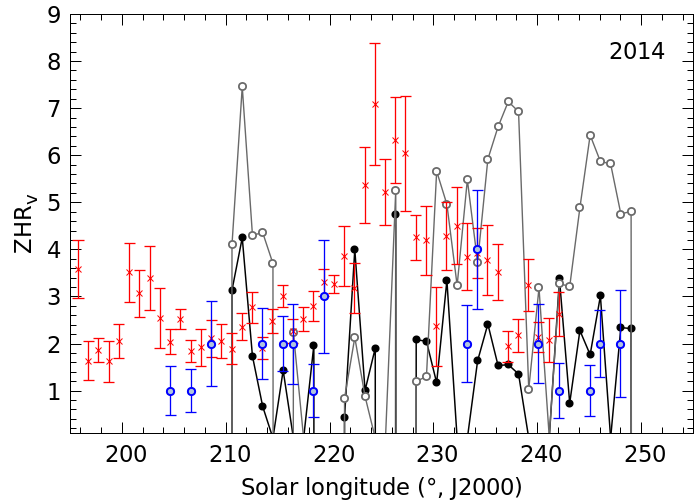}
	\includegraphics[width=.45\textwidth]{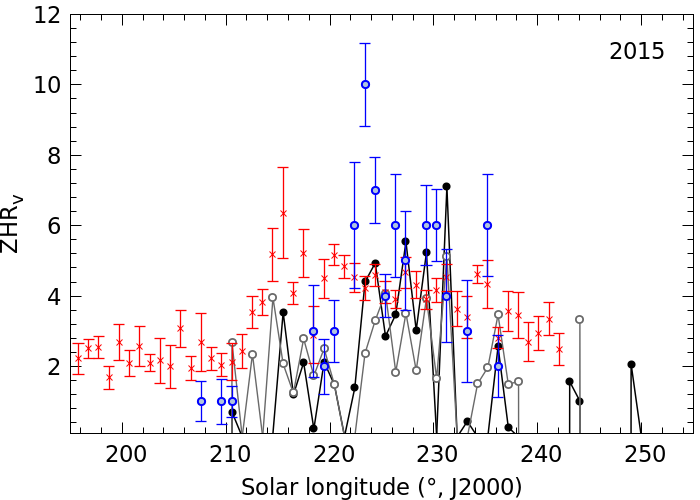}
	\includegraphics[width=.45\textwidth]{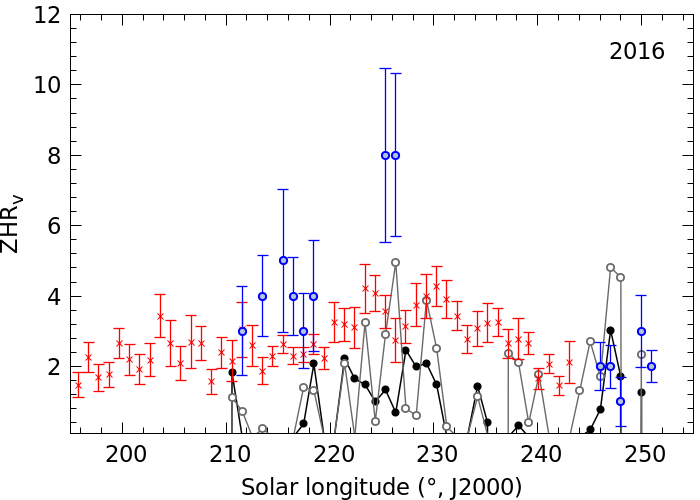}
	\includegraphics[width=.45\textwidth]{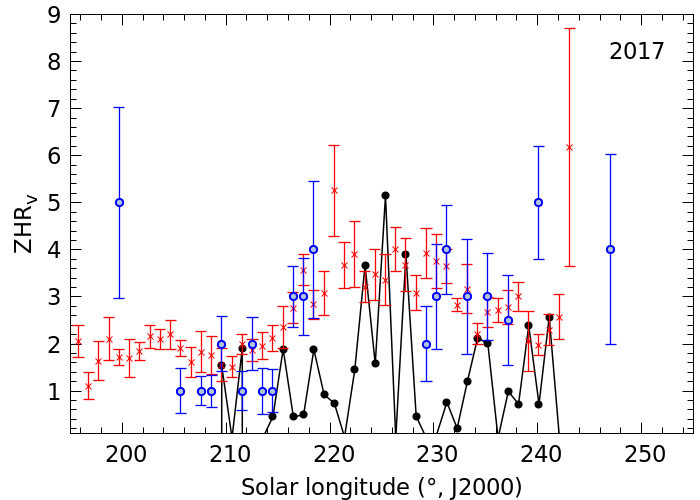}
	\includegraphics[width=.45\textwidth]{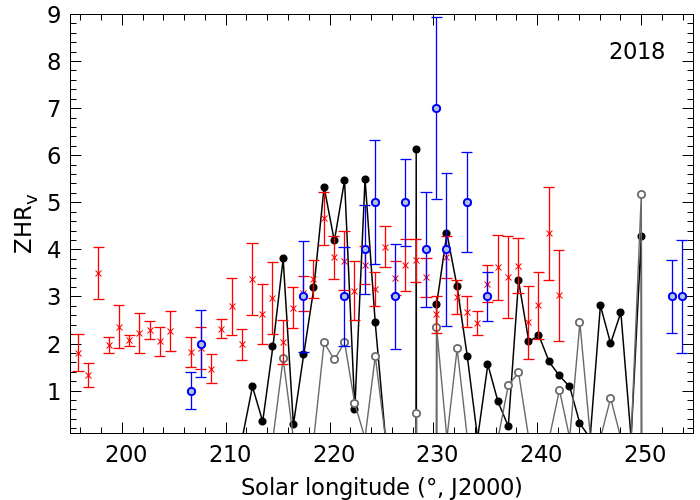}
	\includegraphics[width=.45\textwidth]{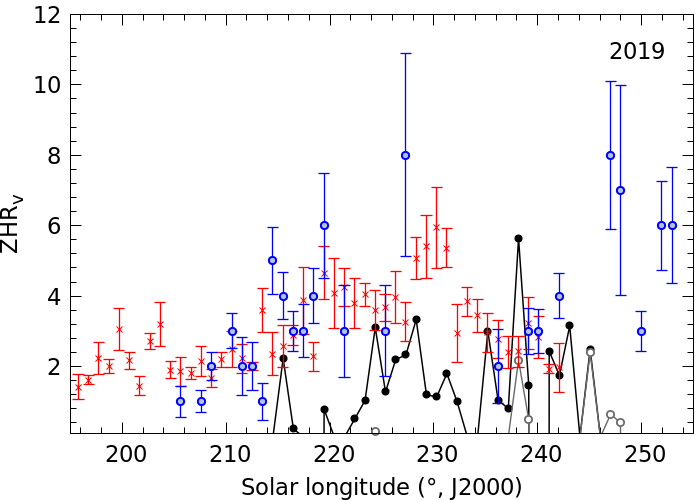}
	\includegraphics[width=.45\textwidth]{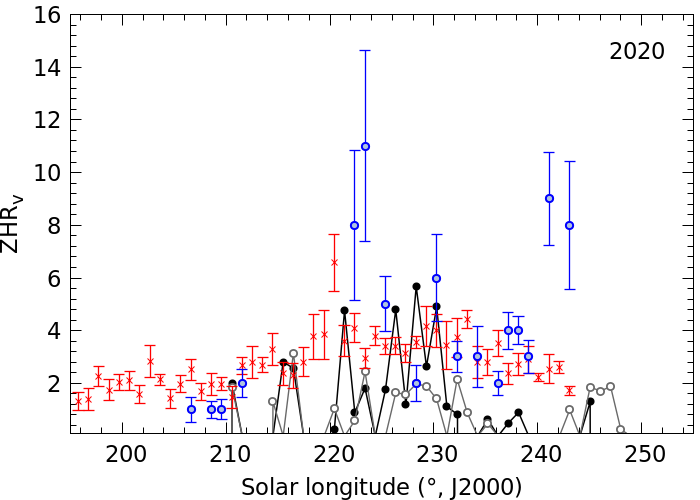}
	\caption{Activity profiles of the NTA between 2014 and 2020. Blue dots: IMO VMDB, red dots: IMO VMN, black line: CMOR 38 MHz, grey line: CMOR 29 MHz.}
\end{figure}

\newpage
\clearpage
\textbf{STA}\\

\begin{figure}[!ht]
	\centering
	\includegraphics[width=.45\textwidth]{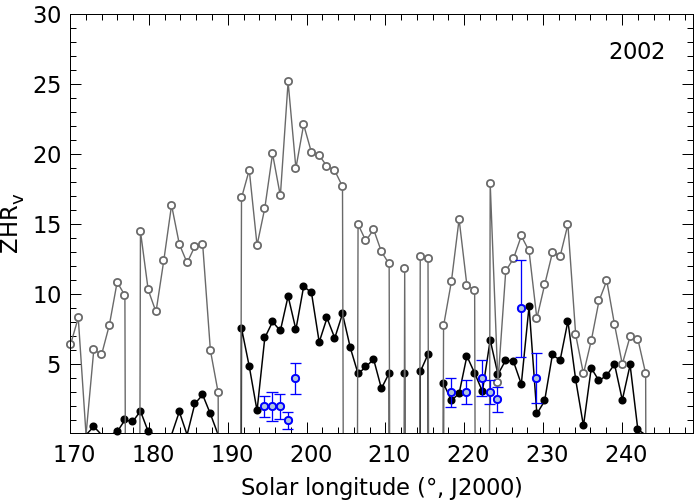}
	\includegraphics[width=.45\textwidth]{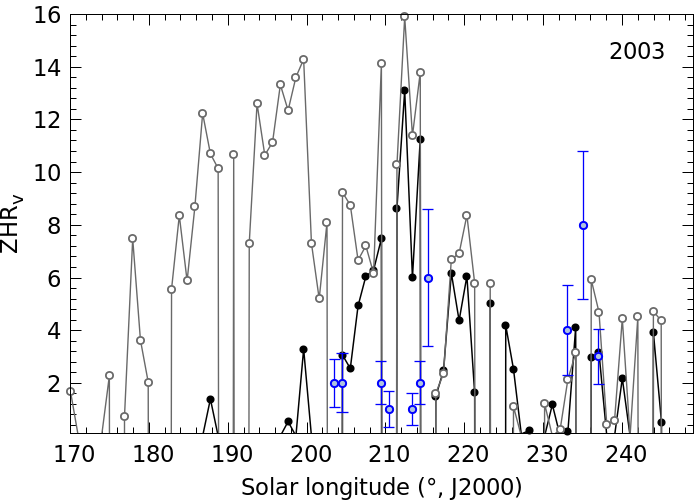}
	\includegraphics[width=.45\textwidth]{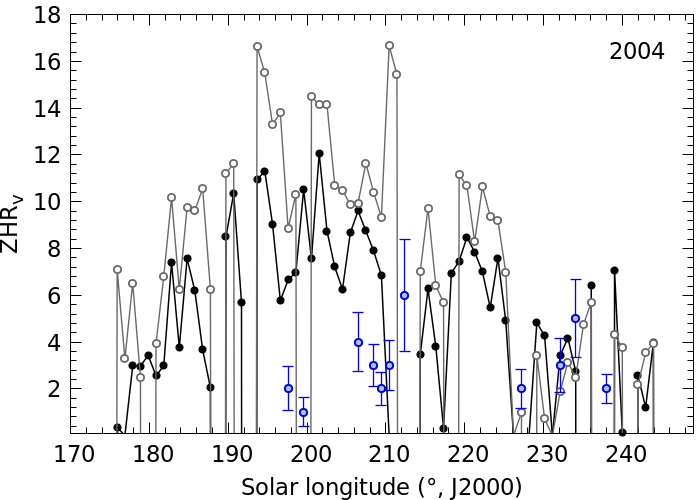}
	\includegraphics[width=.45\textwidth]{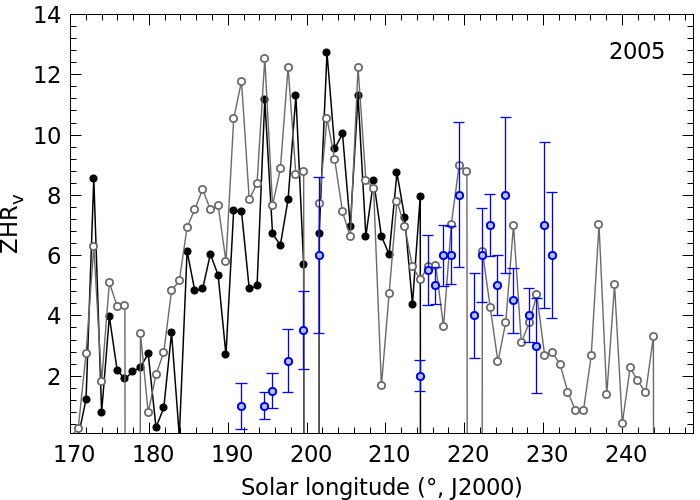}
	\includegraphics[width=.45\textwidth]{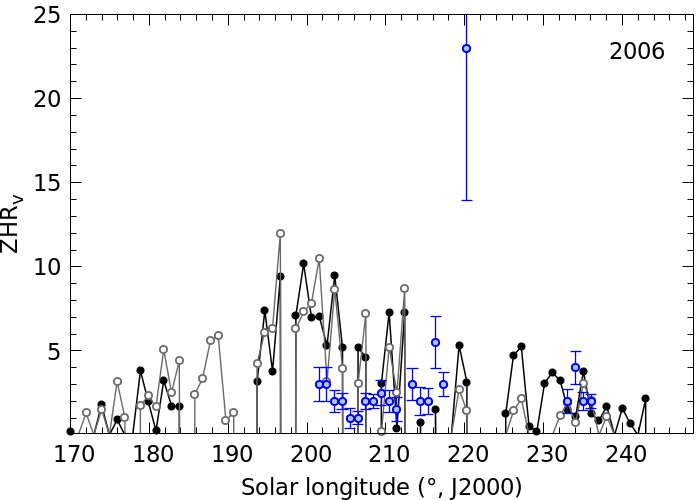}
	\includegraphics[width=.45\textwidth]{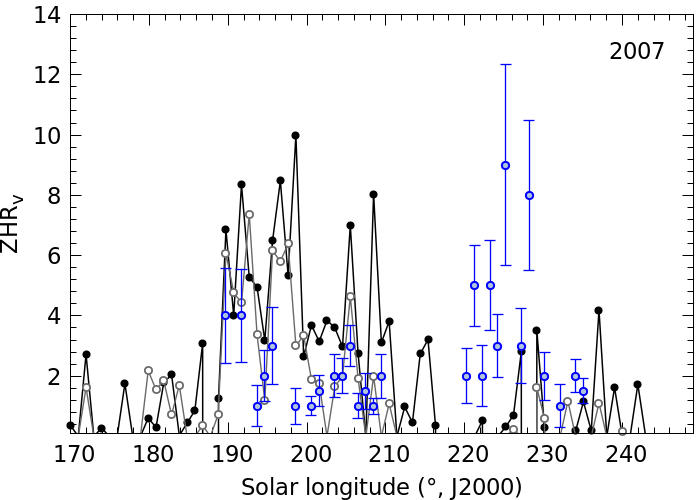}
	\includegraphics[width=.45\textwidth]{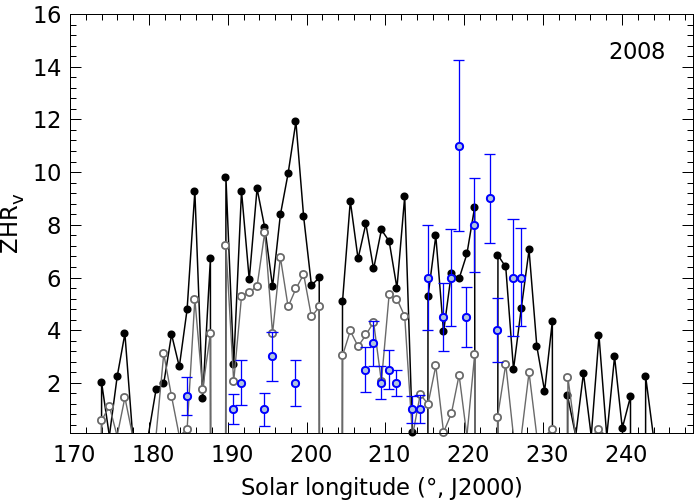}
	\includegraphics[width=.45\textwidth]{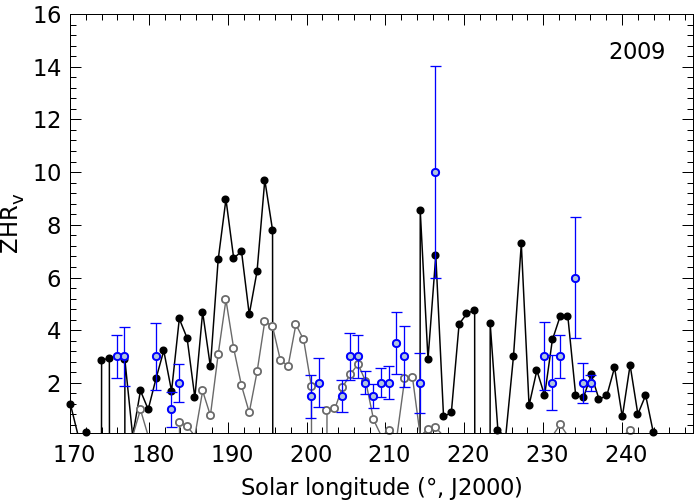}
	\caption{Activity profiles of the STA between 2002 and 2009. Blue dots: IMO VMDB, black line: CMOR 38 MHz, grey line: CMOR 29 MHz.}
\end{figure}

\begin{figure}[!ht]
	\centering
	\includegraphics[width=.45\textwidth]{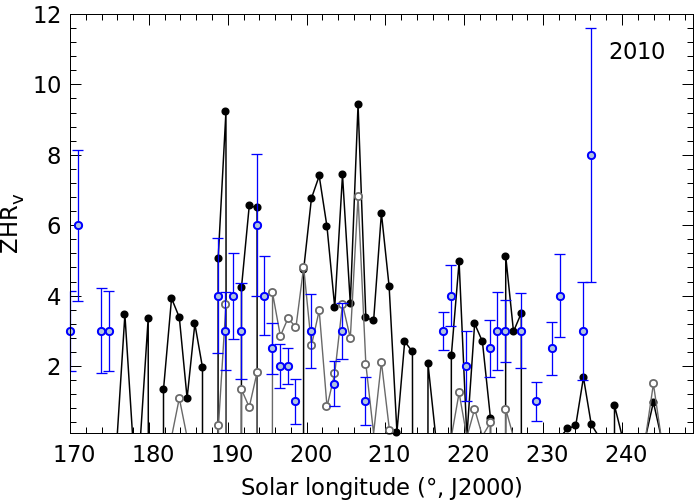}
	\includegraphics[width=.45\textwidth]{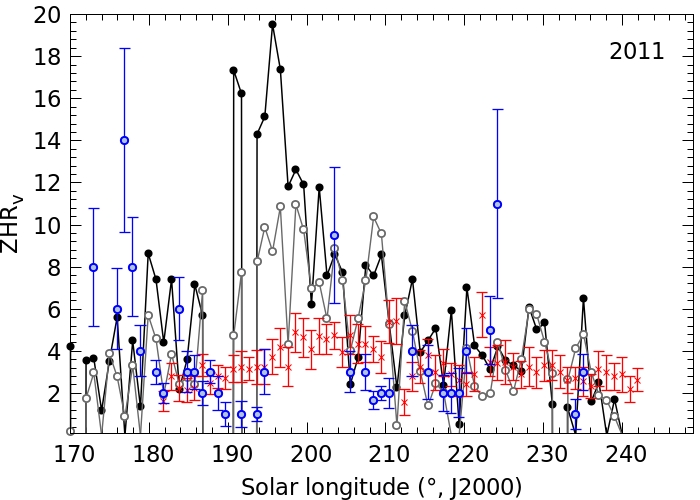}
	\includegraphics[width=.45\textwidth]{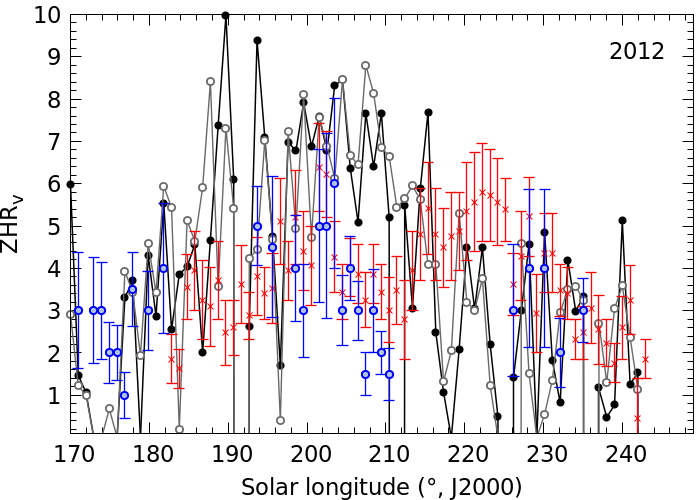}
	\includegraphics[width=.45\textwidth]{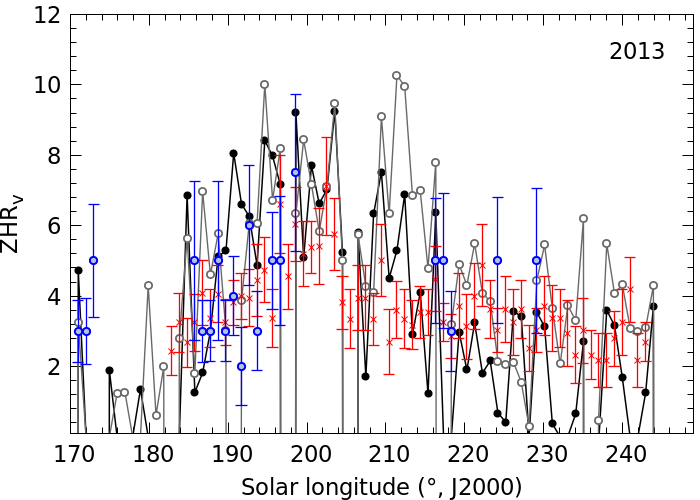}
	\includegraphics[width=.45\textwidth]{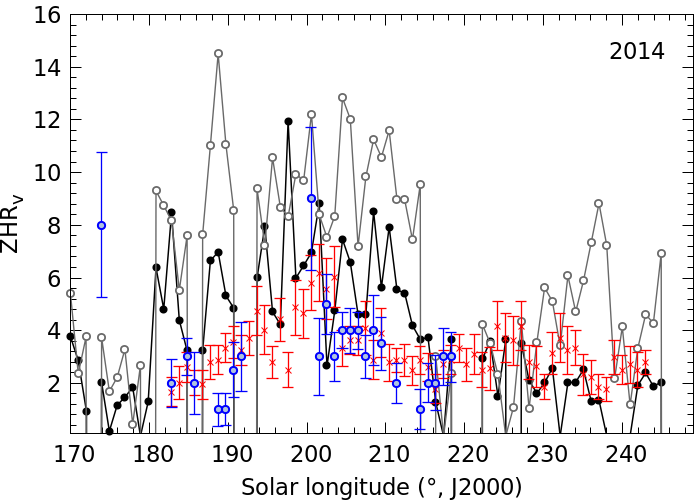}
	\includegraphics[width=.45\textwidth]{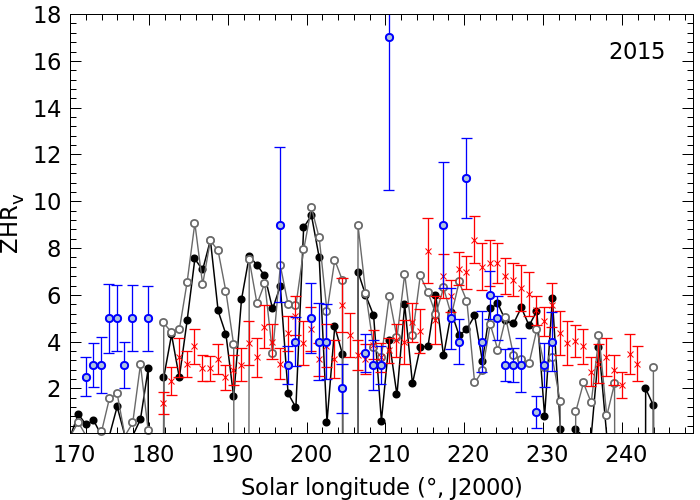}
	\includegraphics[width=.45\textwidth]{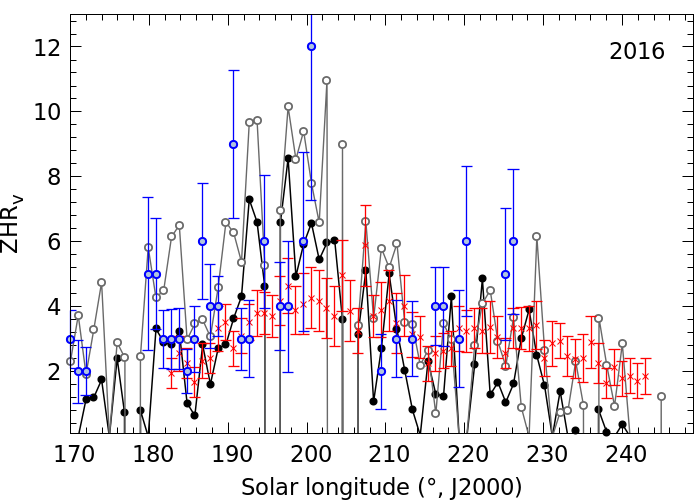}
	\includegraphics[width=.45\textwidth]{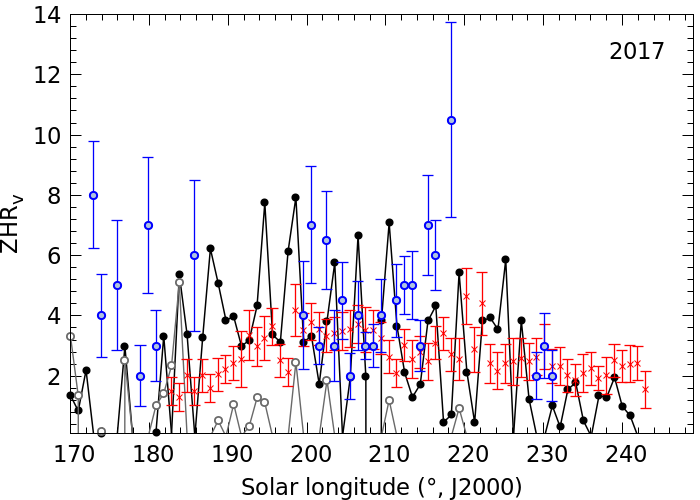}
	\caption{Activity profiles of the STA between 2010 and 2017. Blue dots: IMO VMDB, red dots: IMO VMN, black line: CMOR 38 MHz, grey line: CMOR 29 MHz.}
\end{figure}

\begin{figure}[!ht]
	\centering
	\includegraphics[width=.45\textwidth]{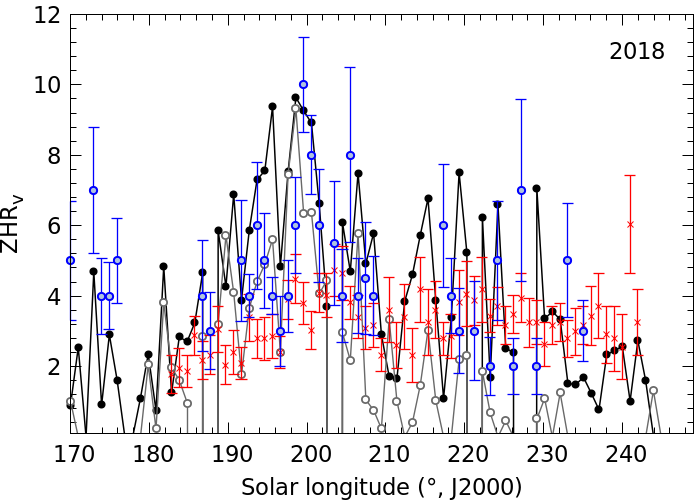}
	\includegraphics[width=.45\textwidth]{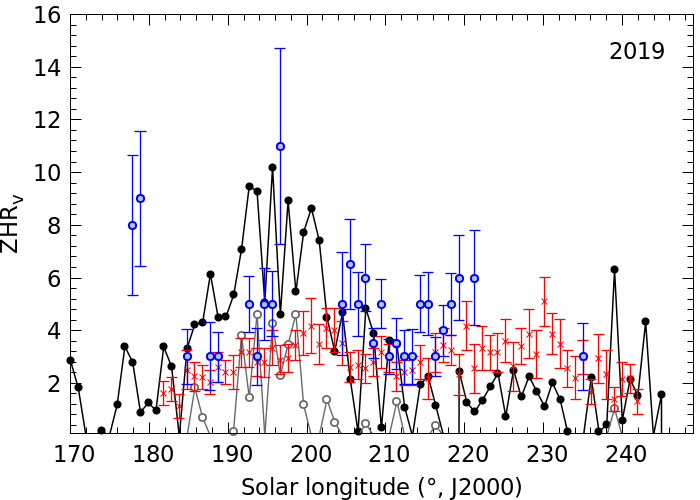}
	\includegraphics[width=.45\textwidth]{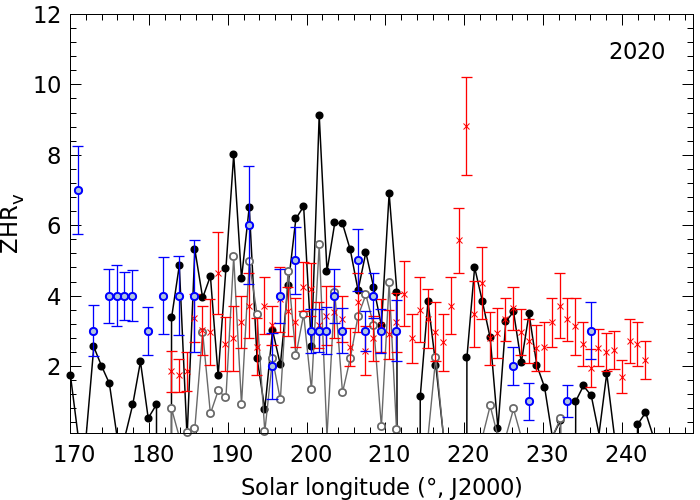}
	\caption{Activity profiles of the STA between 2018 and 2020. Blue dots: IMO VMDB, red dots: IMO VMN, black line: CMOR 38 MHz, grey line: CMOR 29 MHz.}
\end{figure}

\newpage
\textbf{BTA}\\

\begin{figure}[!ht]
	\centering
	\includegraphics[width=.45\textwidth]{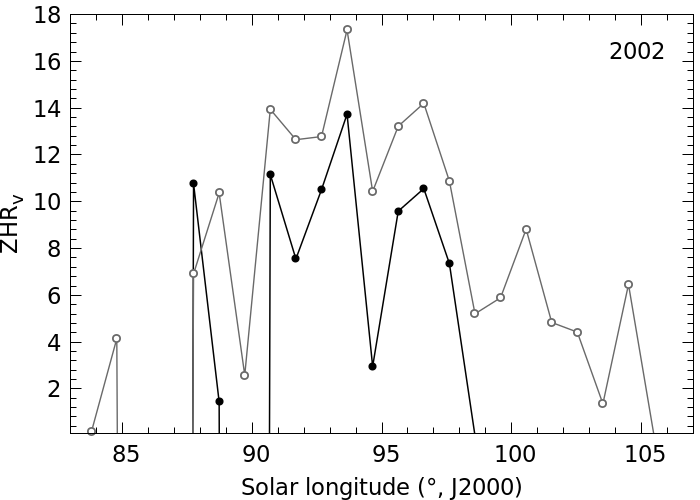}
	\includegraphics[width=.45\textwidth]{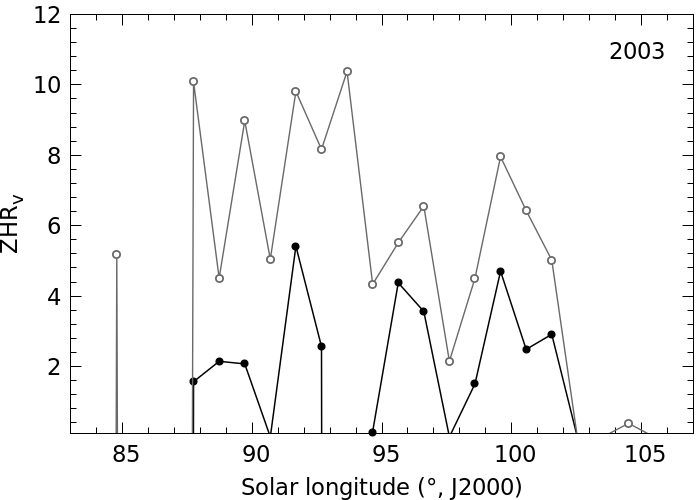}
	\caption{Activity profiles of the BTA in 2002 and 2003, as measured from single station echo rates from the CMOR 29 MHz (grey line) or 38 MHz (black line) systems.}
\end{figure}

\begin{figure}[!ht]
	\centering
	\includegraphics[width=.45\textwidth]{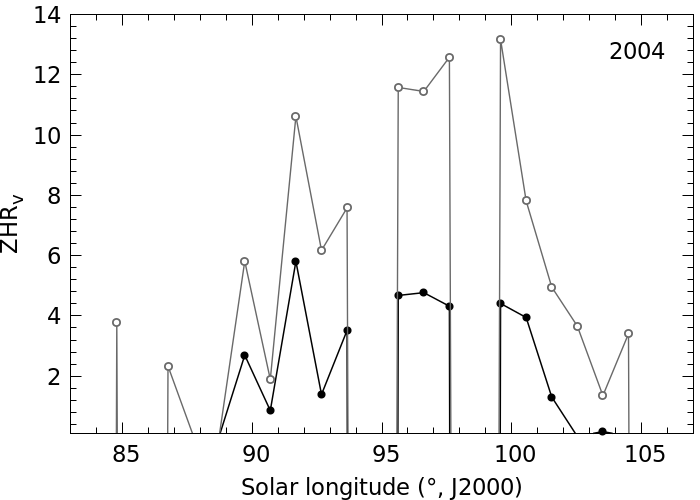}
	\includegraphics[width=.45\textwidth]{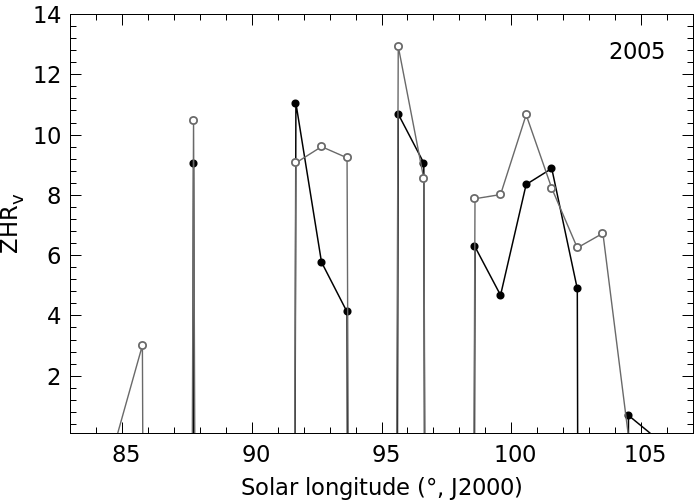}
	\includegraphics[width=.45\textwidth]{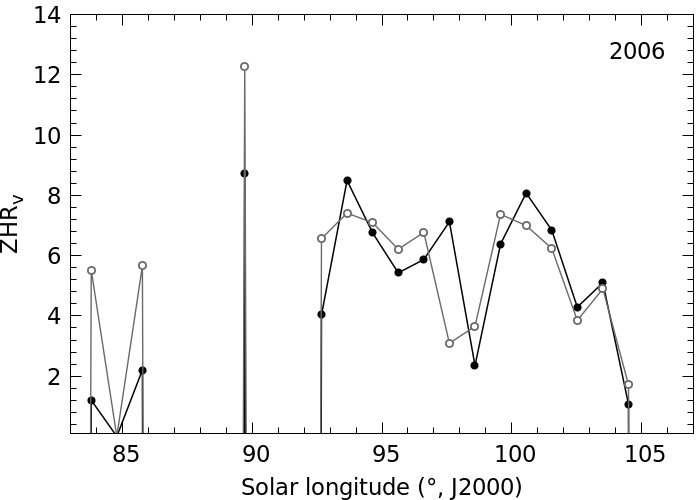}
	\includegraphics[width=.45\textwidth]{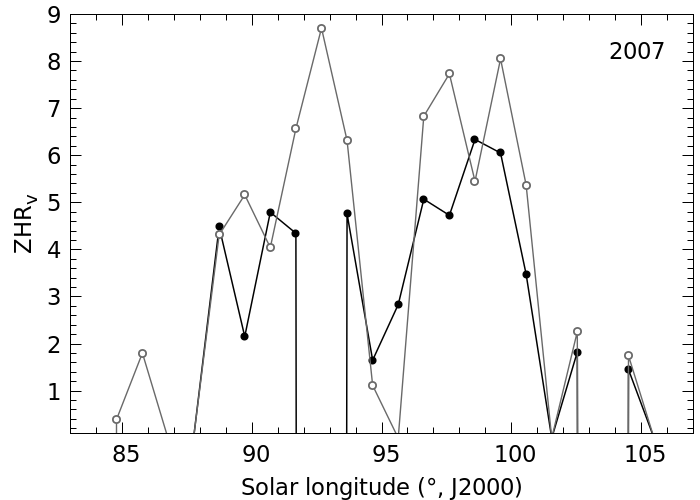}
	\includegraphics[width=.45\textwidth]{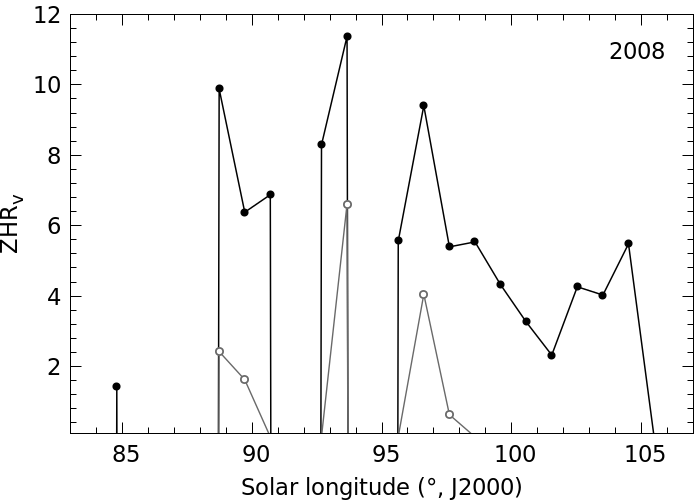}
	\includegraphics[width=.45\textwidth]{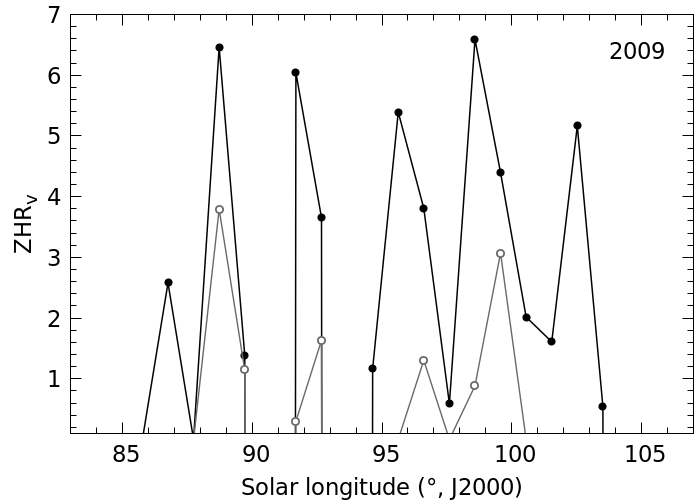}
	\includegraphics[width=.45\textwidth]{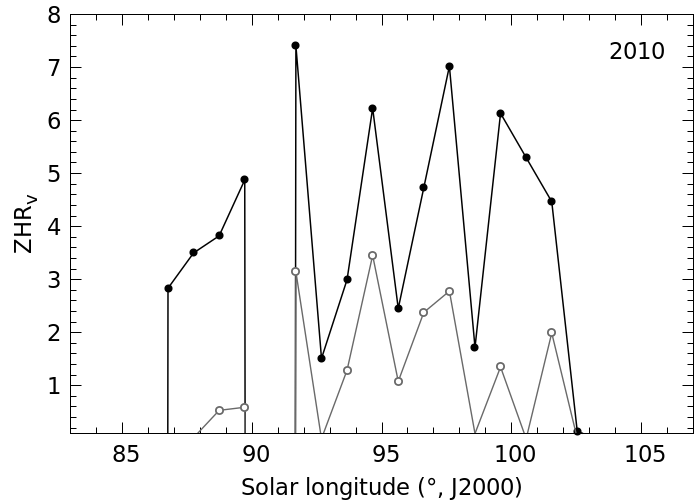}
	\includegraphics[width=.45\textwidth]{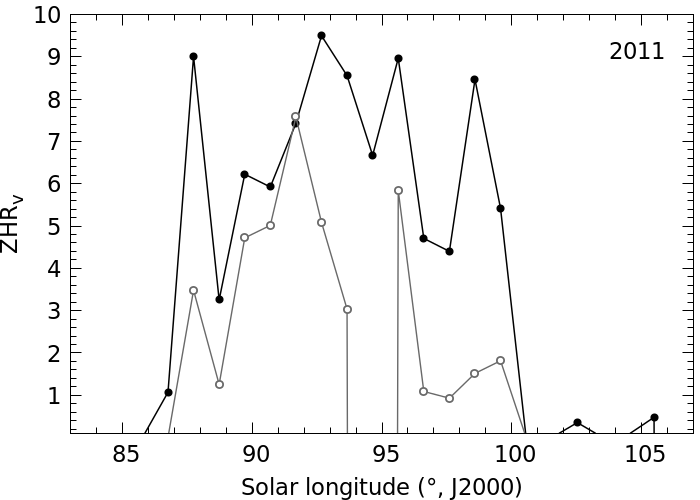}
	\caption{Activity profiles of the BTA between 2004 and 2011, as measured from single station echo rates for the CMOR 29 MHz (grey line) or 38 MHz (black line) systems.}
\end{figure}

\begin{figure}[!ht]
	\centering
	\includegraphics[width=.45\textwidth]{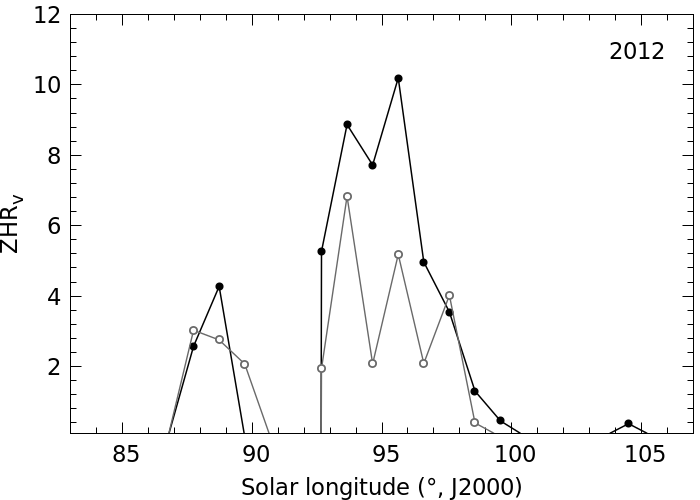}
	\includegraphics[width=.45\textwidth]{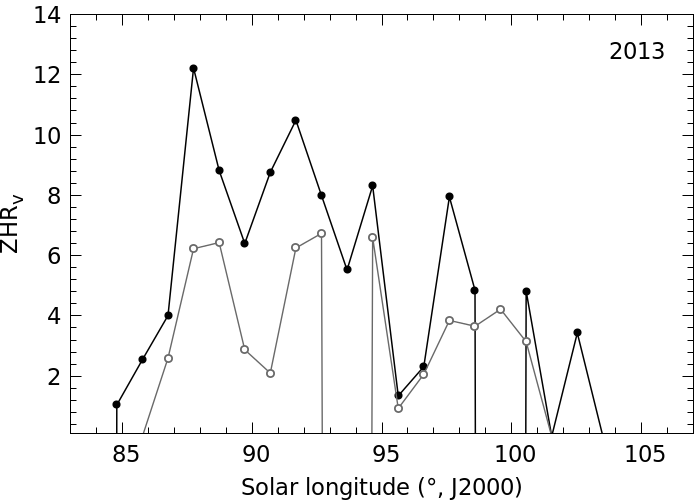}
	\includegraphics[width=.45\textwidth]{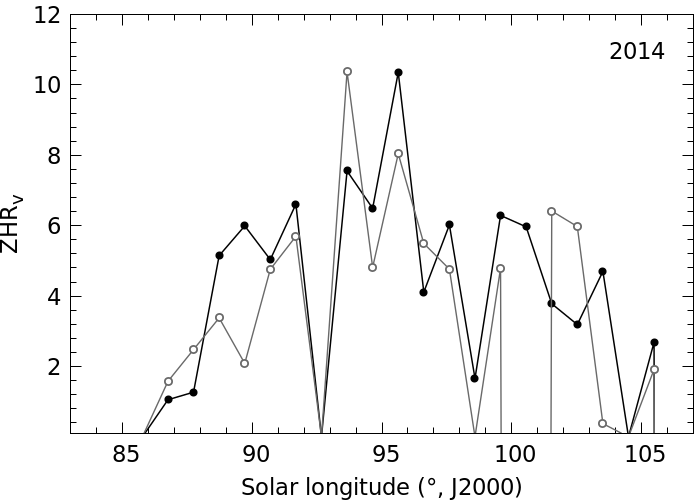}
	\includegraphics[width=.45\textwidth]{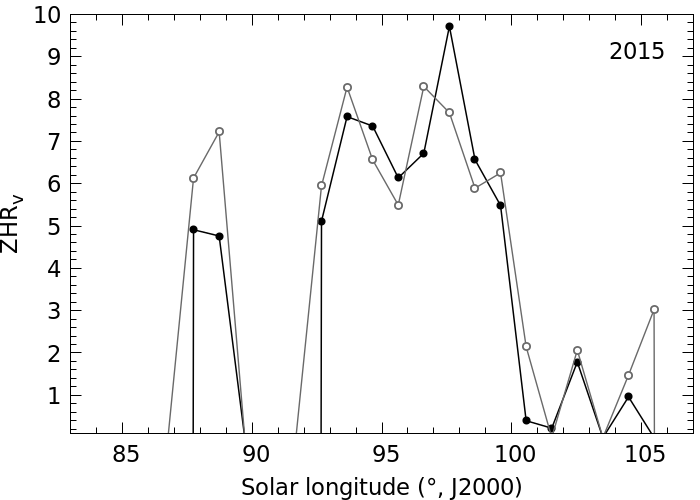}
	\includegraphics[width=.45\textwidth]{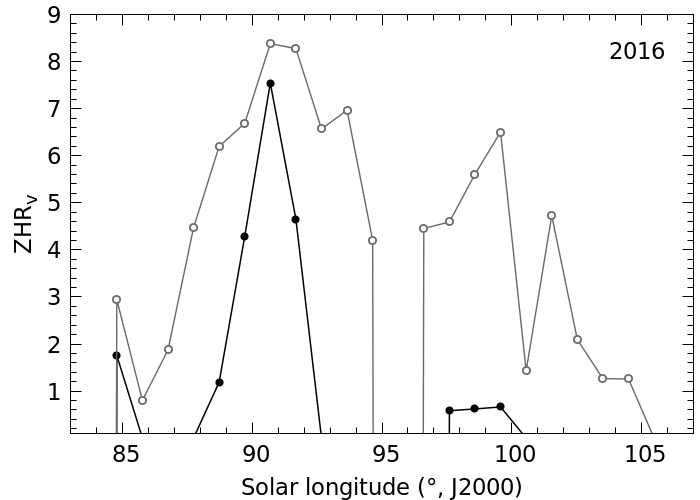}
	\includegraphics[width=.45\textwidth]{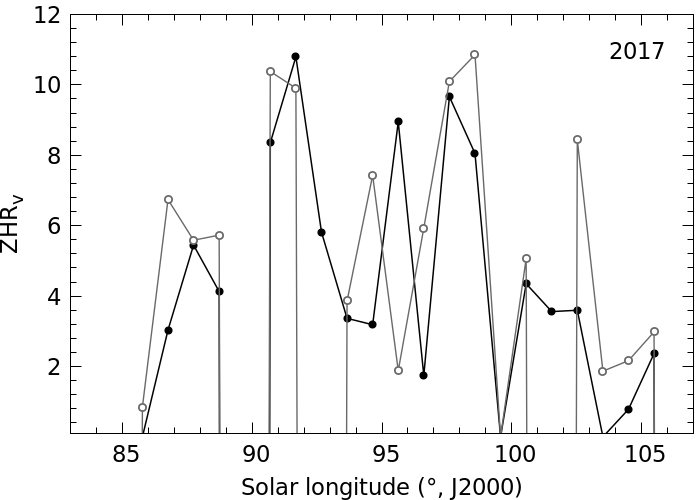}
	\includegraphics[width=.45\textwidth]{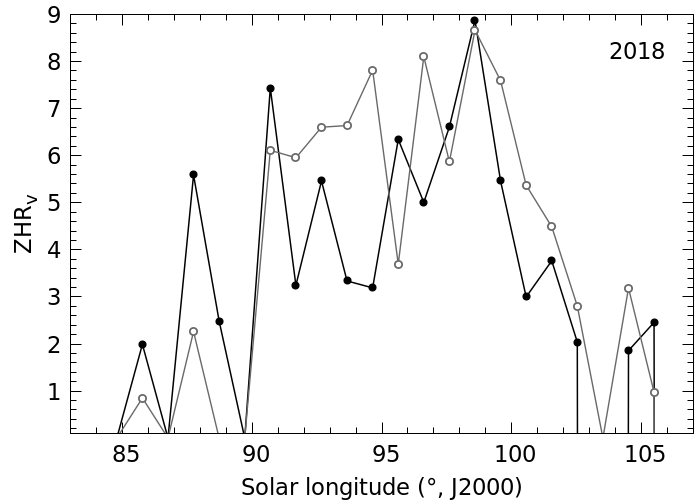}
	\includegraphics[width=.45\textwidth]{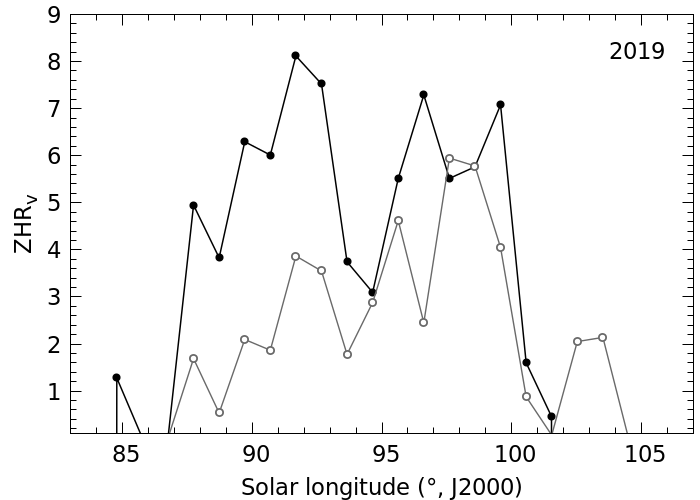}
	\caption{Activity profiles of the BTA between 2012 and 2019, as measured from single station echo rates for the CMOR 29 MHz (grey line) or 38 MHz (black line) systems.}
\end{figure}

\newpage
\clearpage
\begin{figure}[!ht]
	\centering
	\includegraphics[width=.45\textwidth]{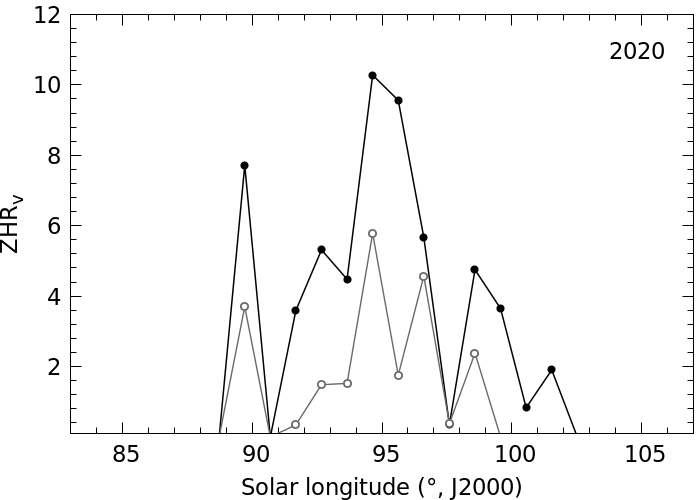}
	\includegraphics[width=.45\textwidth]{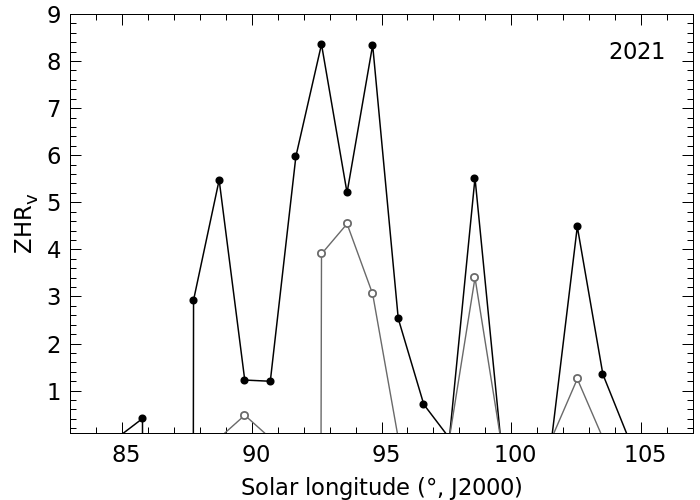}
	\caption{Activity profiles of the BTA in 2020 and 2021, as measured from single station echo rates for the CMOR 29 MHz (grey line) or 38 MHz (black line) systems.}
\end{figure}

\textbf{ZPE}\\

\begin{figure}[!ht]
	\centering
	\includegraphics[width=.45\textwidth]{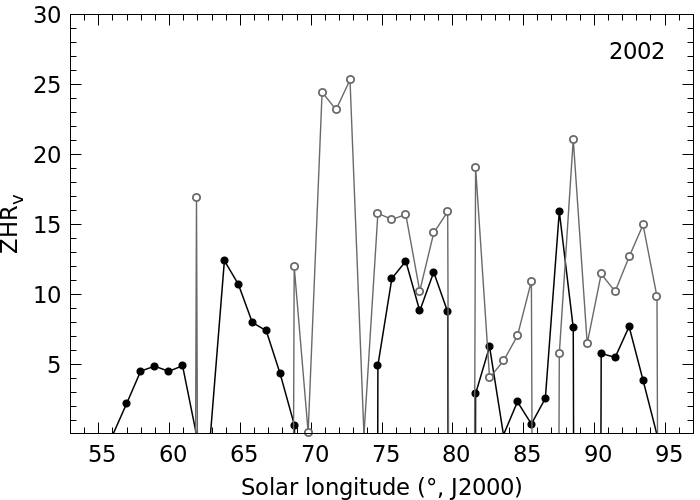}
	\includegraphics[width=.45\textwidth]{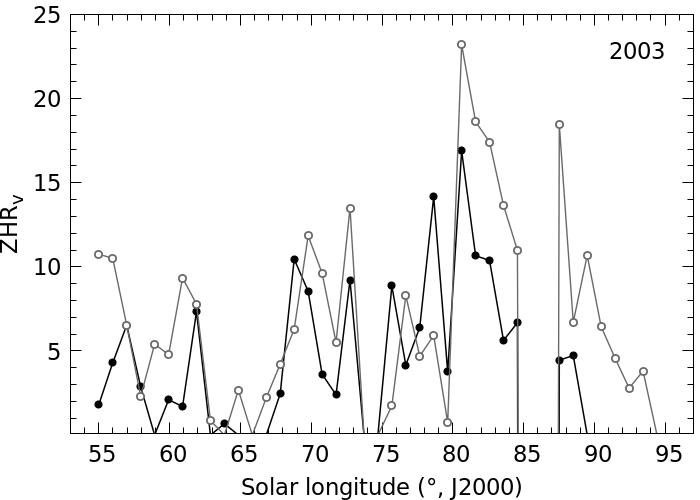}
	\includegraphics[width=.45\textwidth]{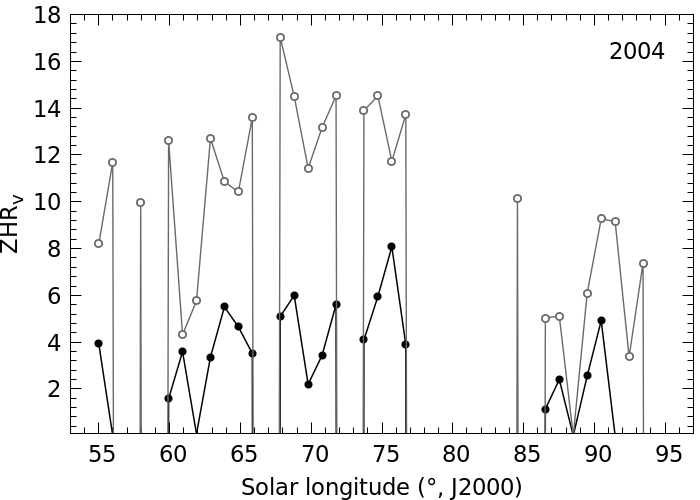}
	\includegraphics[width=.45\textwidth]{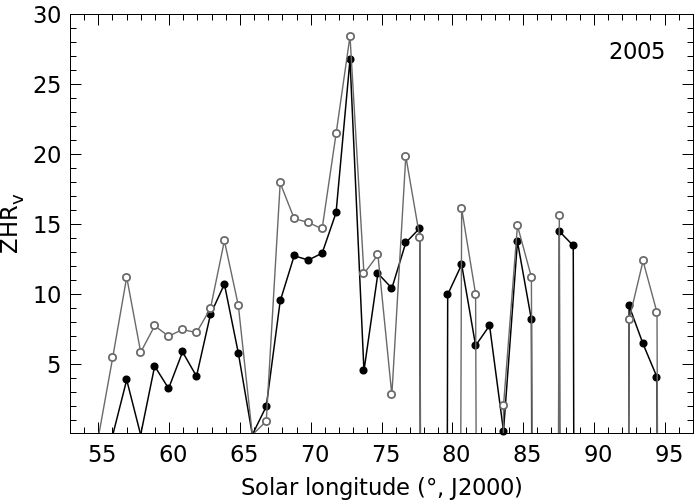}
	\caption{Activity profiles of the ZPE between 2002 and 2005, as measured from single station echo rates for single station echo rates for the CMOR 29 MHz (grey line) or 38 MHz (black line) systems.}
\end{figure}

\begin{figure}[!ht]
	\centering
	\includegraphics[width=.45\textwidth]{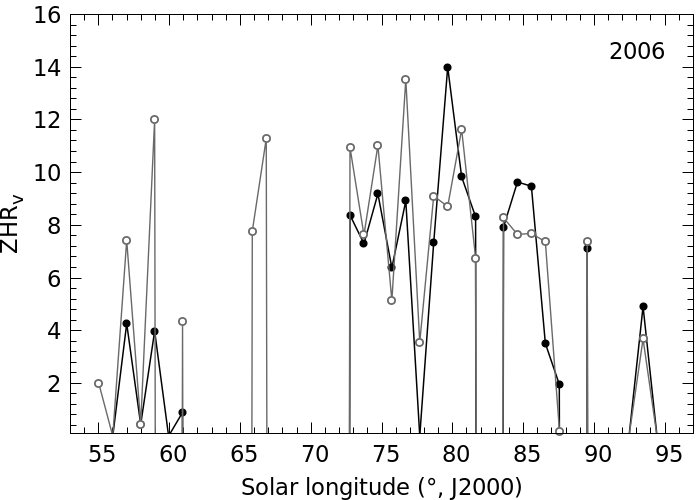}
	\includegraphics[width=.45\textwidth]{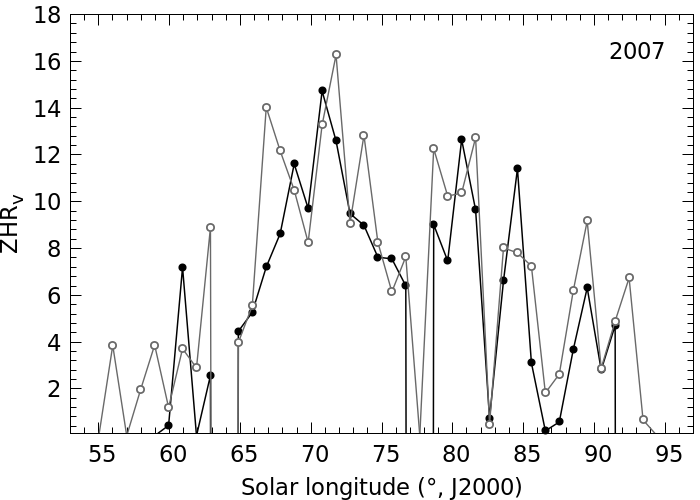}
	\includegraphics[width=.45\textwidth]{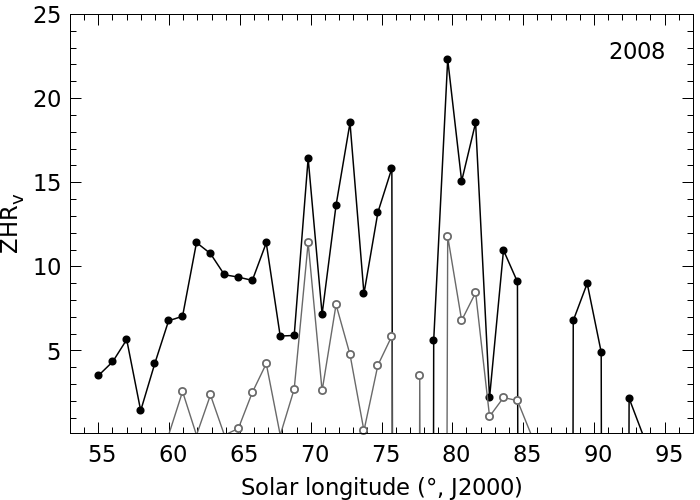}
	\includegraphics[width=.45\textwidth]{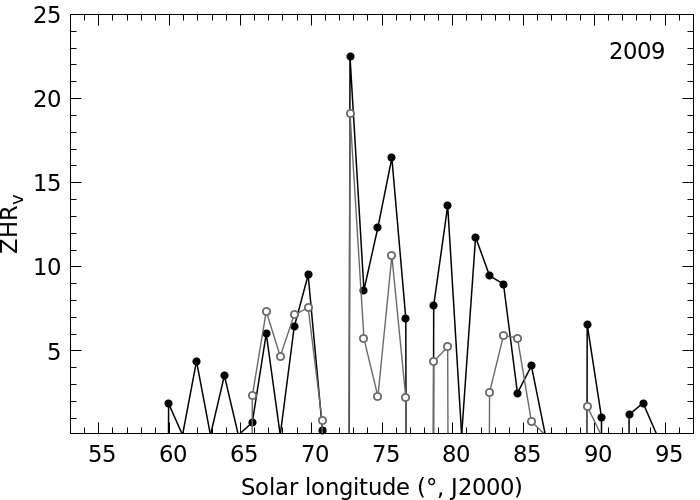}
	\includegraphics[width=.45\textwidth]{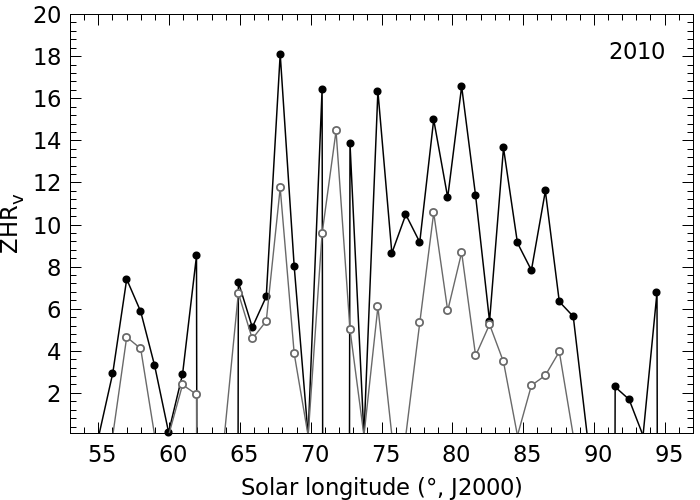}
	\includegraphics[width=.45\textwidth]{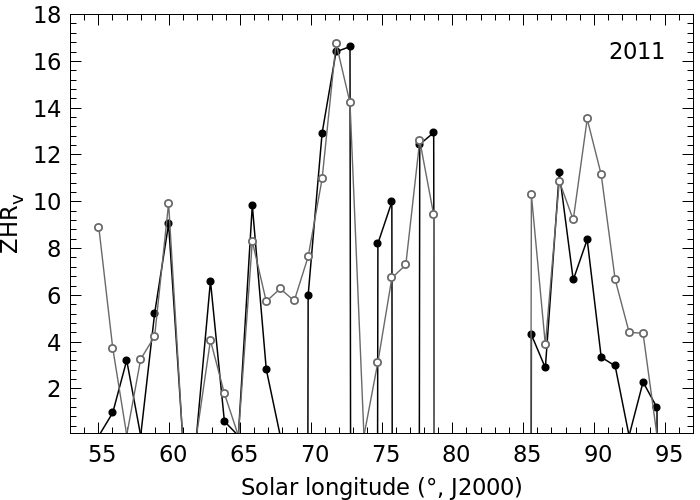}
	\includegraphics[width=.45\textwidth]{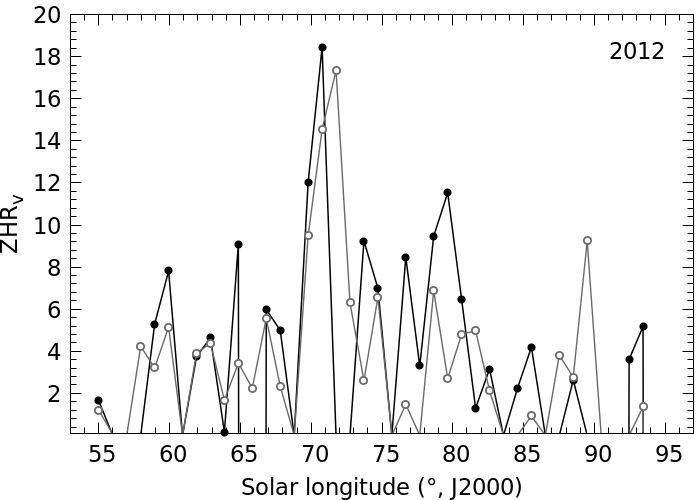}
	\includegraphics[width=.45\textwidth]{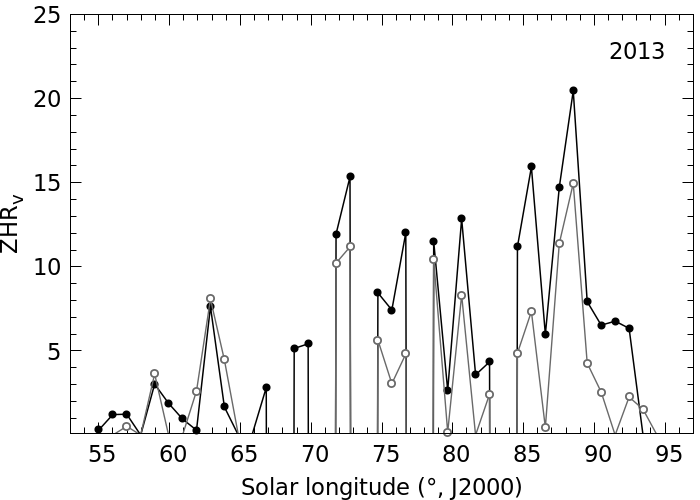}
	\caption{Activity profiles of the ZPE between 2006 and 2013, as measured from single station echo rates for the CMOR 29 MHz (grey line) or 38 MHz (black line) systems.}
\end{figure}

\begin{figure}[!ht]
	\centering
	\includegraphics[width=.45\textwidth]{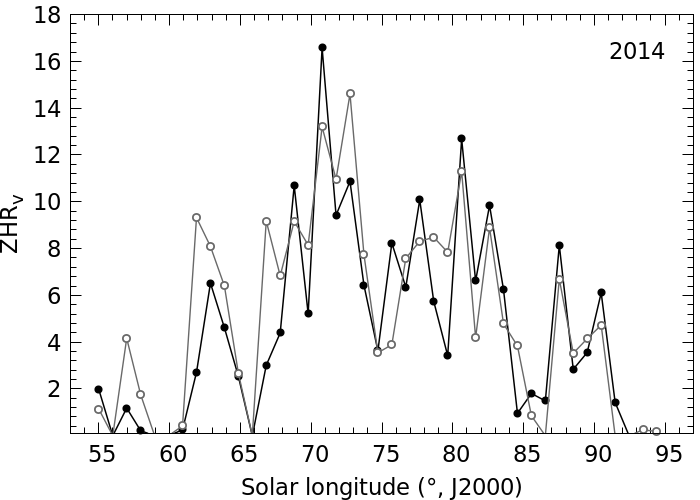}
	\includegraphics[width=.45\textwidth]{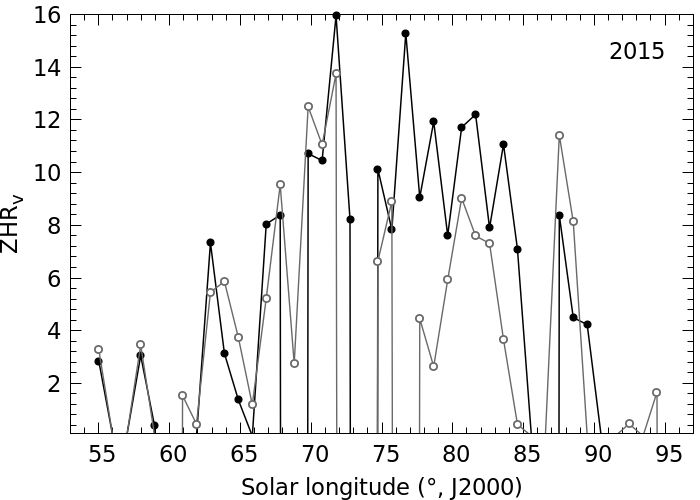}
	\includegraphics[width=.45\textwidth]{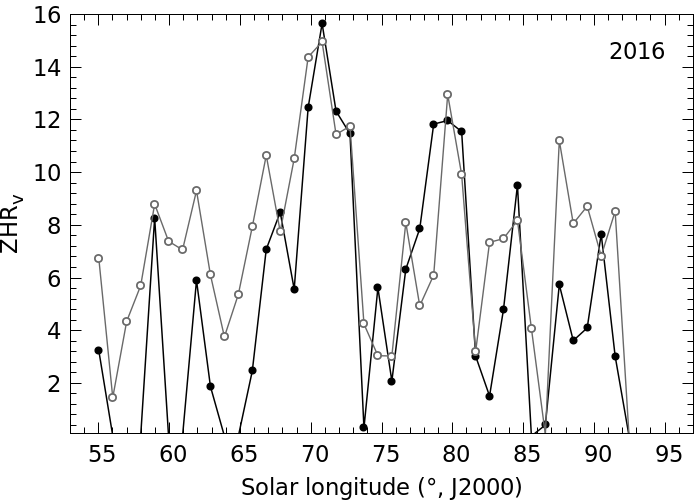}
	\includegraphics[width=.45\textwidth]{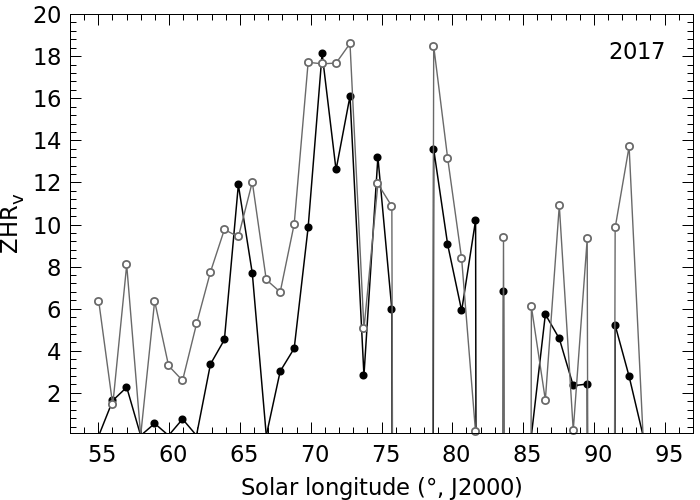}
	\includegraphics[width=.45\textwidth]{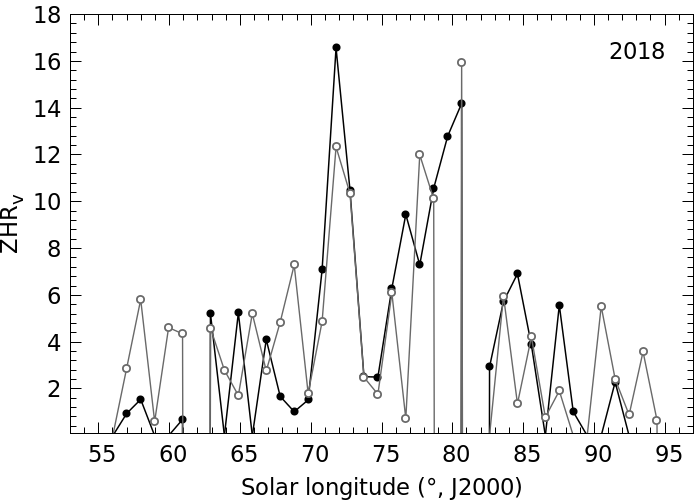}
	\includegraphics[width=.45\textwidth]{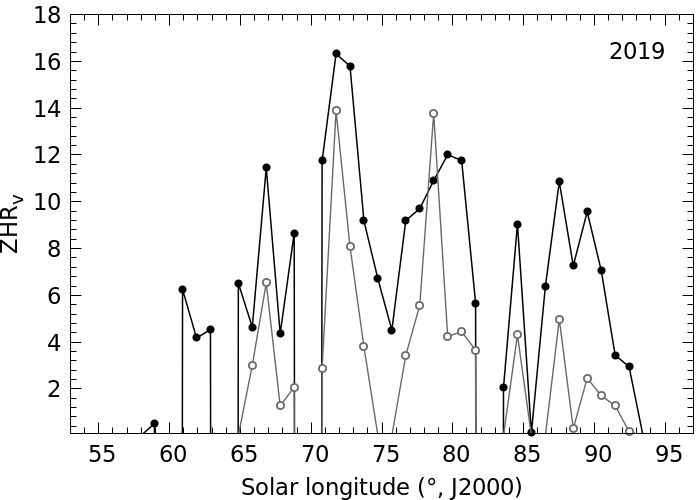}
	\includegraphics[width=.45\textwidth]{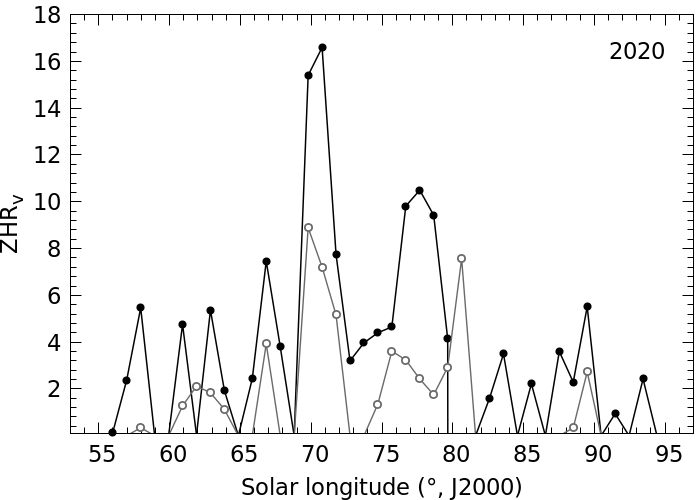}
	\includegraphics[width=.45\textwidth]{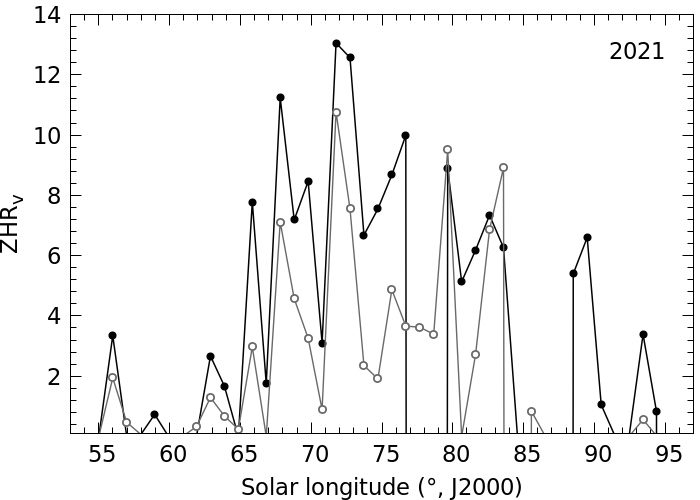}
	\caption{Activity profiles of the ZPE between 2014 and 2021, as measured from  from single station echo rates for the CMOR 29 MHz (grey line) or 38 MHz (black line) systems.}
\end{figure}

\clearpage 
\newpage

\textbf{A3 Maximum annual activity measured by CMOR} \\

\begin{figure}[!ht]
     \centering
     \includegraphics[width=.45\textwidth]{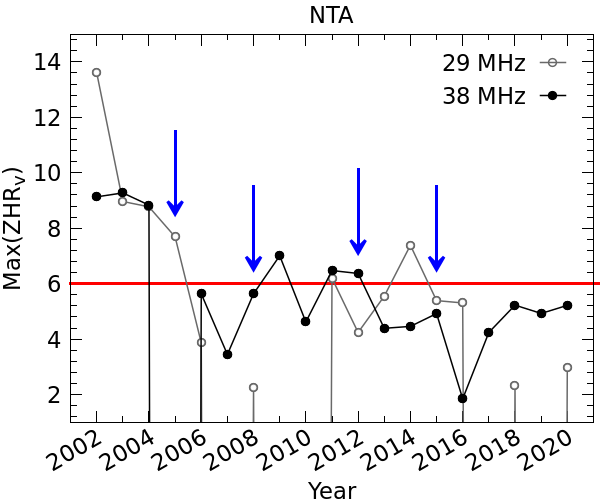}
     \includegraphics[width=.45\textwidth]{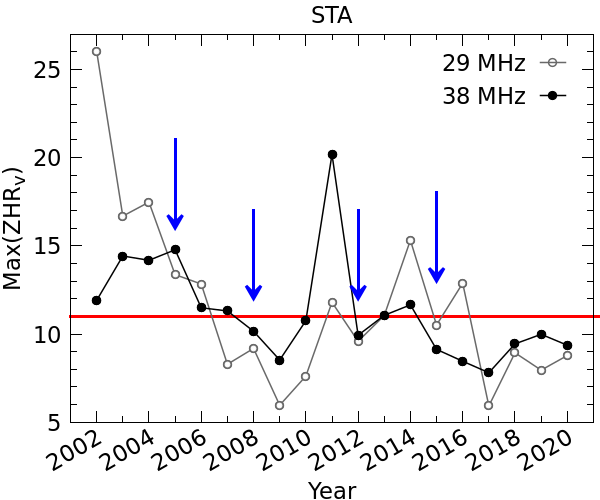}\\[0.1cm]
     \includegraphics[width=.45\textwidth]{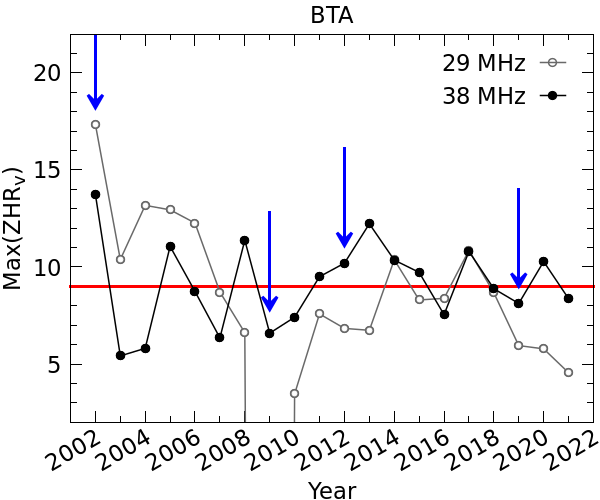}
     \includegraphics[width=.45\textwidth]{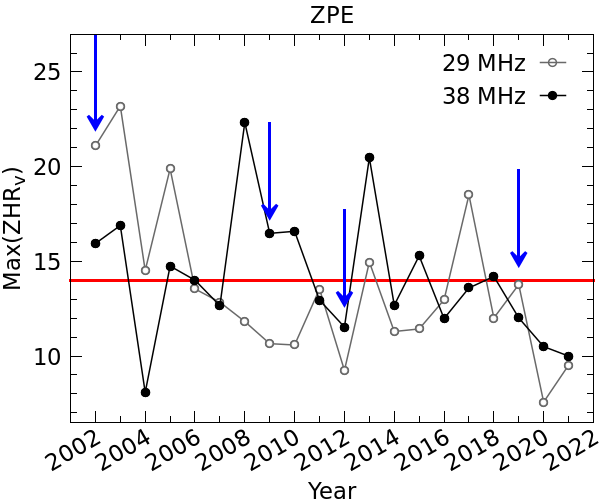}
     \caption{\label{fig:max_cmor}Maximum ZHR$_v$ measured by CMOR 29 MHz and 38 MHz for the NTA, STA, BTA and ZPE. The resonant years, corresponding to the return of the TSC stream, are marked with the blue arrows. The average activity level of the showers measured by each system is shown with a horizontal red line.}
 \end{figure}

Figure \ref{fig:max_cmor} shows the maximum annual activity measured by the 29 and 38 MHz CMOR systems for each apparition of the showers between 2002 and 2021. The maximum was identified by searching over definite intervals in solar longitude, corresponding to the epochs of highest intensity expected for each shower (see Section 7.2). Gaps in the figure are caused by years for which an incomplete activity profile was recorded. Resonant swarm years as predicted by \citet{Asher1993b}, in autumn or spring, are marked with vertical blue arrows in the plots. 

\setfignumprefix{A3.}

\newpage
\textbf{APPENDIX B: RADIANT}\\ 

\begin{figure}[!ht]
	\includegraphics[width=.45\textwidth]{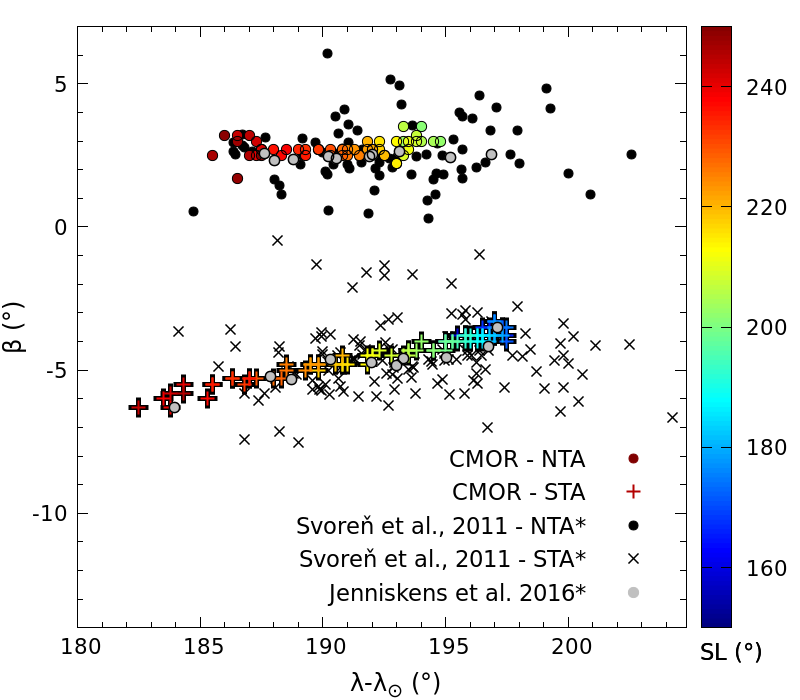}
	\includegraphics[width=.45\textwidth]{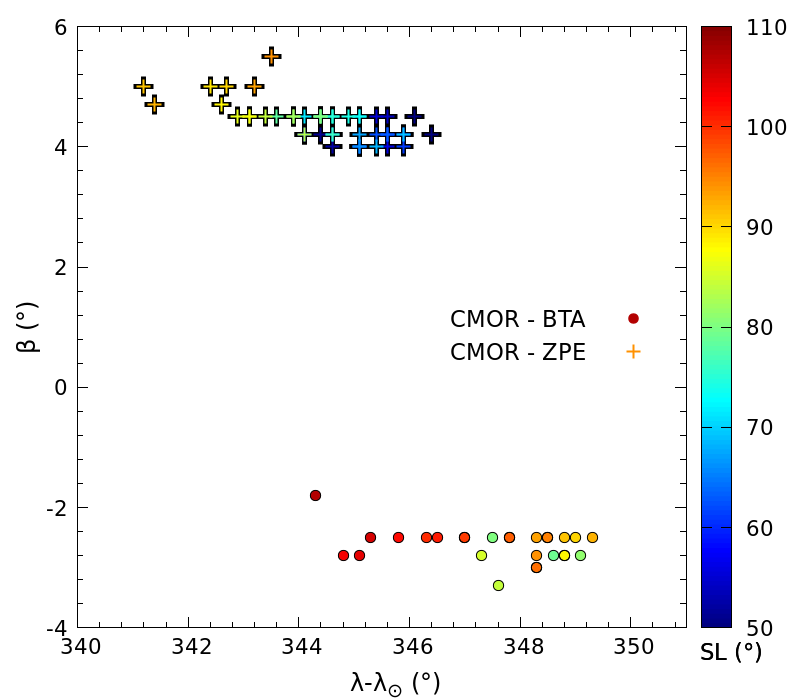}
	\caption{\label{fig:radiants_ecl}Ecliptic sun-centered latitude and longitude of the NTA, STA (left panel) and BTA, ZPE (right panel) radiants as a function of solar longitude (color coded). The average shower radiant location in one degree SL bins determined from the wavelet analysis of 29 MHz CMOR data is identified with circles (NTA \& BTA) and crosses (STA \& ZPE). Ecliptic coordinates computed from the geocentric radiants selected by \protect\cite{Svoren2011} (black circles and crosses) and \protect\cite{Jenniskens2016} (grey circles) are shown for comparison.}
	
\end{figure}

We summarize below information about the Taurid radiants. Figure \ref{fig:radiants_ecl} illustrates the ecliptic coordinates of the radiants, and plots in Section B2 show the radiant location of all four streams for different datasets as a function of solar longitude. \\ 

\textbf{B1 Radiants in Ecliptic coordinates} \\

\setfignumprefix{B1.}

\clearpage
\newpage
\textbf{B2 Radiant drift} \\

\setfignumprefix{B2.}
\begin{figure}[!ht]
 \centering
 \includegraphics[width=.4\textwidth]{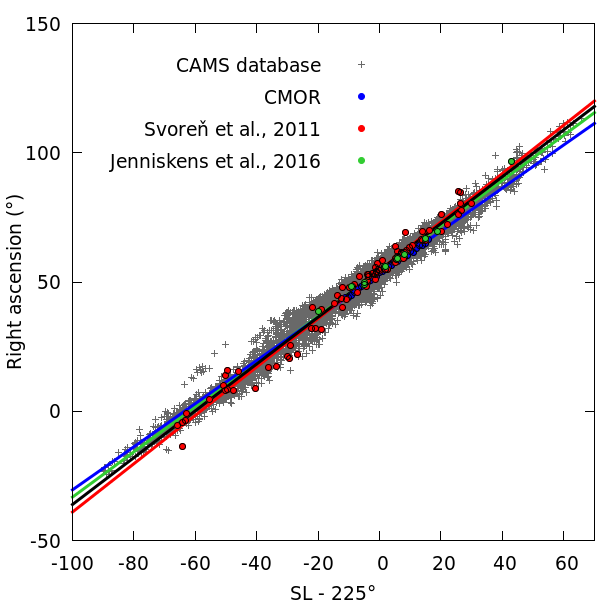}
 \includegraphics[width=.4\textwidth]{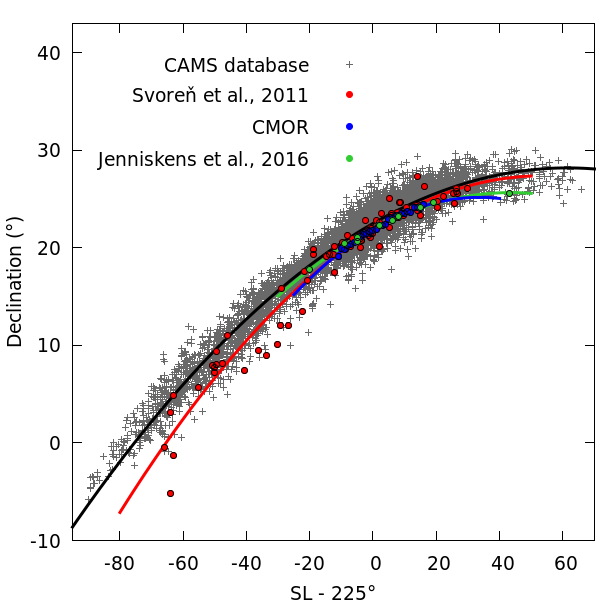}\\
 \includegraphics[width=.4\textwidth]{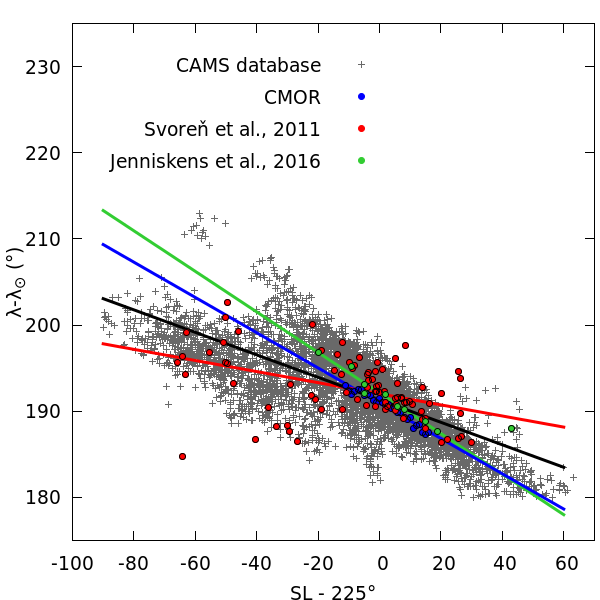}
 \includegraphics[width=.4\textwidth]{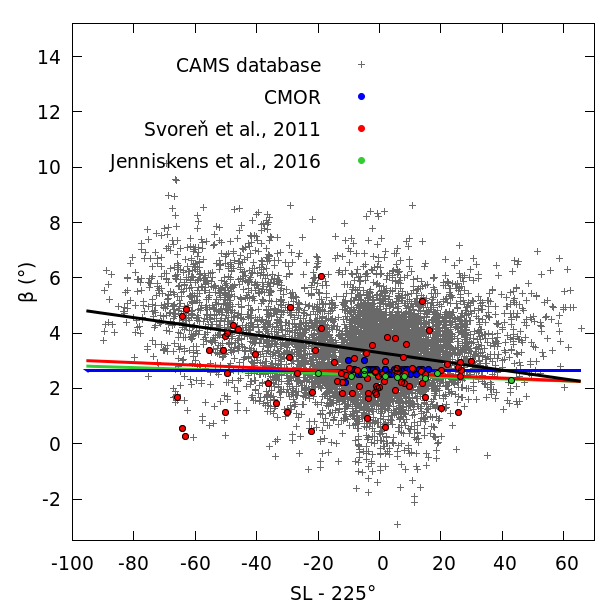}\\
 \includegraphics[width=.4\textwidth]{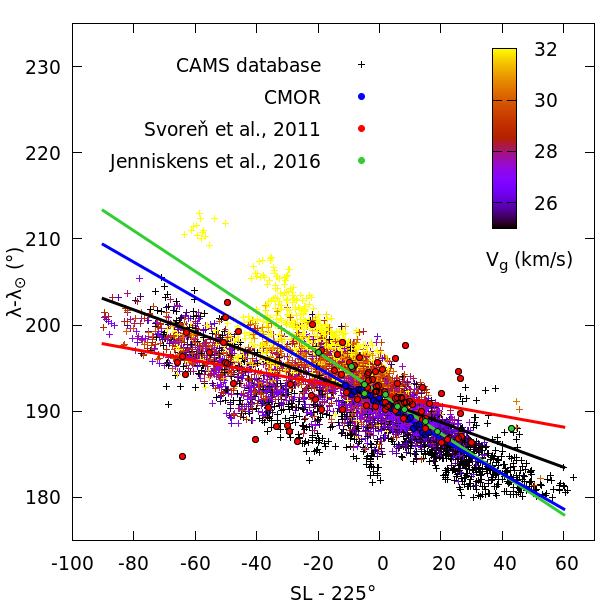}
 \includegraphics[width=.4\textwidth]{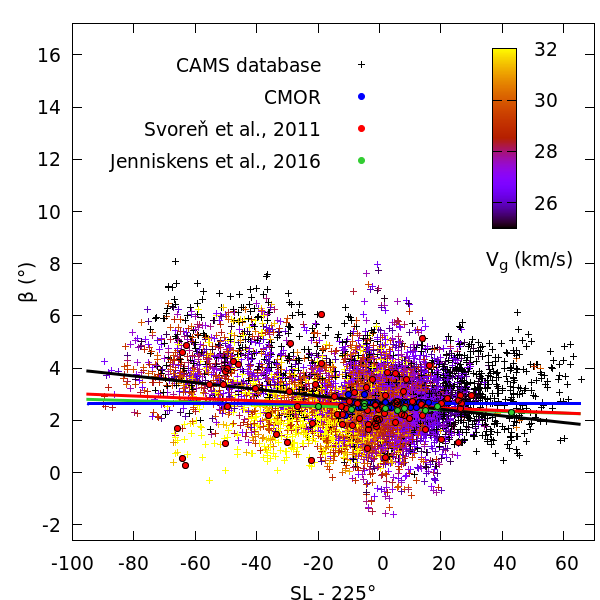}
 \caption{\label{fig:NTA_drift}Radiant drift of the NTA as a function of solar longitude of the right ascension, declination, sun-centered ecliptic latitude and longitude (bottom panels). Shown are measurements by CMOR (wavelet analysis, in blue), CAMS and the IAU MDC \protect\citep[in red, cf.][]{Svoren2011}. The NTA identified within the CAMS database between 2010 and 2016 are represented in grey, while the shower's sub-components identified by \protect\cite{Jenniskens2016} are identified in green. The solid curves illustrate the result of a linear or quadratic fit of each data set, line color corresponding to symbol colors. Fit results can be found in Table 2. The bottom panels of the Figure show the change in the radiants ecliptic coordinates, but with measurements from CAMS (crosses) coloured as a function of the meteor's geocentric velocity.}
\end{figure}

\begin{figure}[!ht]
 \centering
 \includegraphics[width=.48\textwidth]{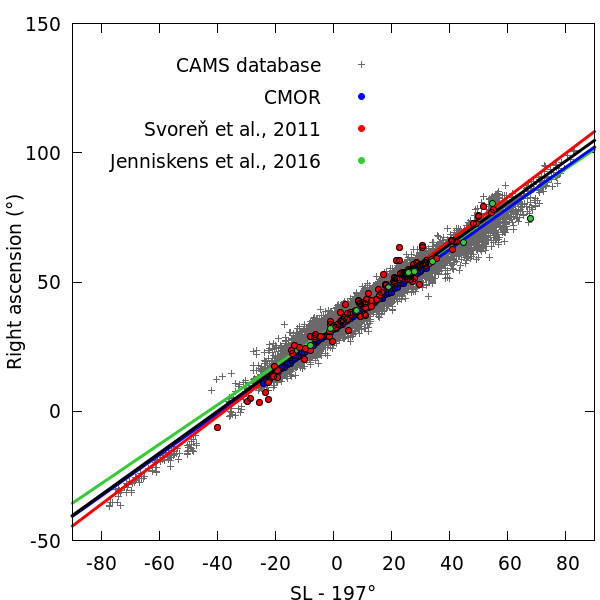}
 \includegraphics[width=.48\textwidth]{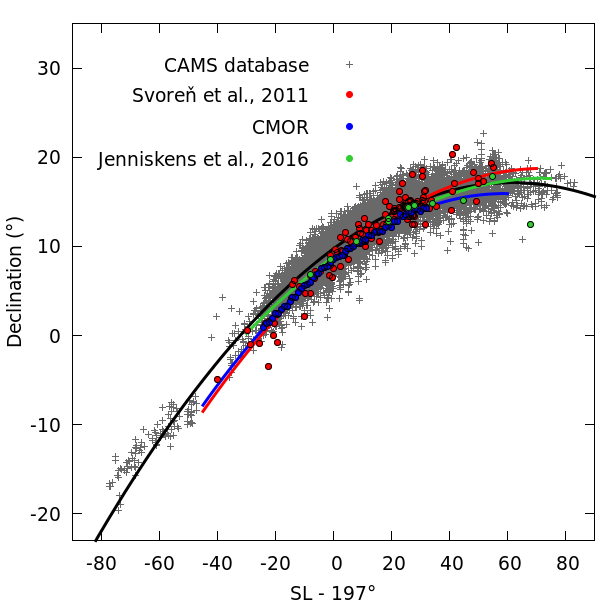}\\
 \includegraphics[width=.48\textwidth]{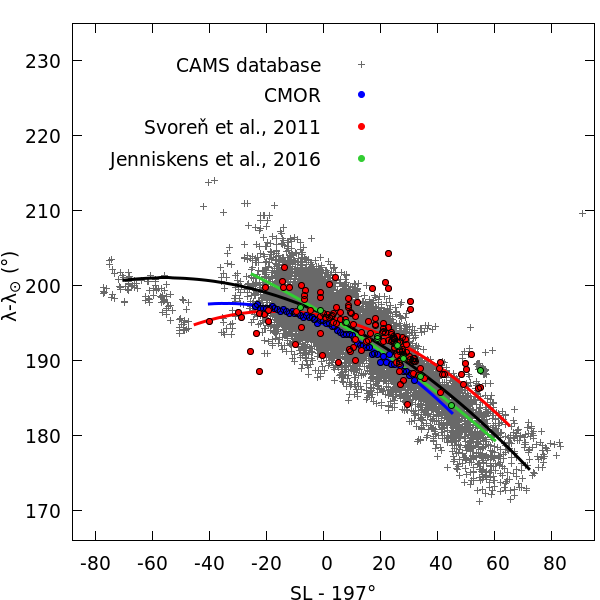}
 \includegraphics[width=.48\textwidth]{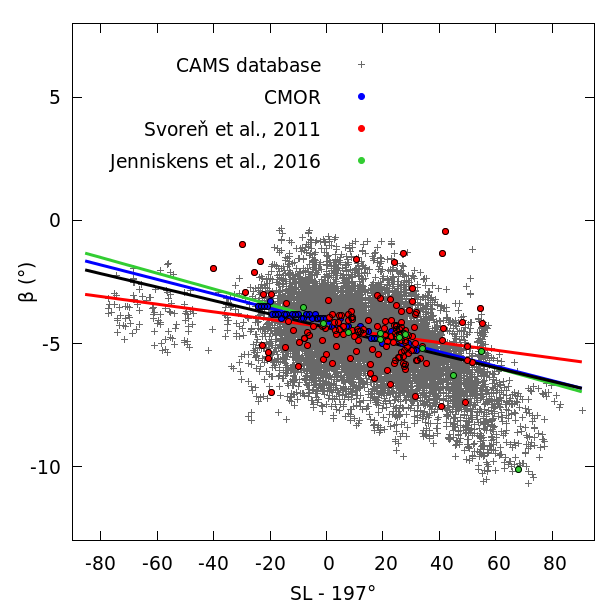}
 \caption{\label{fig:STA_drift}Radiant drift of Taurid streams as a function of solar longitude of the right ascension, declination, sun-centered ecliptic latitude and longitude (bottom panels). Shown are the STA as measured by CMOR (wavelet analysis, in blue), CAMS and the IAU MDC \protect\citep[in red, cf.][]{Svoren2011}. The STA identified within the CAMS database between 2010 and 2016 are represented in grey, while the shower's sub-components identified by \protect\cite{Jenniskens2016} are identified in green. The solid curves illustrate the result of a linear or quadratic fit of each data set, line color corresponding to symbol colors. Fit results can be found in Table 2.}
\end{figure}

\begin{figure}[!ht]
 \centering
 \includegraphics[width=.48\textwidth]{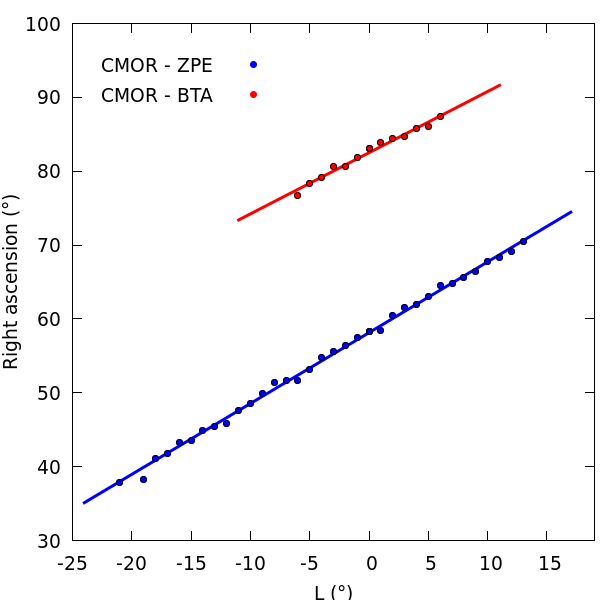}
 \includegraphics[width=.48\textwidth]{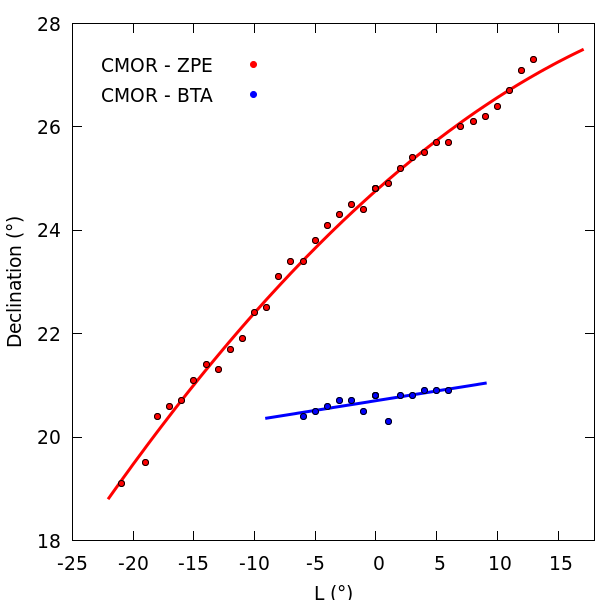}\\
 \includegraphics[width=.48\textwidth]{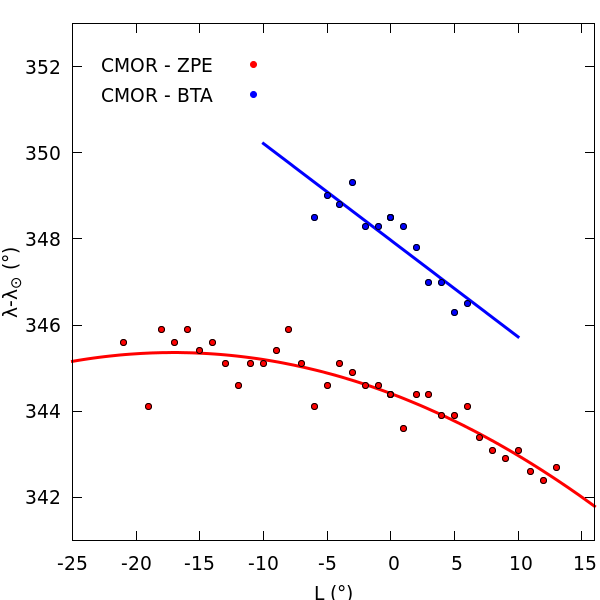}
 \includegraphics[width=.48\textwidth]{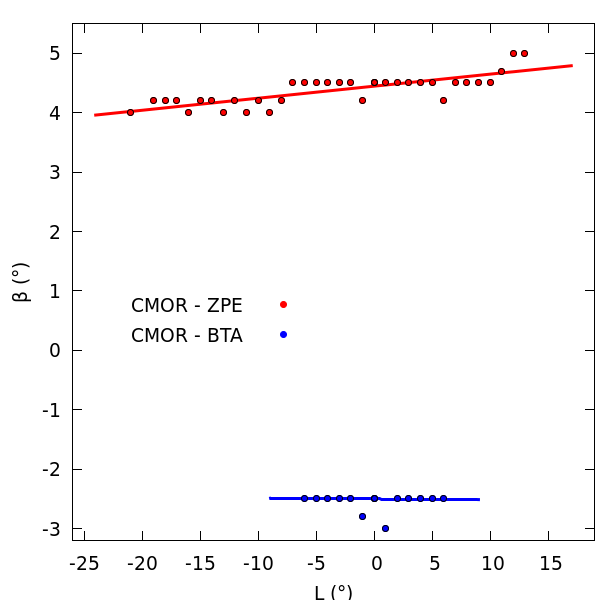}
 \caption{Change in the right ascension, declination, sun-centered ecliptic latitude and longitude (bottom panels) of the BTA (blue) and ZPE (red) as measured by CMOR (wavelet analysis) as a function of solar longitude. The solid curves illustrate the result of a linear or quadratic fit of each shower, line color corresponding to symbol colors. Fit results can be found in Table 2. The solar longitude of reference is $L_m$=95$\degree$ for the BTA and $L_m$=77$\degree$ for the ZPE.} 
\end{figure}


\clearpage
\newpage
\textbf{APPENDIX C: ORBITAL ELEMENTS OF TAURID STREAMS}\\ 

The evolution of the Taurids orbital elements with solar longitude, as measured from different datasets are shown below. We provide plots of possible correlations between the orbital elements of the showers that provide constraining information about the TMC formation.\\

\textbf{C1 Evolution with solar longitude} \\

\textbf{NTA}\\

\setfignumprefix{C1.}
\begin{figure}[!ht]
    \centering
    \includegraphics[width=.32\textwidth]{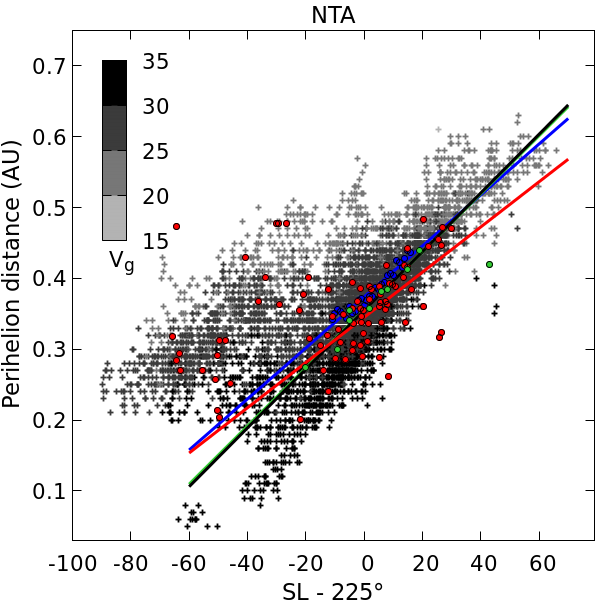}
    \includegraphics[width=.32\textwidth]{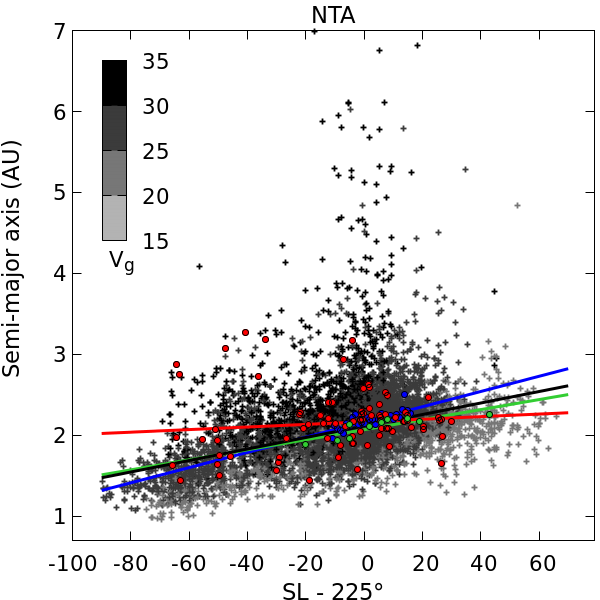}
    \includegraphics[width=.32\textwidth]{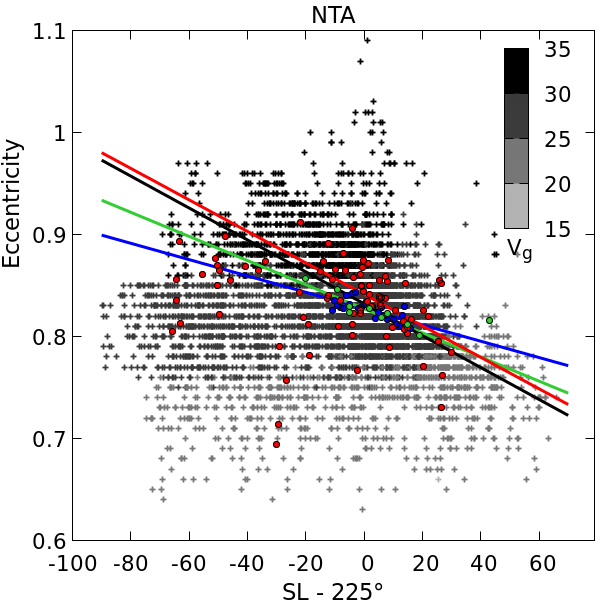}\\
    \includegraphics[width=.32\textwidth]{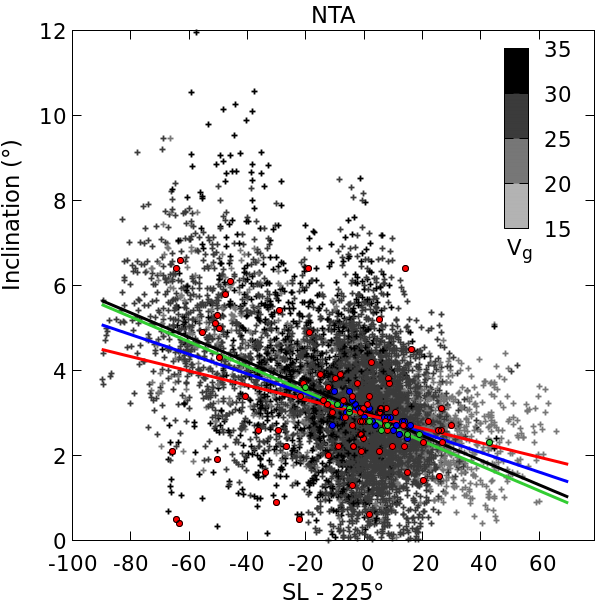}
    \includegraphics[width=.32\textwidth]{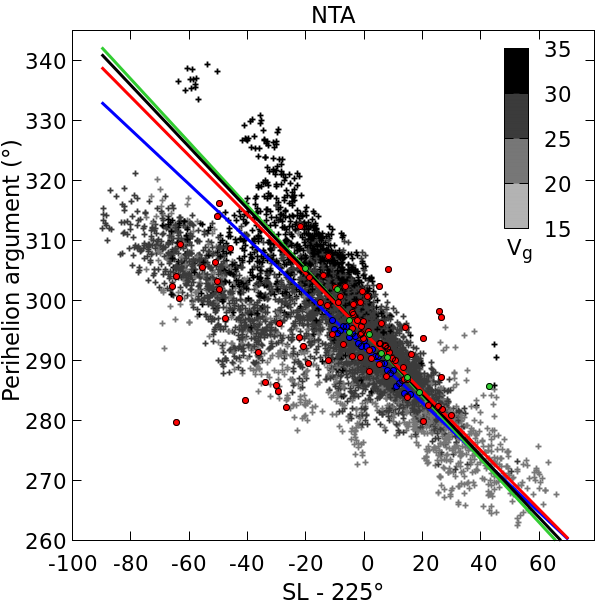}
    \includegraphics[width=.32\textwidth]{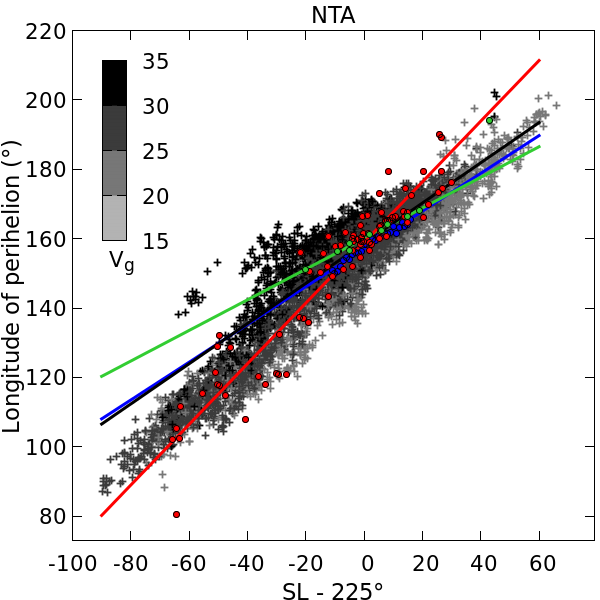}\\
    \includegraphics[width=.32\textwidth]{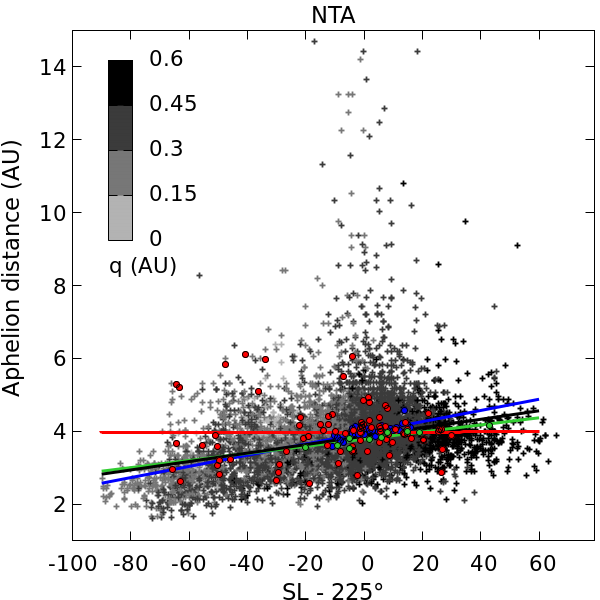}
    \includegraphics[width=.32\textwidth]{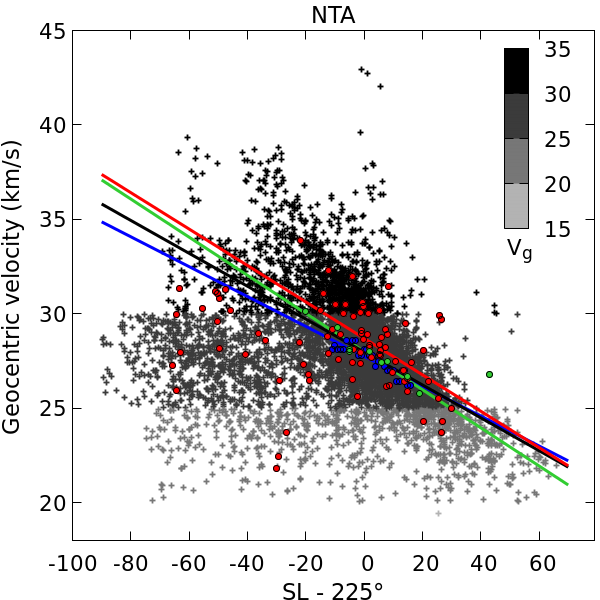}
    \caption{Evolution of the NTA perihelion distance, semi-major axis, eccentricity, inclination, perihelion argument, longitude of perihelion, aphelion distance and geocentric velocity with the solar longitude centered using a reference $L_m$=225$\degree$. Grey crosses represent the meteors contained in the CAMS database, colour-coded as a function of the geocentric velocity $V_g$ or the perihelion distance $q$. Red and green dots correspond to the meteors and sub-components identified by \cite{Svoren2011} and \cite{Jenniskens2016} respectively, and blue dots refer to the wavelet analysis applied to CMOR data. Lines illustrate fitting the measurements with a linear or quadratic function, line color corresponding to symbol colors. Fit coefficients are presented in Table C1.4. }
    \label{fig:NTA_sl}
\end{figure}

\newpage
\textbf{STA}\\

\begin{figure}[!ht]
    \centering
    \includegraphics[width=.32\textwidth]{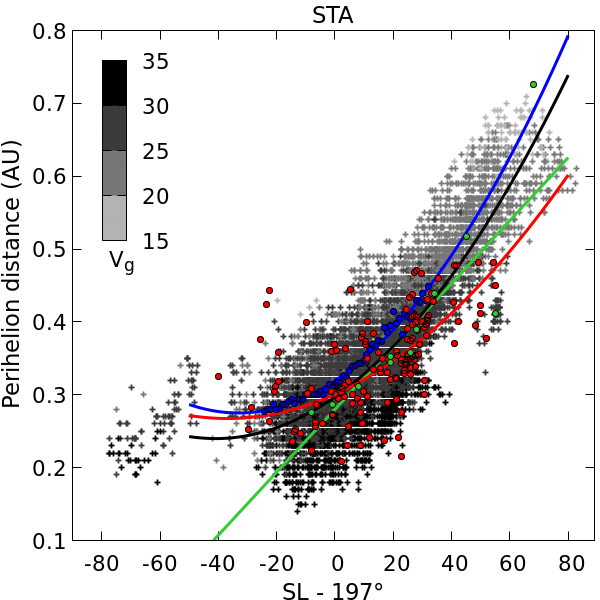}
    \includegraphics[width=.32\textwidth]{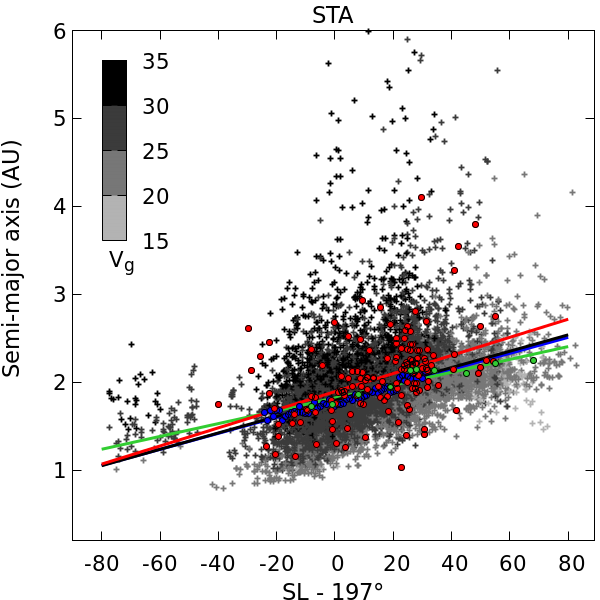}
    \includegraphics[width=.32\textwidth]{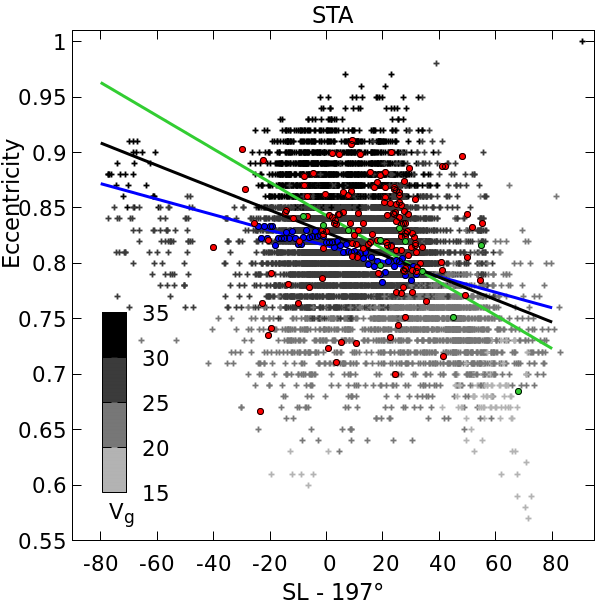}\\
    \includegraphics[width=.32\textwidth]{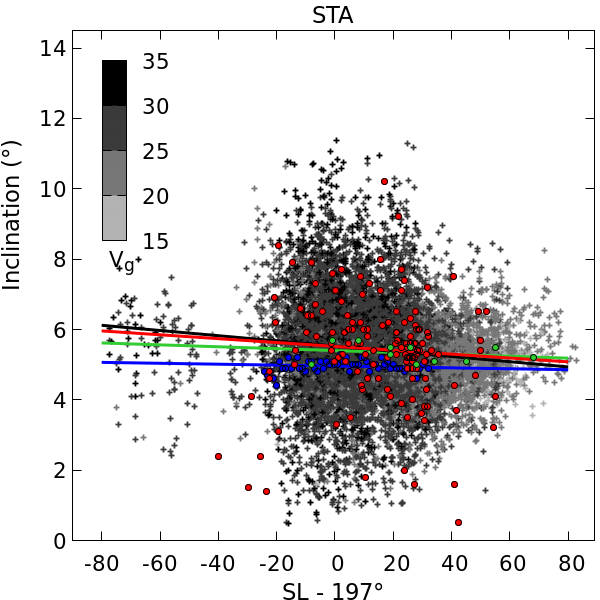}
    \includegraphics[width=.32\textwidth]{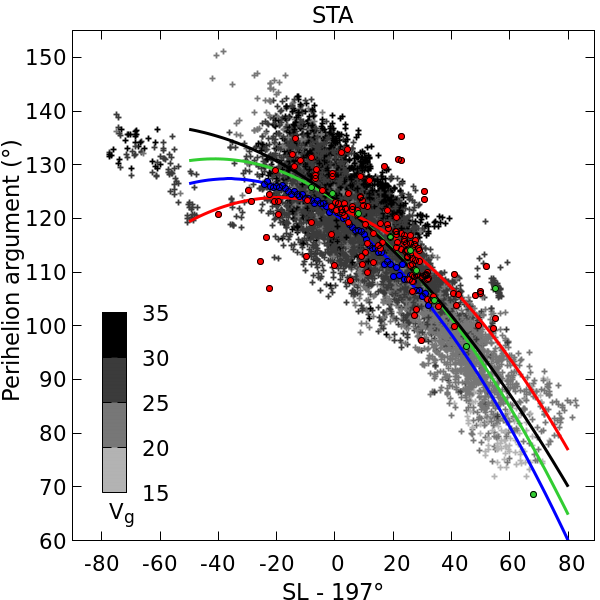}
    \includegraphics[width=.32\textwidth]{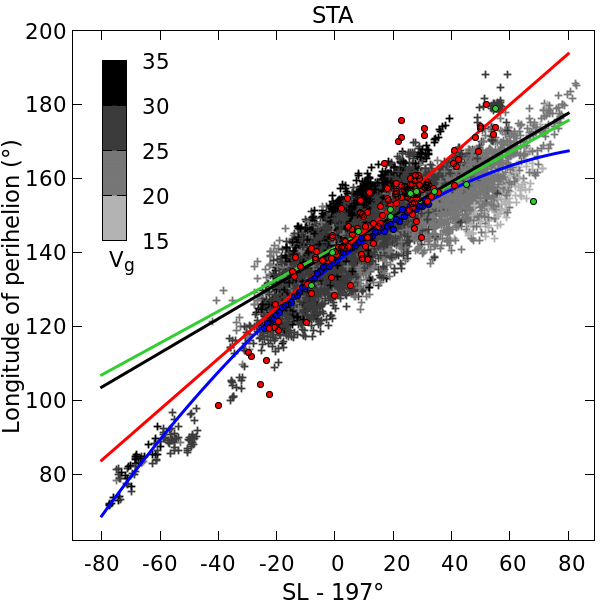}\\
    \includegraphics[width=.32\textwidth]{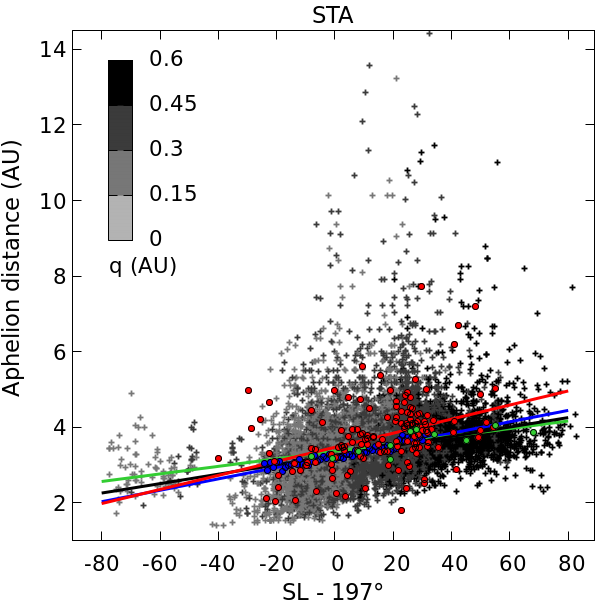}
    \includegraphics[width=.32\textwidth]{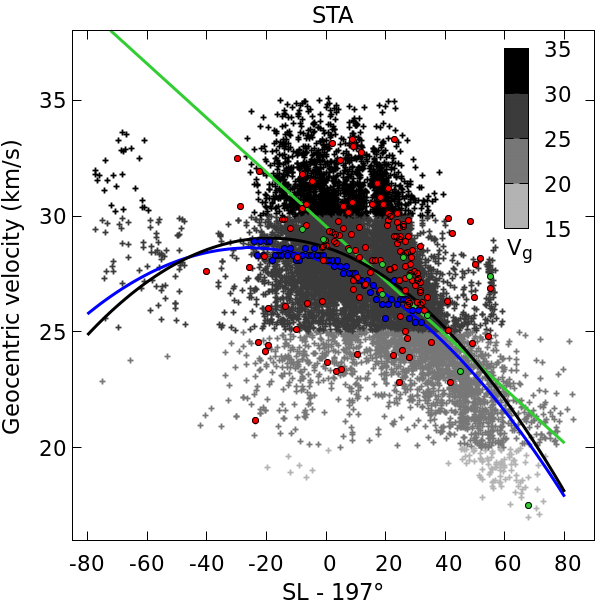}
    \caption{Evolution of the STA perihelion distance, semi-major axis, eccentricity, inclination, perihelion argument, longitude of perihelion, aphelion distance and geocentric velocity with the solar longitude centered on the reference $L_m$=197$\degree$. Grey crosses represent the meteors contained in the CAMS database, colour-coded in function of the geocentric velocity $V_g$ or the perihelion distance $q$. Red and green dots correspond to the meteors and sub-components identified by \cite{Svoren2011} and \cite{Jenniskens2016} respectively, and blue dots refer to the wavelet analysis applied to CMOR data. Lines illustrate attempts of fitting the measurements with a linear or quadratic function, line color corresponding to symbol colors. Fits coefficients are presented in Table C1.4. }
    \label{fig:STA_sl}
\end{figure}

\newpage
\textbf{BTA \& ZPE}\\

\begin{figure}[!ht]
    \centering
    \includegraphics[width=.32\textwidth]{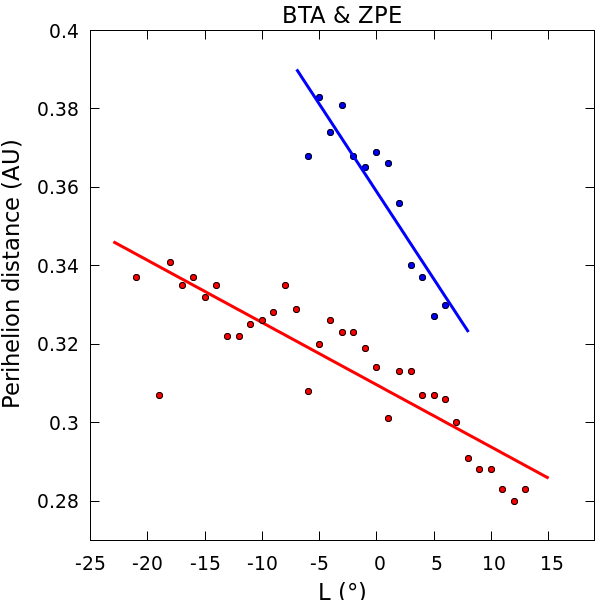}
    \includegraphics[width=.32\textwidth]{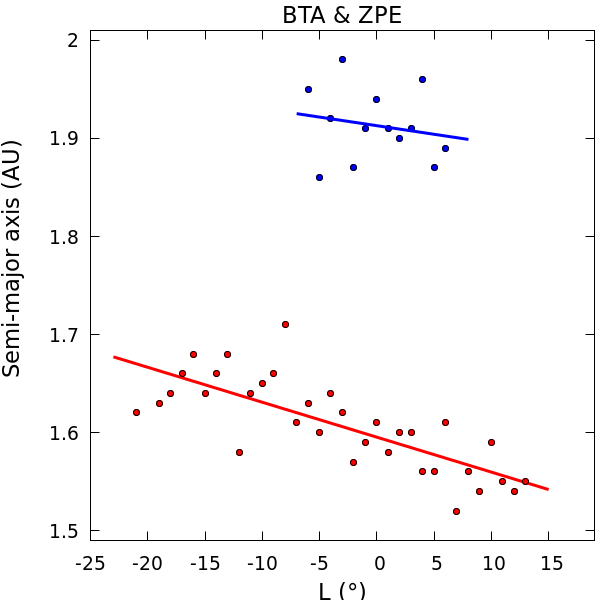}
    \includegraphics[width=.32\textwidth]{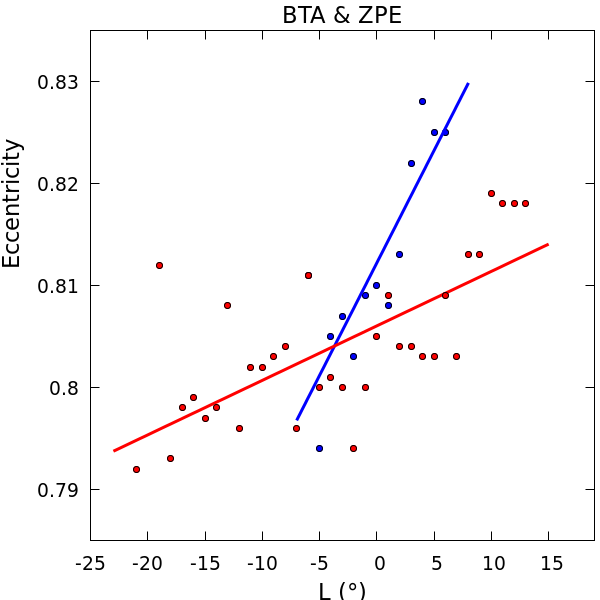}\\
    \includegraphics[width=.32\textwidth]{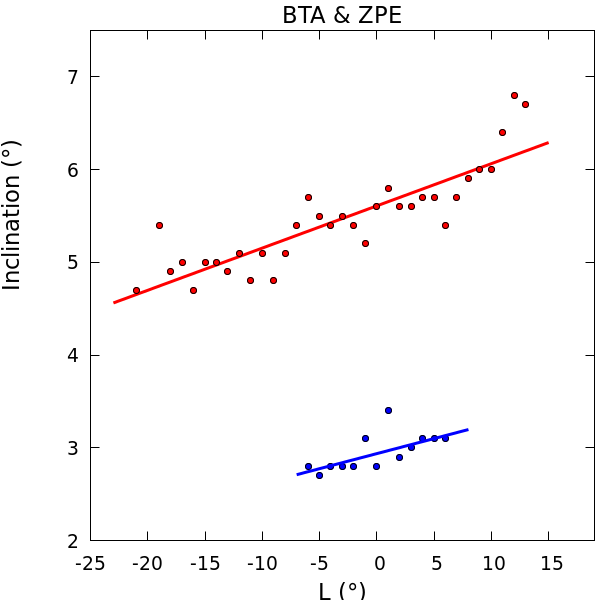}
    \includegraphics[width=.32\textwidth]{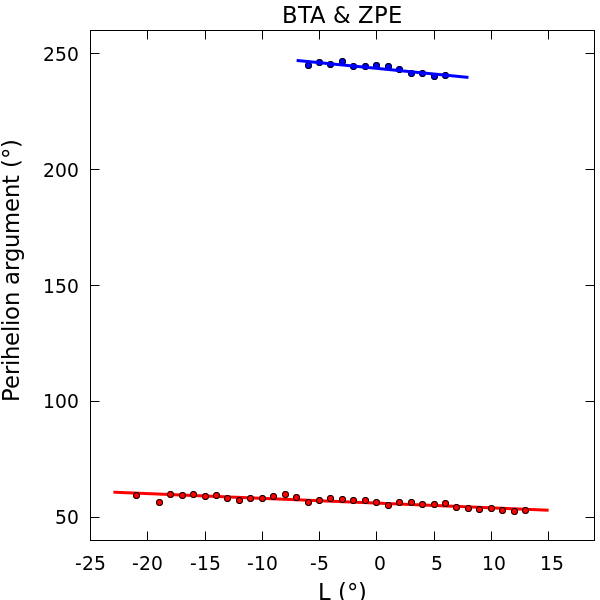}
    \includegraphics[width=.32\textwidth]{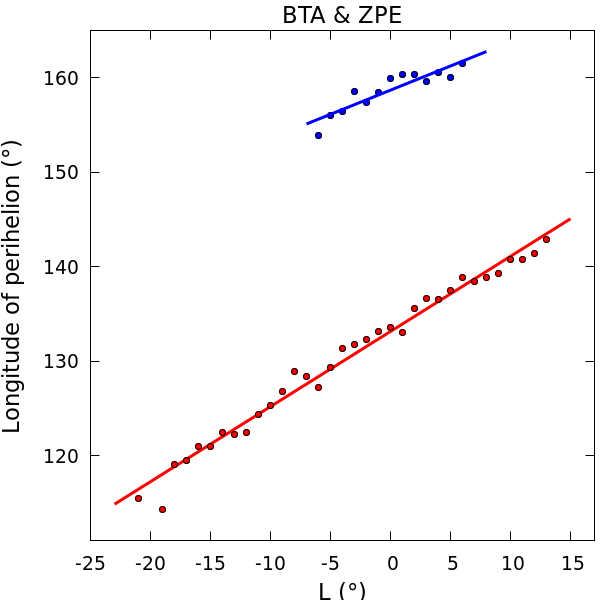}\\
    \includegraphics[width=.32\textwidth]{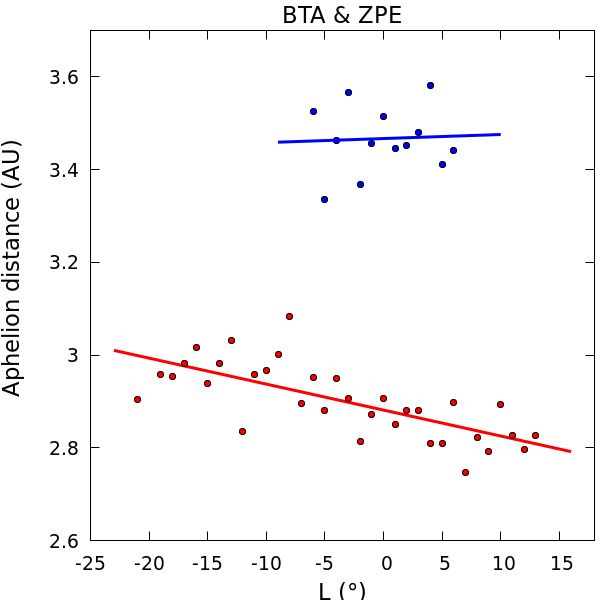}
    \includegraphics[width=.32\textwidth]{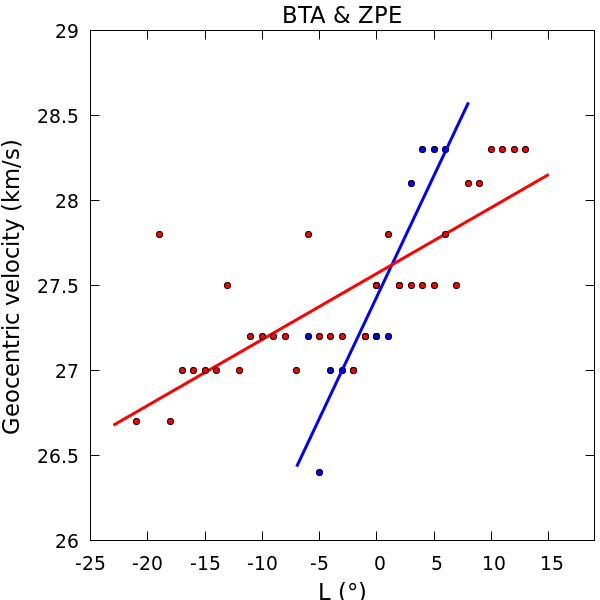}
    \caption{Evolution of the BTA (in blue) and ZPE (in red) perihelion distance, semi-major axis, eccentricity, inclination, perihelion argument, longitude of perihelion, aphelion distance and geocentric velocity with $L= SL-L_m$ from the wavelet analysis of CMOR data. The solar longitude of reference is $L_m$=95$\degree$ for the BTA and $L_m$=77$\degree$ for the ZPE. Straight lines illustrate attempts of fitting the measurements with a linear or quadratic function, line color corresponding to symbol colors. Fit coefficients are presented in Table C1.4. }
    \label{fig:BZ_sl}
\end{figure}

\newpage
\textbf{Summary: orbital drift}\\

 \settabnumprefix{C1.}{3}
\begin{table*}[!ht]
\resizebox{\textwidth}{!}{
  \begin{tabular}{rrrrrr}
     & NTA\hspace{0.5cm}\mbox{ } & STA\hspace{0.5cm}\mbox{ } & BTA\hspace{0.5cm}\mbox{ } & ZPE\hspace{0.5cm}\mbox{ } & \\
    \cline{2-5} \\[-0.3cm]
    \hline
      & & & &\\[-0.3cm]
     q & 0.3739+0.0036*L & 0.3194+0.0027*L+4.05*$10^{-5}$*L$^2$ & 0.3588-0.0044*L & 0.3095-0.0015*L & [C]\\
      & (0.3450+0.0032*L) & (0.3007+0.0018*L) & -- & -- & [S] \\
      & 0.3555+0.0041*L & (0.2801+0.0043*L) & -- & -- & [J] \\
      & 0.3548+0.0041*L & 0.2971+0.0028*L+3.41*$10^{-5}$*L$^2$ & -- & -- & [Cd] \\[0.1cm]
      \cline{3-6} \\[-0.3cm]
      & 0.325+0.00413*L & 0.363+0.0041*L & -- & -- & [P] \\
      & 0.346+0.0041*L  & 0.269+0.0041*L & -- & -- & [P*] \\
      \hline
      & & & &\\[-0.3cm]
    a & 2.1646+0.0094*L & 1.7825+0.0091*L & 1.9130-0.0018*L & 1.5954-0.0036*L & [C]\\
      & (2.1661+0.0016*L) & (1.8930+0.0103*L) & -- & -- & [S]\\
      & 2.0688+0.0062*L & 1.8216+0.0073*L & -- & -- & [J]\\
      & 2.1143+0.0071*L & 1.7950+0.0093*L & -- & -- & [Cd]\\[0.1cm]
      \cline{3-6} \\[-0.3cm]
      & 2.072+0.00872*L & 2.106+0.01443*L & -- & -- & [P] \\
      & 2.116+0.00872*L & 1.774+0.01443*L & & & [P*] \\
      \hline
      & & & &\\[-0.3cm]
    e & 0.8274-0.0008*L & 0.8159-0.0007*L & 0.8122+0.0022*L & 0.8060+0.0005*L & [C]\\
      & (0.8413-0.0015*L) & -- & -- & -- & [S]\\
      & 0.8272-0.0012*L & (0.8431-0.0015*L) & -- & -- & [J]\\
      & 0.8321-0.0016*L & (0.8278-0.0010*L) & -- & -- & [Cd]\\ [0.1cm]
      \cline{3-6} \\[-0.3cm]
      & 0.838-0.00102*L & 0.823-0.00062*L & -- & -- & [P] \\ 
      & 0.833-0.00102*L & 0.837-0.00062*L & & & [P*] \\
      \hline
      & & & &\\[-0.3cm]
    i & 2.9970-0.0231*L & 4.9609-0.0013*L & 2.9370+0.0325*L & 5.6092+0.0455*L & [C] \\
      & 2.9751-0.0289*L & 5.5193-0.0055*L & -- & -- & [S] \\
      & 2.9250-0.0292*L & (5.4010-0.0027*L) & -- & -- & [J] \\
      & 3.0446-0.0289*L & 5.5273-0.0074*L & -- & -- & [Cd] \\[0.1cm]
      \cline{3-6} \\[-0.3cm]
      & 3.35-0.0388*L   & 5.34-0.0204*L & -- & -- & [P] \\ 
      & 3.16-0.0388*L   & 5.81-0.0204*L & & & [P*] \\
      \hline
      & & & &\\[-0.3cm]
    $\omega$ & 292.0564-0.4553*L & 120.7675-0.3626*L & 243.6865-0.4886*L & 56.1352-0.2047*L & [C]\\
             & 294.6706-0.4910*L & (122.0958-0.1845*L-4.76*$10^{-3}$*L$^2$) & -- & -- & [S] \\
             & 294.6935-0.5278*L & 123.3959-0.3723*L-4.97*$10^{-3}$*L$^2$ & -- & -- & [J] \\
             & 294.7196-0.5146*L & 123.6068-0.4176*L-3.15*$10^{-3}$*L$^2$ & -- & -- & [Cd] \\[0.1cm]
      \cline{3-6} \\[-0.3cm]
             & 298.29-0.5408*L   & 113.89-0.5658*L & -- & -- & [P] \\ 
             & 295.59-0.5408*L   & 126.90-0.5658*L   & & & [P*] \\
      \hline
      & & & &\\[-0.3cm]
      $\Pi$ & 157.0569+0.5444*L & 137.0133+0.6174*L-2.97*$10^{-3}$*L$^2$) & 158.6872+0.5113*L & 133.1349+0.7954*L & [C]\\
            & 158.9121+0.8741*L & 138.7082+0.6875*L & -- & -- & [S] \\
            & 160.0333+0.4413*L & 141.2315+0.4302*L & -- & -- & [J] \\
            & 158.7030+0.5792*L & 140.5464+0.4632*L & -- & -- & [Cd] \\[0.1cm]
      \cline{3-6} \\[-0.3cm]
            & 158.3393+0.4583*L & 153.9500+0.4346*L & -- & -- & [P] \\ 
            & 160.6308+0.4583*L & 143.9542+0.4346*L & -- & -- & [P*] \\ 
     \hline
    & & & &\\[-0.3cm]
    $Q$ & 3.9555+0.0154*L & 3.2355+0.0151*L & (3.4670+0.0009*L) & 2.8813-0.0056*L & [C]\\
          & (3.9733+0.0002*L) & (3.4684+0.0186*L) & -- & -- & [S] \\
          & 3.7826+0.0098*L & 3.3627+0.0100*L & -- & -- & [J] \\
          & 3.8643+0.0116*L & 3.2546+0.0125*L & -- & -- & [Cd] \\ 
    \hline
    & & & &\\[-0.3cm]
    $V_g$ & 27.7533-0.0791*L & (27.9922-0.0492*L-9.63*$10^{-4}$*L$^2$) & (27.4342+0.1428*L) & (27.5709+0.0388*L) & [C]\\
          & (28.7037-0.0964*L) & -- & -- & -- & [S] \\
          & 28.0044-0.1009*L & (29.5536-0.1169*L) & -- & -- & [J] \\
          & 27.9800-0.0871*L & (28.6449-0.0423*L) & -- & -- & [Cd] \\ 
    \hline
    & & & &\\[-0.3cm]
   L & SL-225$\degree$ & SL-197$\degree$ & SL-95$\degree$ & SL-77$\degree$ &  [C,S,J,Cd,P*]\\
    & SL-220$\degree$ & SL-220$\degree$ & -- & -- &  [P]\\
    \hline
    \hline
  \end{tabular}}
  \caption{Variation of the orbital elements of the NTA, STA, BTA and ZPE from different observations as a function of solar longitude (L).  The orbital drifts were computed using C: CMOR measurements (wavelet analysis), S: the selection of photographic meteors of \protect\cite{Svoren2011}, J: the showers' sub-components identified by \protect\cite{Jenniskens2016} and Cd: the NTA and STA contained in the CAMS database between 2010 and 2016. Estimates published P: \protect\cite{Porubcan2002} were modified to correct for the updated $L$ values (P*). Drifts estimates were obtained by fitting a linear or quadratic function to the observations, using the least-squares method. Expressions between parenthesis indicate highly uncertain values, due to the significant dispersion in the measured orbital elements. The results are plotted along with the observations in Figures \ref{fig:NTA_sl}, \ref{fig:STA_sl} and \ref{fig:BZ_sl}. }
\end{table*}

\newpage
\textbf{C2 Density maps} \\

\textbf{C2.1 CAMS} \\[-0.1cm]

The following figures are similar to the orbital evolution of the NTA and STA presented in Figures \ref{fig:NTA_sl} and \ref{fig:STA_sl} but with individual meteors of the CAMS database colour-coded by density. Here black areas represent regions of the orbital space where almost no Taurid meteor has been observed with CAMS, while coloured regions indicate higher density of meteors. The core of the streams is illustrated by the red and yellow areas. \\

\setfignumprefix{C2.1.}
\begin{figure}[!ht]
    \centering
    \includegraphics[width=.32\textwidth]{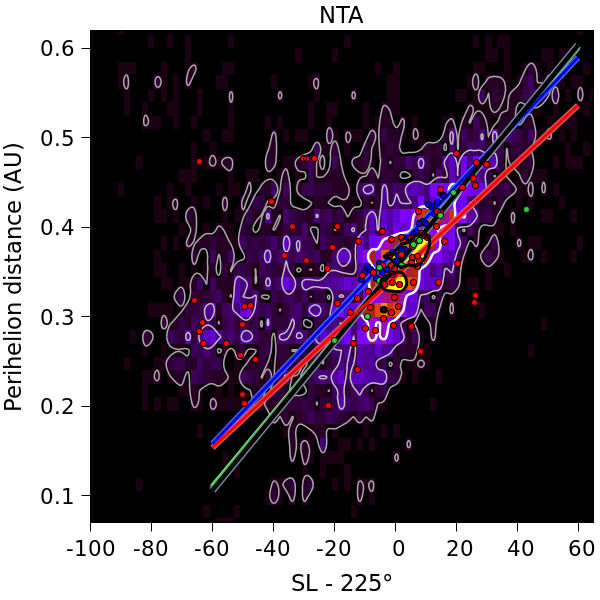}
    \includegraphics[width=.32\textwidth]{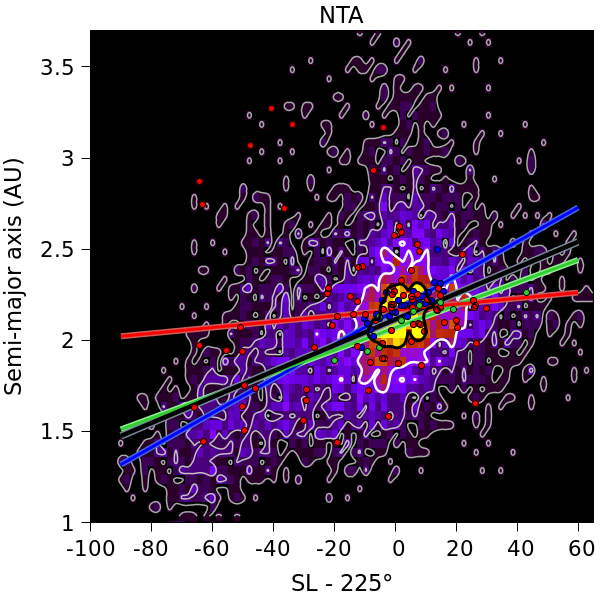}
    \includegraphics[width=.32\textwidth]{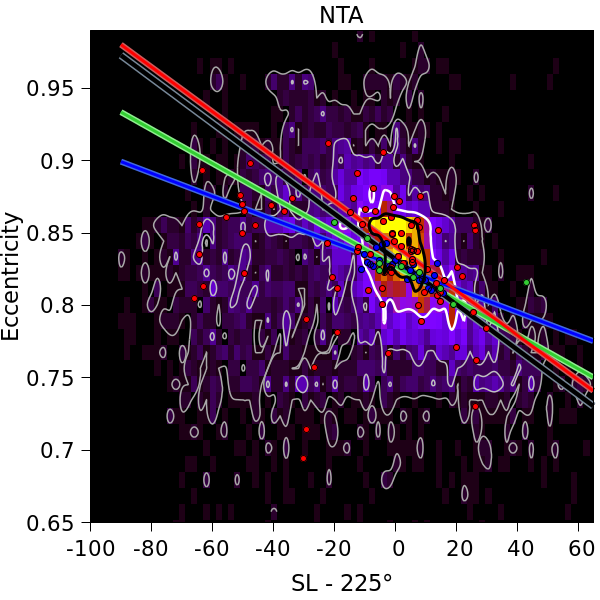}\\
    \includegraphics[width=.32\textwidth]{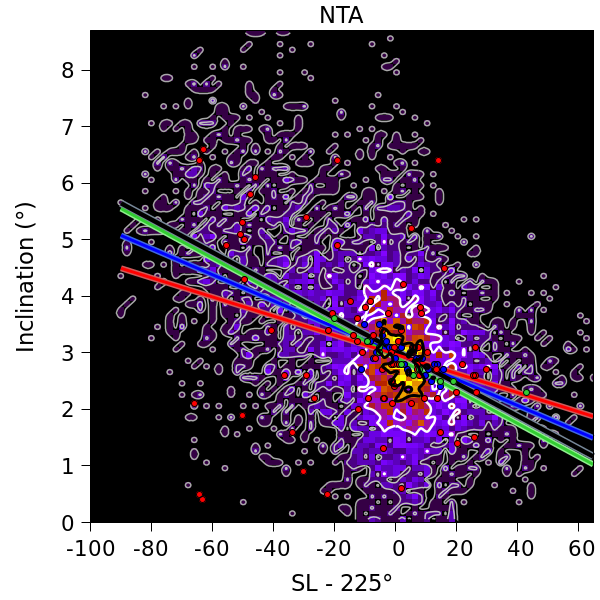}
    \includegraphics[width=.32\textwidth]{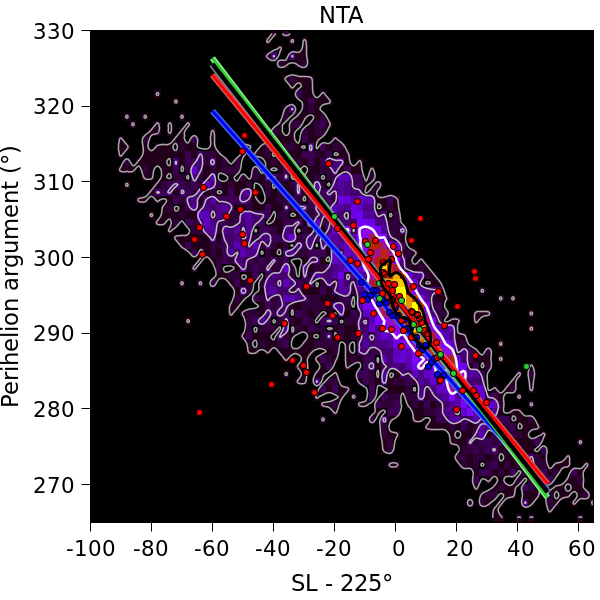}
    \includegraphics[width=.32\textwidth]{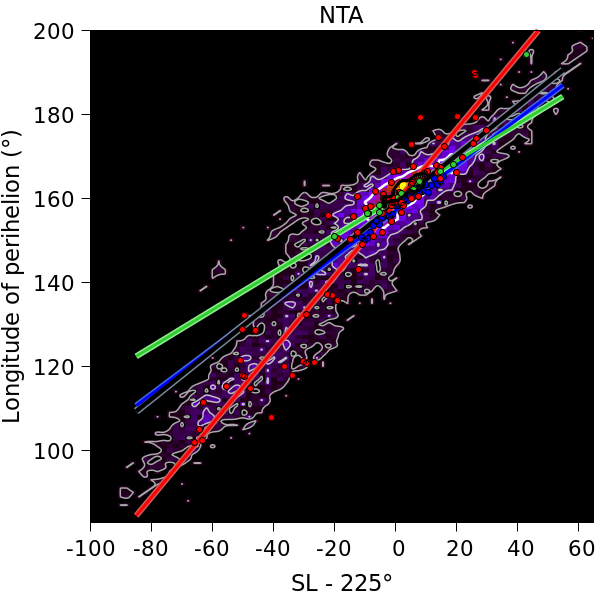}\\
    \includegraphics[width=.32\textwidth]{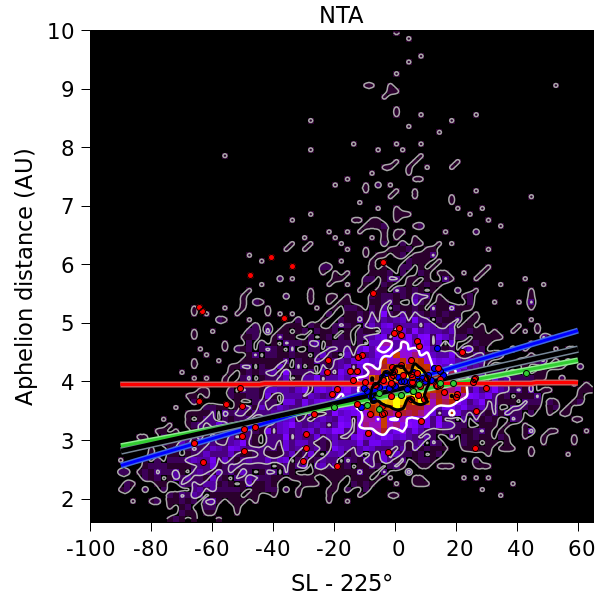}
    \includegraphics[width=.32\textwidth]{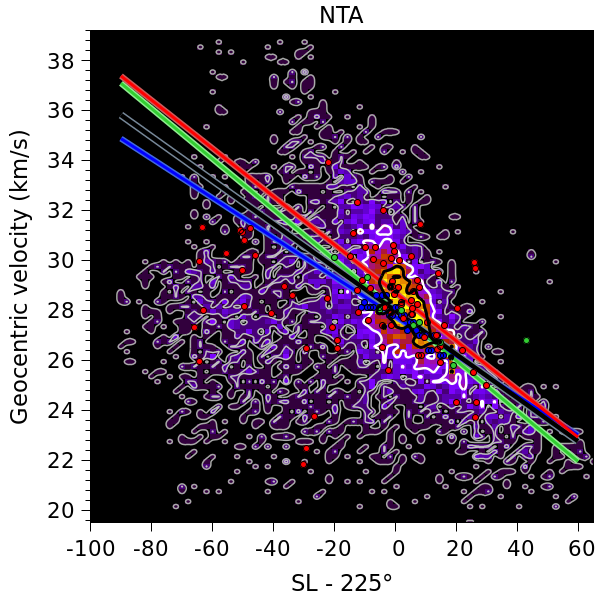}\hspace{0.2cm}
    \includegraphics[height=5.5cm]{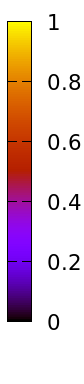}
    \caption{Evolution of the NTA perihelion distance, semi-major axis, eccentricity, inclination, perihelion argument, longitude of perihelion, aphelion distance and geocentric velocity with the solar longitude centered on the reference $L_m$=225$\degree$. Coloured areas represent the density of meteors recorded in the CAMS database, from low (black) to high (yellow). Red and green dots correspond to the meteors and sub-components identified by \cite{Svoren2011} and \cite{Jenniskens2016} respectively, and blue dots are from the wavelet analysis of CMOR data. Lines illustrate the attempts of fitting the measurements with a linear or quadratic function presented in section C1. }
    \label{fig:NTA_map}
\end{figure}

\begin{figure}[!ht]
    \centering
    \includegraphics[width=.32\textwidth]{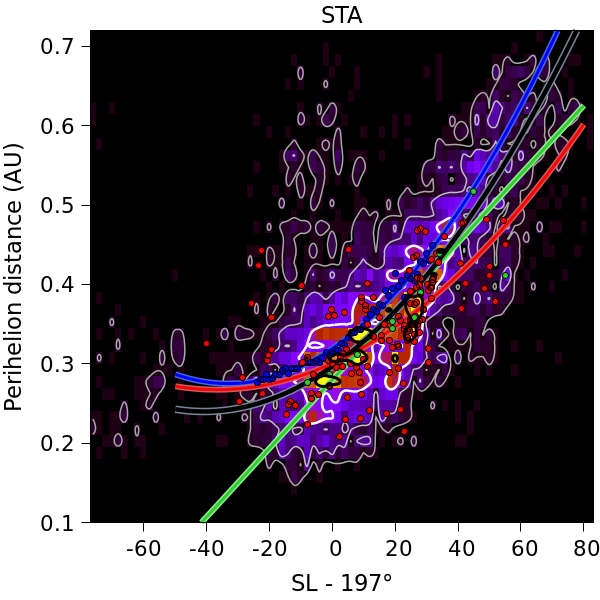}
    \includegraphics[width=.32\textwidth]{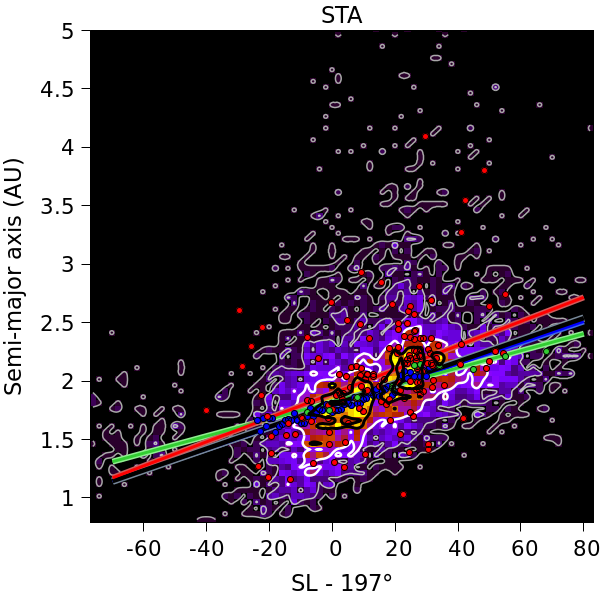}
    \includegraphics[width=.32\textwidth]{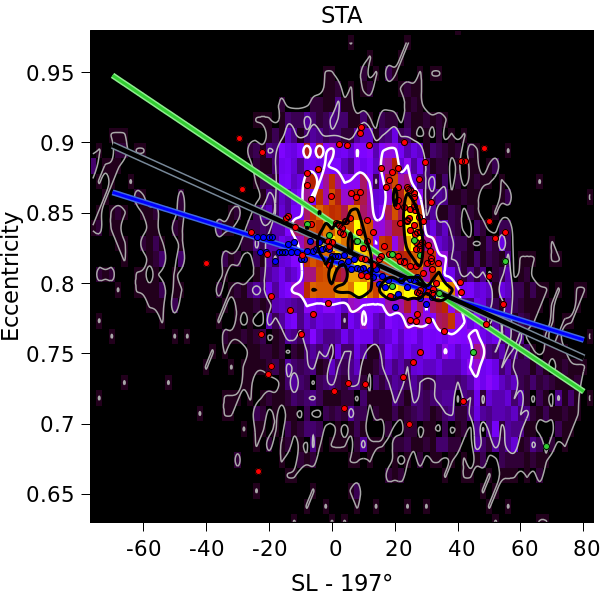}\\
    \includegraphics[width=.32\textwidth]{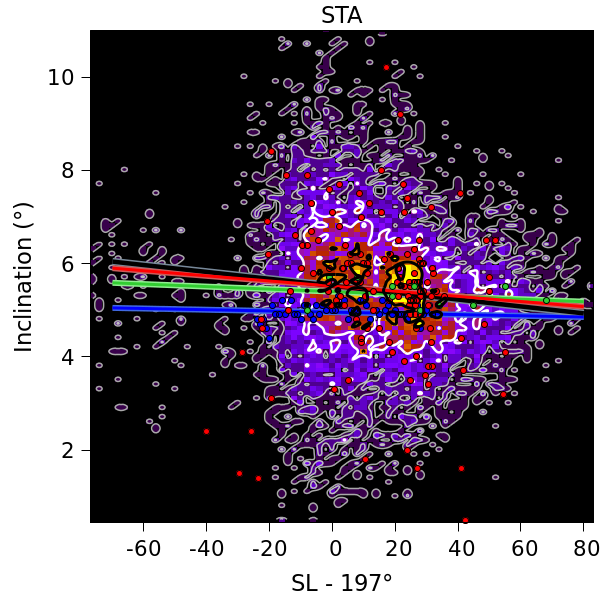}
    \includegraphics[width=.32\textwidth]{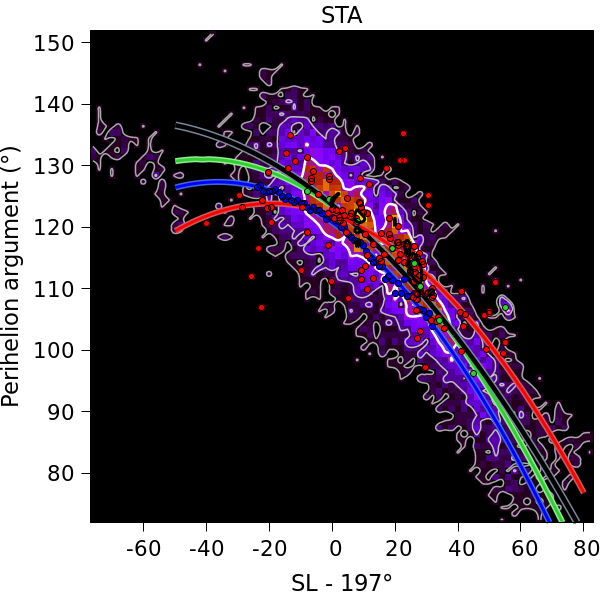}
    \includegraphics[width=.32\textwidth]{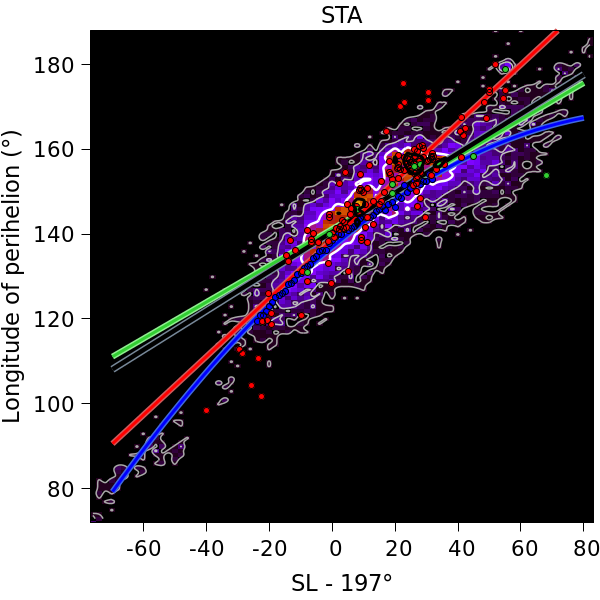}\\
    \includegraphics[width=.32\textwidth]{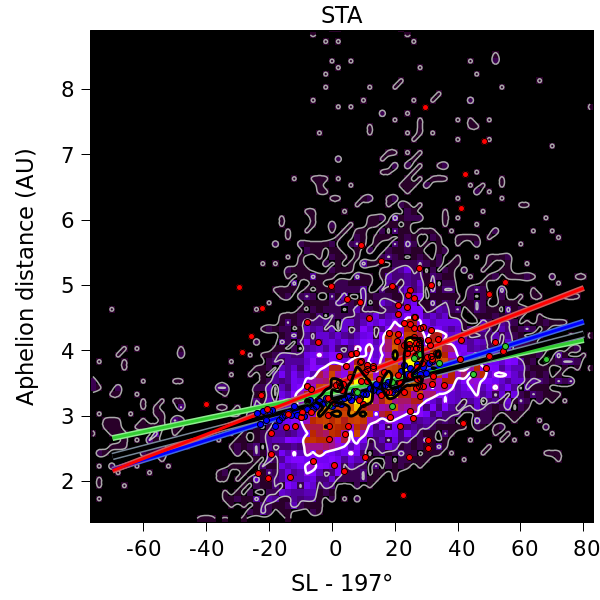}
    \includegraphics[width=.32\textwidth]{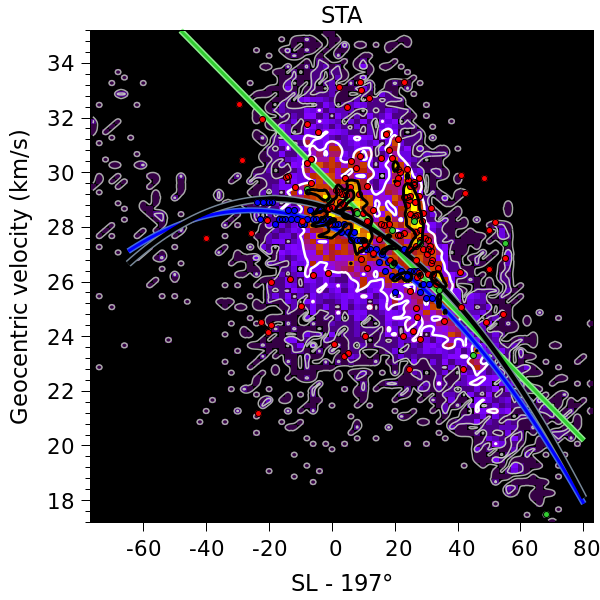} \hspace{0.2cm}
    \includegraphics[height=5.5cm]{colorbar.png}
    \caption{Evolution of the STA perihelion distance, semi-major axis, eccentricity, inclination, perihelion argument, longitude of perihelion, aphelion distance and geocentric velocity with the solar longitude centered on the reference $L_m$=197$\degree$. Coloured areas represent the density of meteors recorded in the CAMS database, from low (black) to high (yellow). Red and green dots correspond to the meteors and sub-components identified by \cite{Svoren2011} and \cite{Jenniskens2016} respectively, and blue dots refer to the wavelet analysis of CMOR data. Lines illustrate the attempts of fitting the measurements with a linear or quadratic function presented in section C1. }
    \label{fig:STA_map}
\end{figure}

\newpage

\textbf{C2.2 CMOR} \\[-0.1cm]

The following plots represent the density maps computed from individual meteor orbits extracted from the CMOR database using the Convex Hull approach. This new method allows identification of individual meteor orbits of a given meteor shower with a confidence level of 95\% for large datasets. The principle of the Convex Hull approach is described in Section 2.4.3. The parameters used to identify the radar meteors of each Taurid shower are summarized in Table \ref{tab:CH_parameters}. In these maps, black areas represent orbital regions where no shower meteor was detected by CMOR, while coloured regions indicate a higher density of meteors. The core of the streams is illustrated by the red and yellow areas. Offsets between the wavelet analysis (cyan dots and lines) and concentration of individual orbits are due to different corrections applied for meteor deceleration in the atmosphere. The wavelet analysis used the original deceleration correction for CMOR \citep{Brown2005}, leading to slightly higher geocentric velocities than the Convex Hull method that considered the more recent correction found by \citet{Froncisz2020}.    \\

\newpage

 \settabnumprefix{C2.}{1}
\begin{table}[!ht]
    \centering
    \begin{tabular}{ccccccc}
     Shower  & $\lambda-\lambda_\odot$ ($\degree$) & $\beta$ ($\degree$) & $V_g$ & Background SL ($\degree$) & Shower duration ($\degree$) & Number of voxels\\
     \hline
     \hline
     \mbox{}\\[-0.3cm]
     NTA & [180,204] & [-1,-15] & 10\% & [199,203] & [211,238] & 8x8x8 \\
     STA & [183,214] & [-20,-1] & 10\% & [161,165] \& [235,239] & [169,226] & 6x6x6 \\
     BTA & [335,360] & [-15,-3] & 10\% & [108,112] & [82,101] & 5x5x5 \\ 
     ZPE & [335,360] & [0,15]   & 10\% & [43,47] & [58,89] & 6x6x6 \\[0.1cm]
     \hline
     \end{tabular}
    \caption{\label{tab:CH_parameters} Parameters considered for the Convex Hull identification of Taurid orbits. Columns 2 to 4 represent the limiting range of sun-centered ecliptic longitude ($\lambda-\lambda_\odot$), latitude ($\beta$) and geocentric velocity ($V_g$). Column 5 indicates the epochs used to estimate the average sporadic background density (i.e., when the shower is not active). Column 6 represent the effective shower duration estimated from the Convex Hull results. The number of voxels used to divide the velocity vectors space is indicated in column 7.}
    \end{table}
    
 \noindent\textbf{NTA} \\[-0.1cm]
  
\setfignumprefix{C2.2.}
\begin{figure}[!ht]
    \centering
    \includegraphics[width=.32\textwidth]{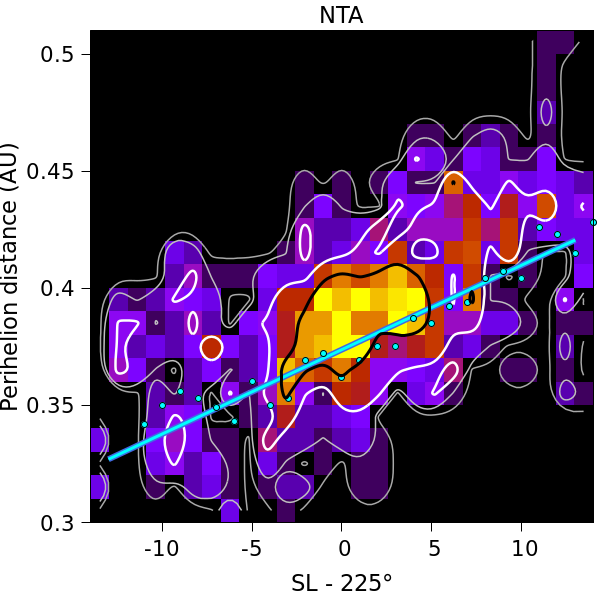}
    \includegraphics[width=.32\textwidth]{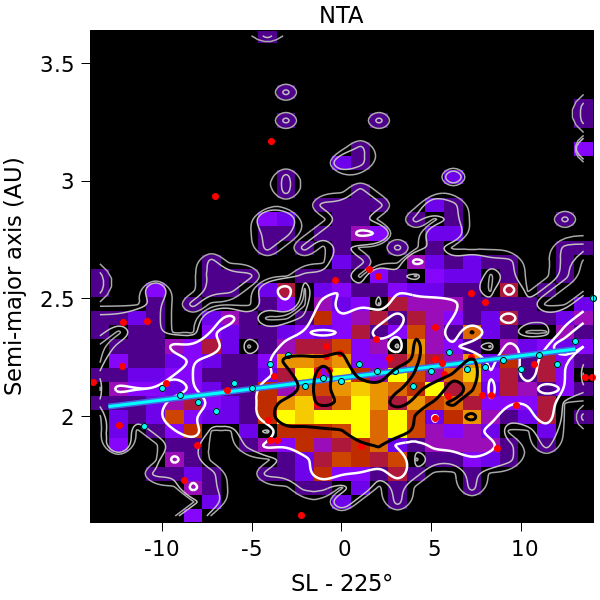}
    \includegraphics[width=.32\textwidth]{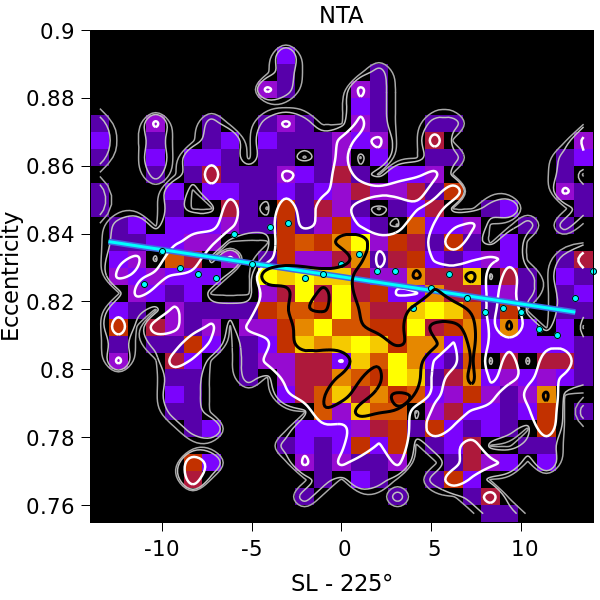}\\
    \includegraphics[width=.32\textwidth]{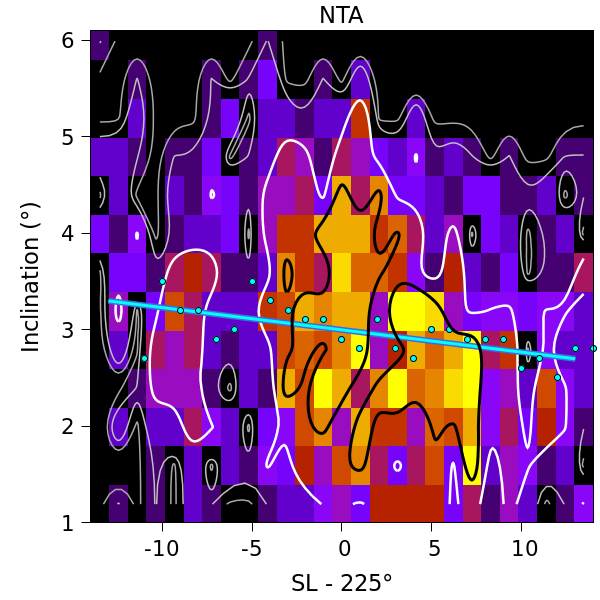}
    \includegraphics[width=.32\textwidth]{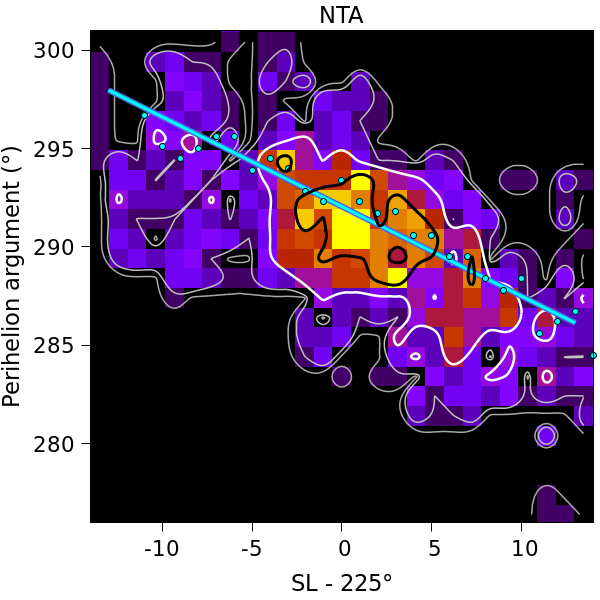}
    \includegraphics[width=.32\textwidth]{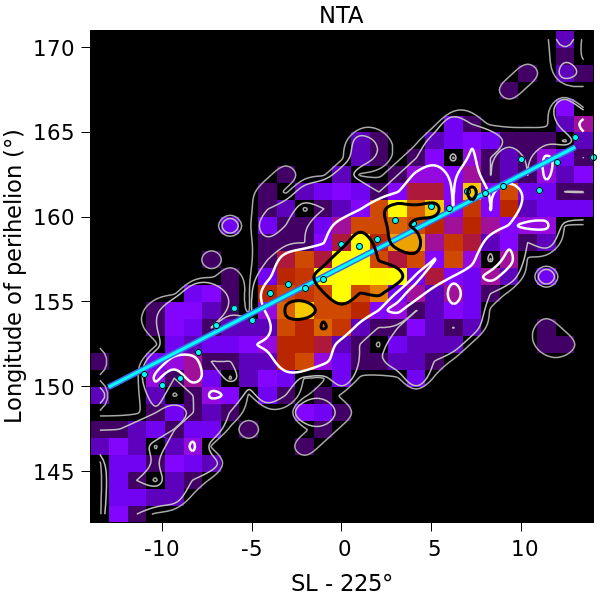}\\
    \includegraphics[width=.32\textwidth]{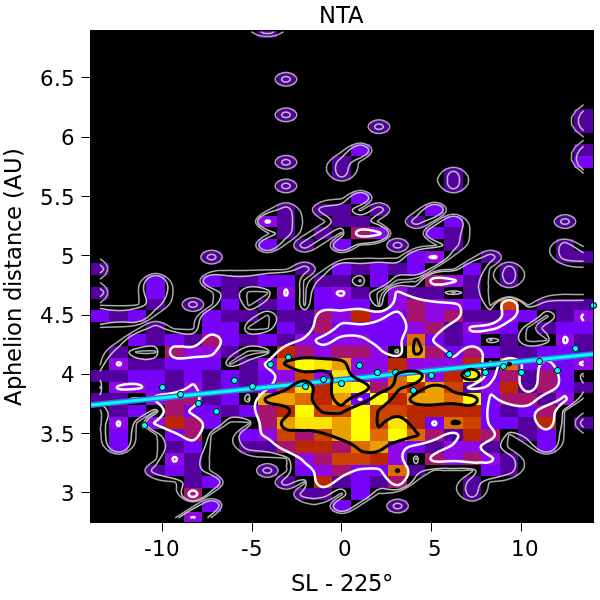}
    \includegraphics[width=.32\textwidth]{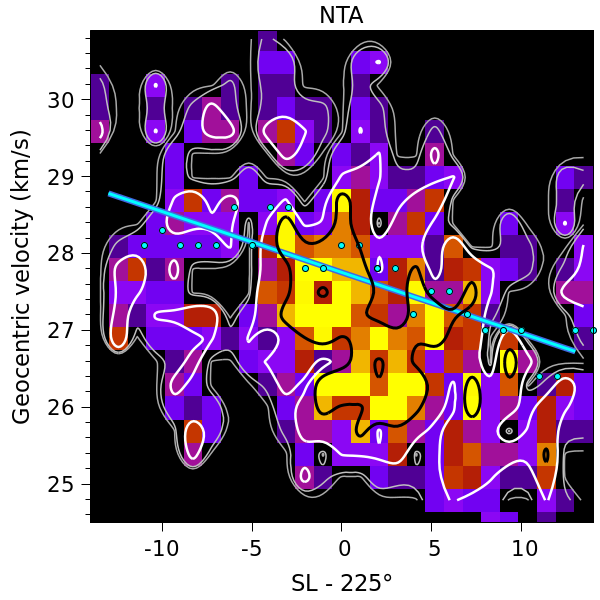} \hspace{0.2cm}
    \includegraphics[height=5.5cm]{colorbar.png}
    \caption{Evolution of the NTA perihelion distance, semi-major axis, eccentricity, inclination, perihelion argument, longitude of perihelion, aphelion distance and geocentric velocity with the solar longitude centered on the reference $L_m$=225$\degree$. Coloured areas represent the density of CMOR meteors extracted with the Convex Hull approach, from low (black) to high (yellow). Blue dots correspond to the result of the wavelet analysis. The solid blue line illustrate the best fit of the wavelet measurements using the linear or quadratic function presented in section C1. }
    \label{fig:CMOR_NTA_map}
\end{figure}

\newpage
\noindent\textbf{STA} \\[-0.1cm]

\begin{figure}[!ht]
    \centering
    \includegraphics[width=.32\textwidth]{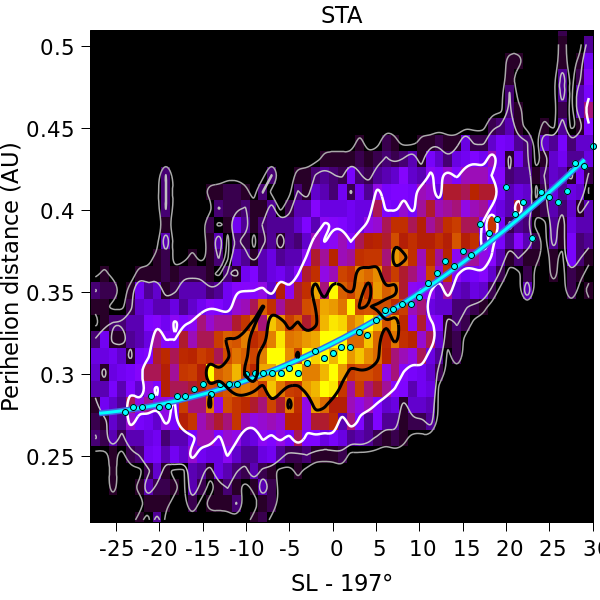}
    \includegraphics[width=.32\textwidth]{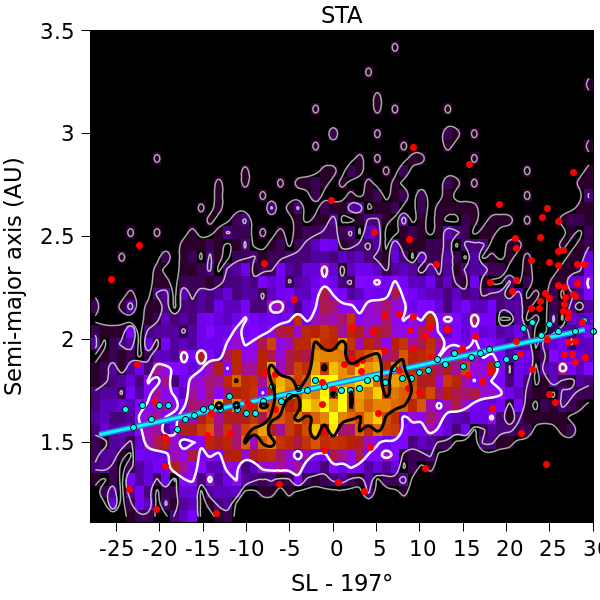}
    \includegraphics[width=.32\textwidth]{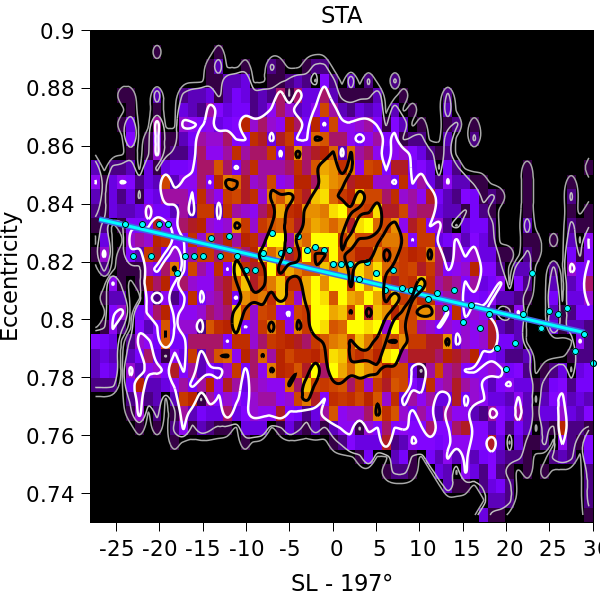}\\
    \includegraphics[width=.32\textwidth]{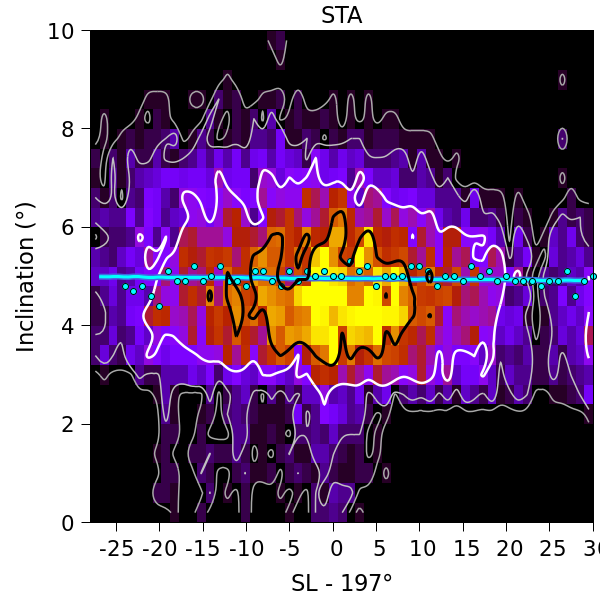}
    \includegraphics[width=.32\textwidth]{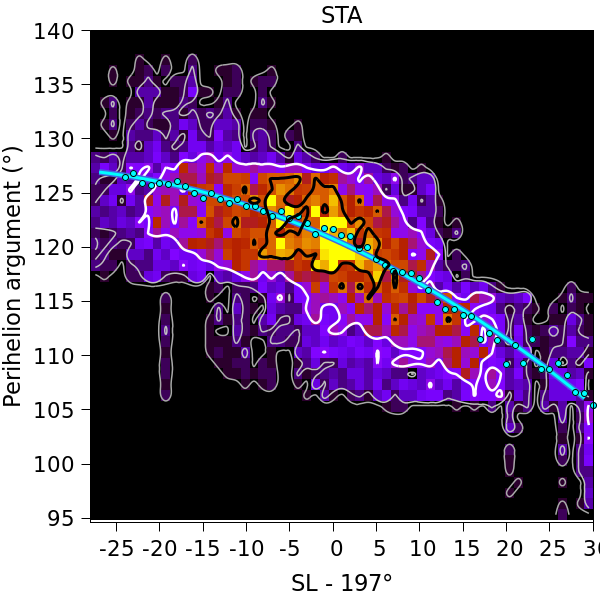}
    \includegraphics[width=.32\textwidth]{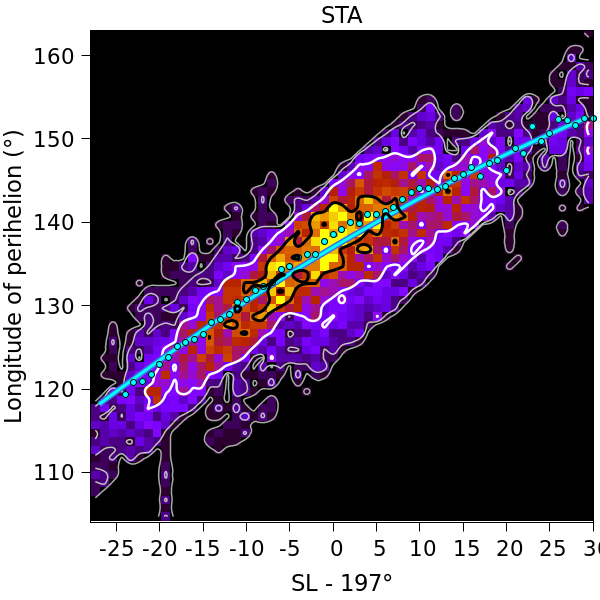}\\
    \includegraphics[width=.32\textwidth]{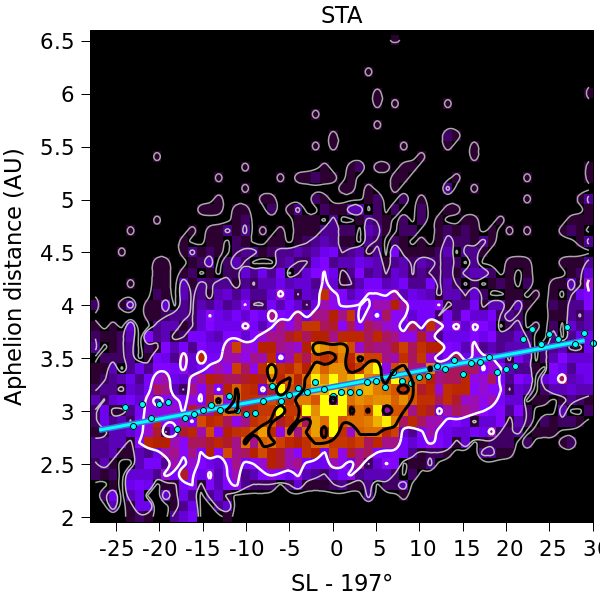}
    \includegraphics[width=.32\textwidth]{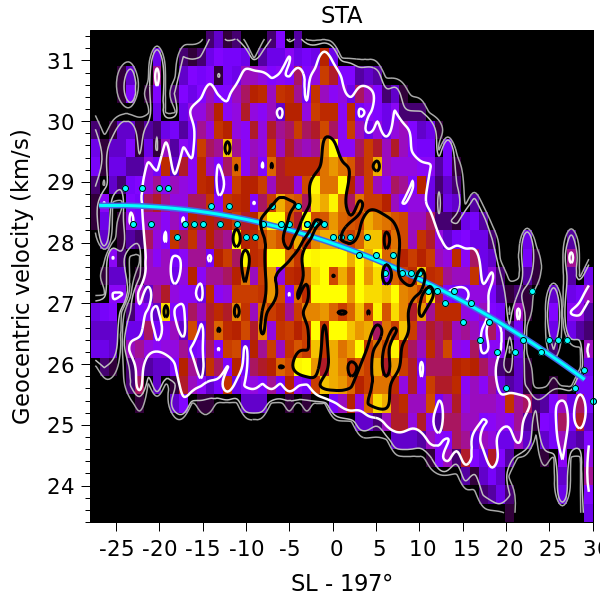} \hspace{0.2cm}
    \includegraphics[height=5.5cm]{colorbar.png}
    \caption{Evolution of the STA perihelion distance, semi-major axis, eccentricity, inclination, perihelion argument, longitude of perihelion, aphelion distance and geocentric velocity with the solar longitude centered on the reference $L_m$=197$\degree$. Coloured areas represent the density of CMOR meteors extracted with the Convex Hull approach, from low (black) to high (yellow). Blue dots correspond to the result of the wavelet analysis. The solid blue line illustrate the best fit of the wavelet measurements using the linear or quadratic function presented in section C1. }
    \label{fig:CMOR_STA_map}
\end{figure}

\newpage
\noindent\textbf{BTA} \\[-0.1cm]

\begin{figure}[!ht]
    \centering
    \includegraphics[width=.32\textwidth]{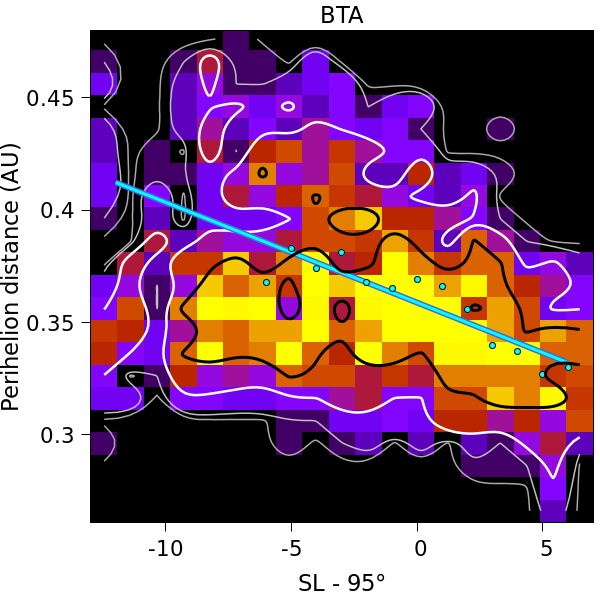}
    \includegraphics[width=.32\textwidth]{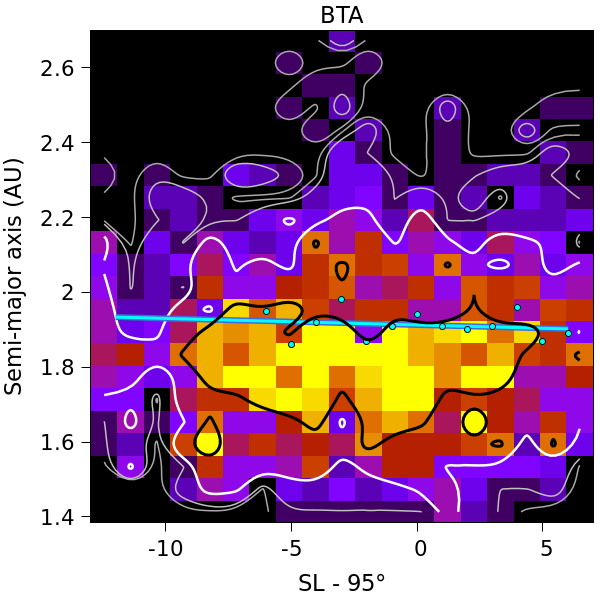}
    \includegraphics[width=.32\textwidth]{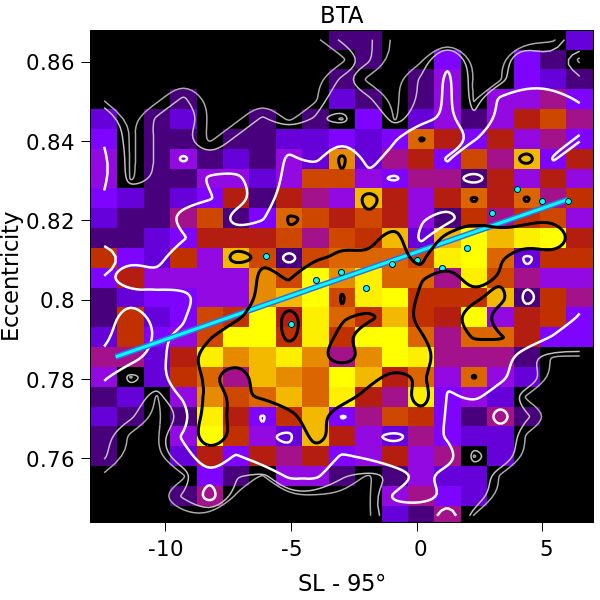}\\
    \includegraphics[width=.32\textwidth]{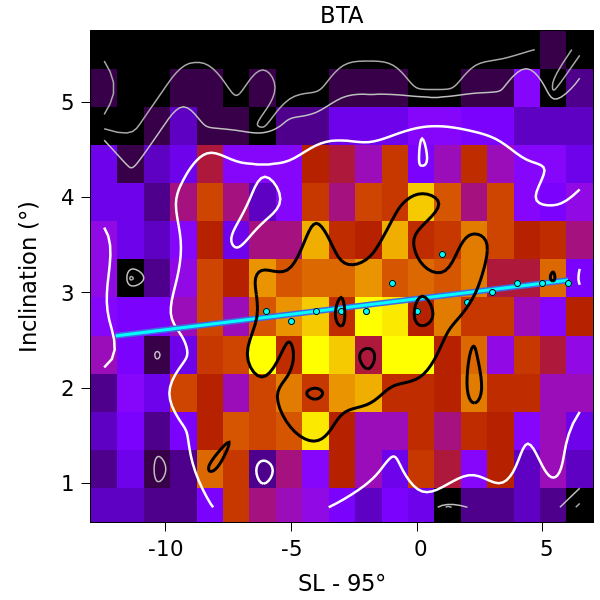}
    \includegraphics[width=.32\textwidth]{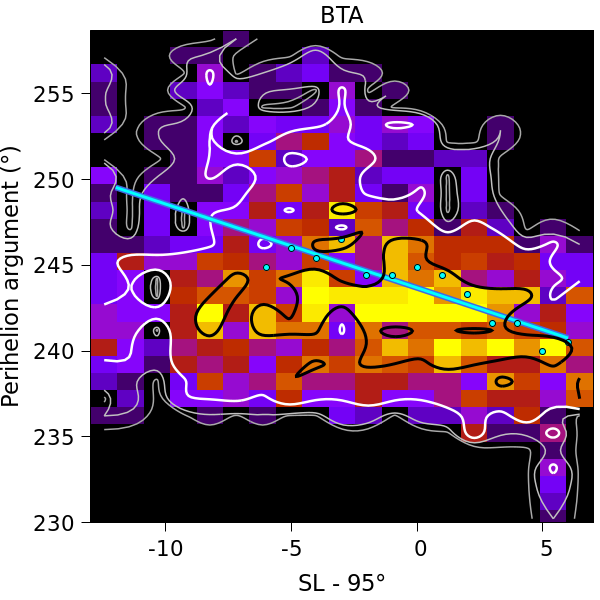}
    \includegraphics[width=.32\textwidth]{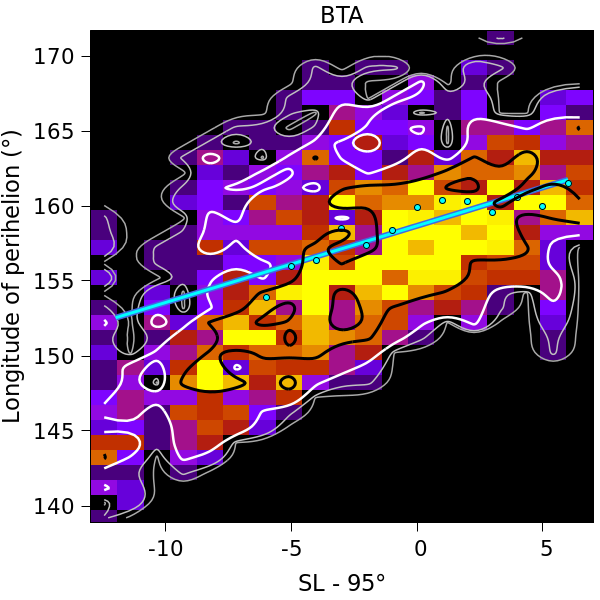}\\
    \includegraphics[width=.32\textwidth]{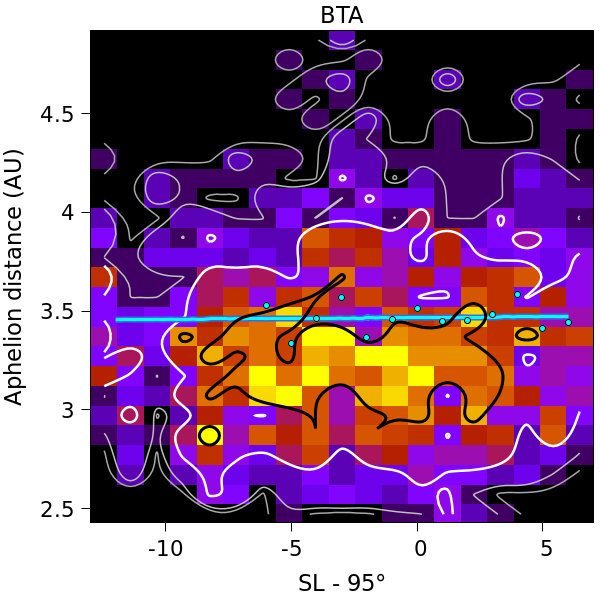}    \includegraphics[width=.32\textwidth]{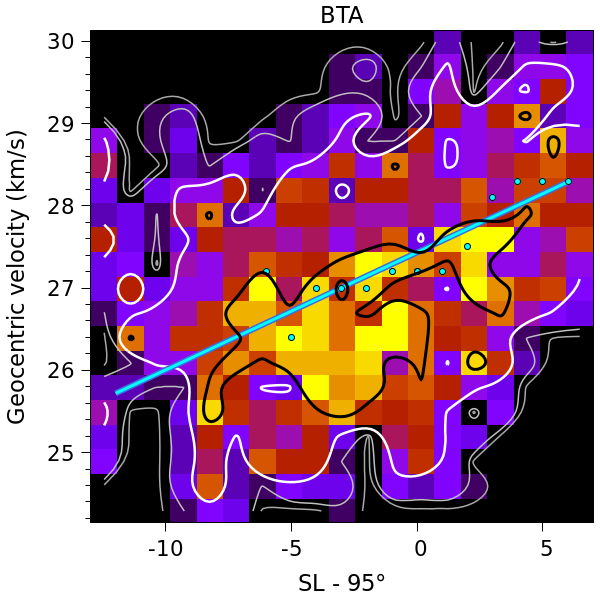} \hspace{0.2cm}
    \includegraphics[height=5.5cm]{colorbar.png}
    \caption{Evolution of the BTA perihelion distance, semi-major axis, eccentricity, inclination, perihelion argument, longitude of perihelion, aphelion distance and geocentric velocity with the solar longitude centered on the reference $L_m$=95$\degree$. Coloured areas represent the density of CMOR meteors extracted with the Convex Hull approach, from low (black) to high (yellow). Blue dots correspond to the result of the wavelet analysis. The solid blue line illustrate the best fit of the wavelet measurements using the linear or quadratic function presented in section C1.}
    \label{fig:CMOR_BTA_map}
\end{figure}

\newpage
\noindent\textbf{ZPE} \\[-0.1cm]

\begin{figure}[!ht]
    \centering
    \includegraphics[width=.32\textwidth]{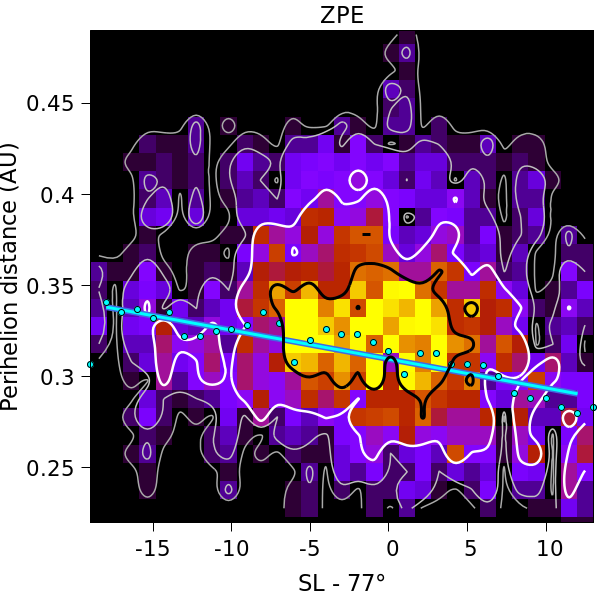}
    \includegraphics[width=.32\textwidth]{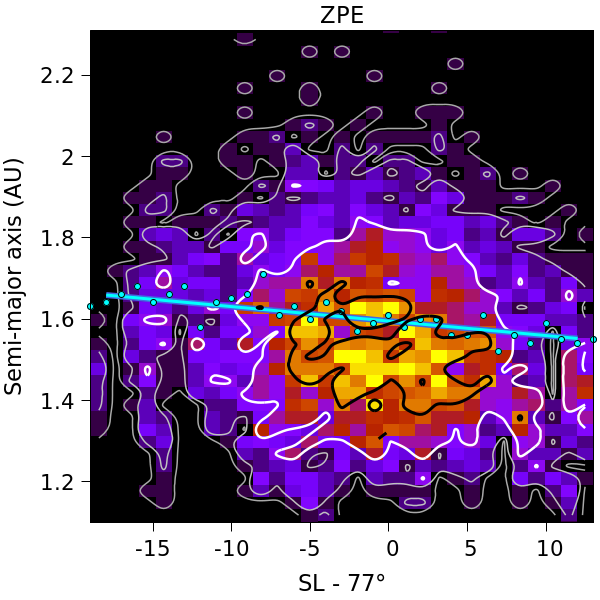}
    \includegraphics[width=.32\textwidth]{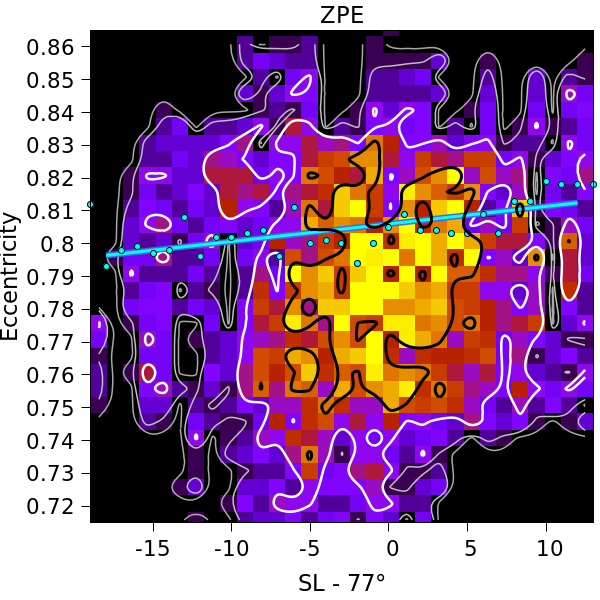}\\
    \includegraphics[width=.32\textwidth]{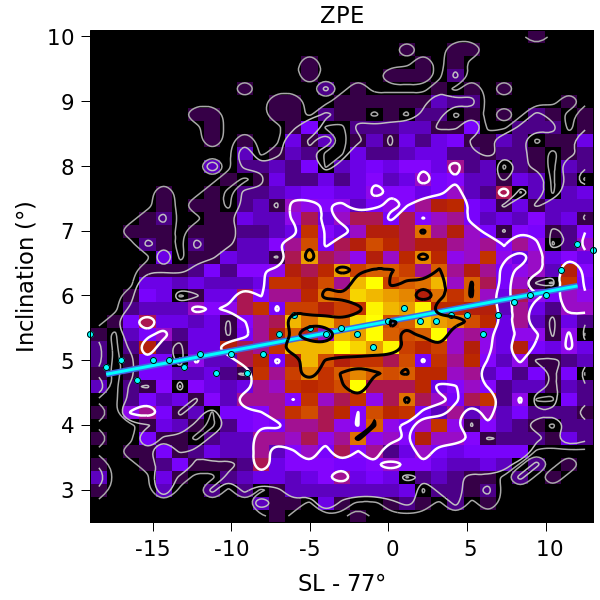}
    \includegraphics[width=.32\textwidth]{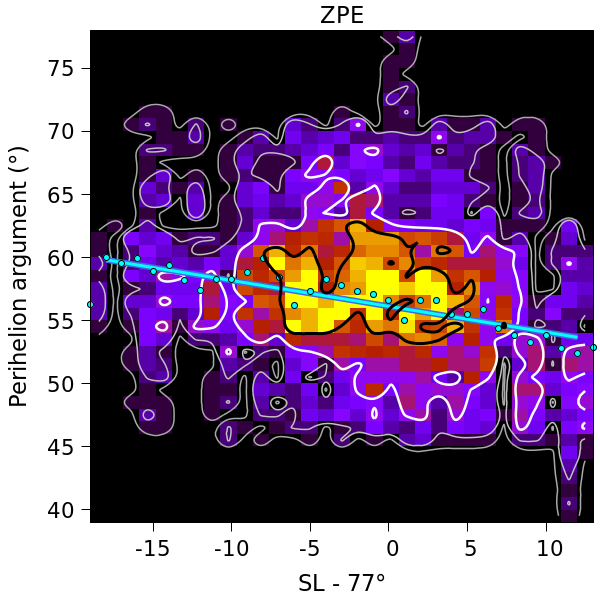}
    \includegraphics[width=.32\textwidth]{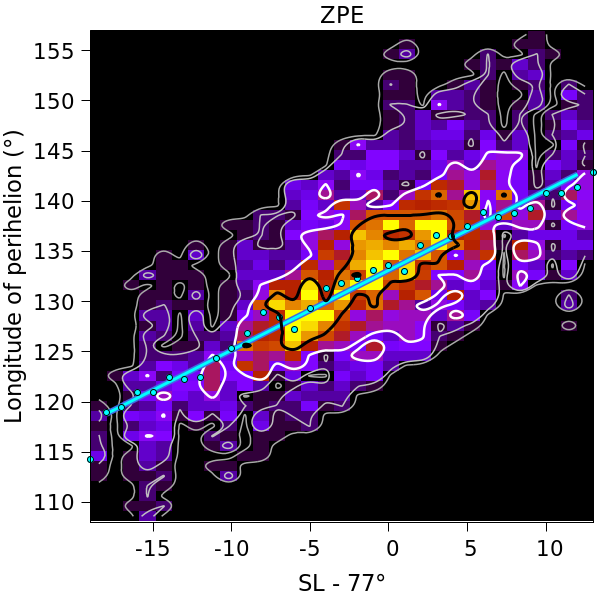}\\
    \includegraphics[width=.32\textwidth]{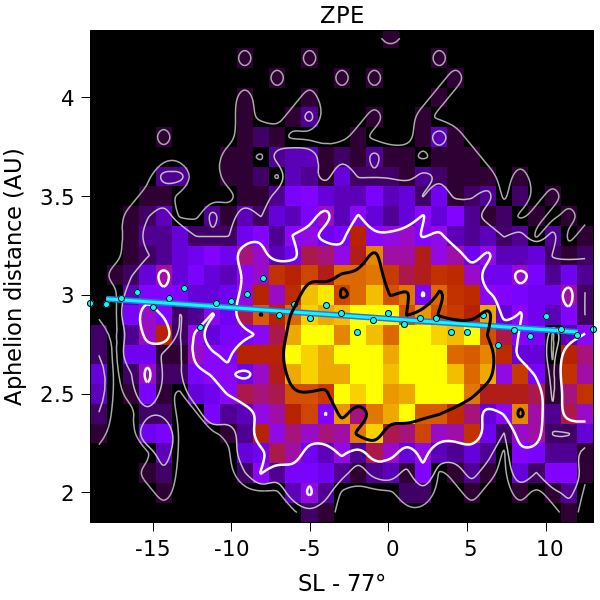}
    \includegraphics[width=.32\textwidth]{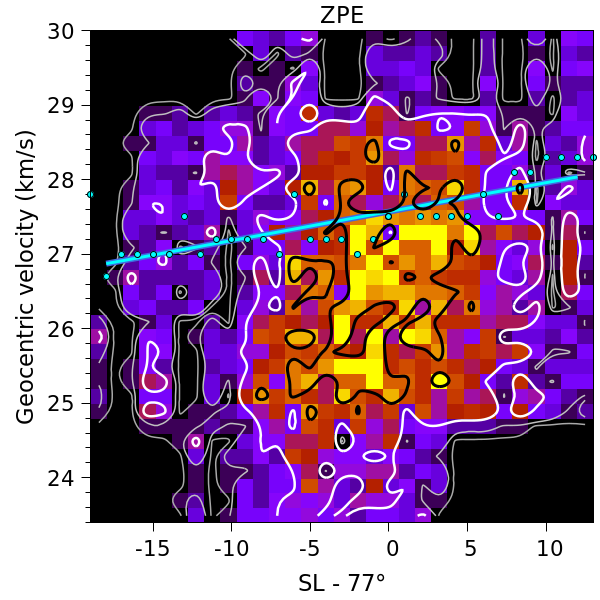} \hspace{0.2cm}
    \includegraphics[height=5.5cm]{colorbar.png}
    \caption{Evolution of the ZPE perihelion distance, semi-major axis, eccentricity, inclination, perihelion argument, longitude of perihelion, aphelion distance and geocentric velocity with the solar longitude centered on the reference $L_m$=77$\degree$. Coloured areas represent the density of CMOR meteors extracted with the Convex Hull approach, from low (black) to high (yellow). Blue dots correspond to the result of the wavelet analysis. The solid blue line illustrate the best fit of the wavelet measurements using the linear or quadratic function presented in section C1. }
    \label{fig:CMOR_ZPE_map}
\end{figure}

\clearpage
\newpage
\textbf{C3 Orbital elements correlations} \\

\textbf{C3.1 Orbital variation of $a$, $\Omega$} \\

\setfignumprefix{C3.}
\begin{figure}[!ht]
\centering
	\includegraphics[trim={0.7cm 0 3.45cm 0},clip,height=0.31\textwidth]{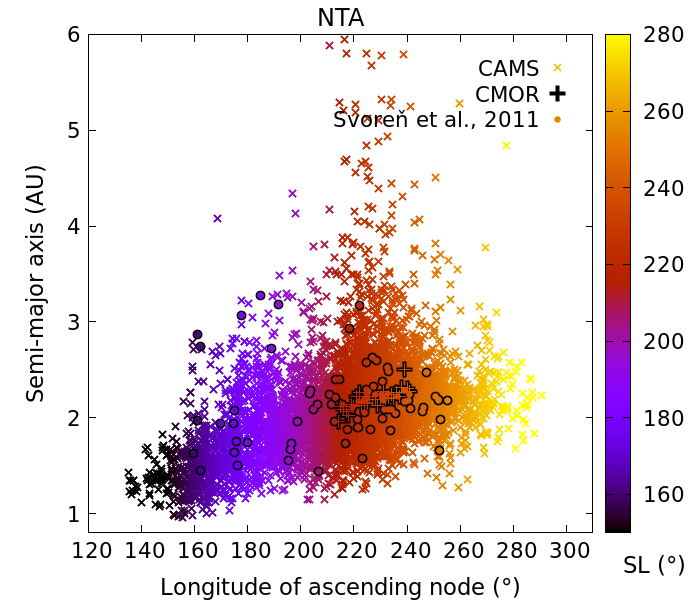}\hspace{-0.1cm}
	\includegraphics[trim={1.7cm 0 0 0},clip,height=0.31\textwidth]{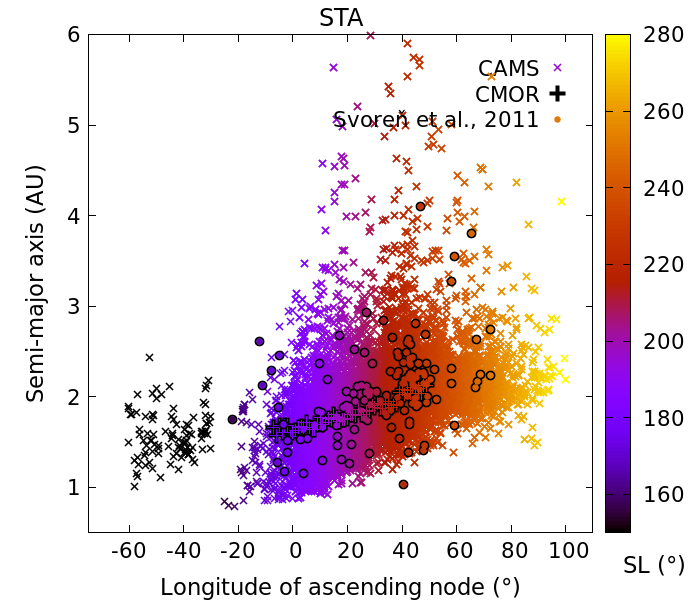}\hspace{-0.1cm}
	\includegraphics[trim={0 0 0.5cm 0},clip,height=0.31\textwidth]{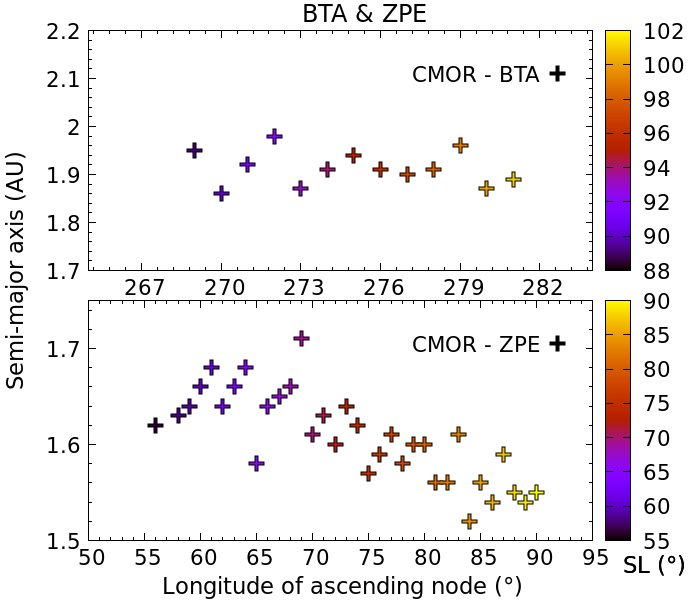}	
	\caption{\label{fig:aW} Semi-major axis and longitude of ascending node of the NTA (left), STA (middle), BTA and ZPE (right) as measured by CAMS, CMOR (wavelet analysis) or selected by \protect\cite{Svoren2011} from the IAU MDC database. }
\end{figure}

\textbf{C3.2 Orbital variation of $q$, $i$} \\

\begin{figure}[!ht]
\centering
	\includegraphics[trim={0.7cm 0 3.45cm 0},clip,height=0.31\textwidth]{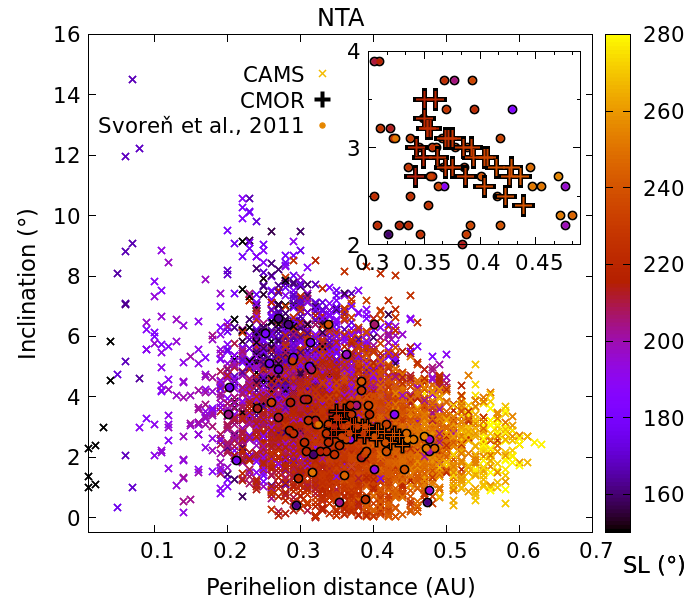}\hspace{-0.17cm}
	\includegraphics[trim={1.5cm 0 0 0},clip,height=0.31\textwidth]{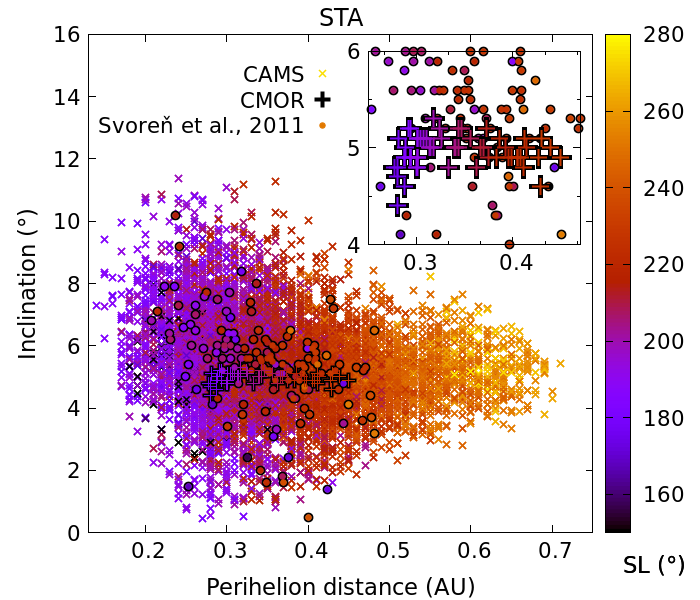}\hspace{-0.25cm}
	\includegraphics[trim={0 0 0.5cm 0},clip,height=0.31\textwidth]{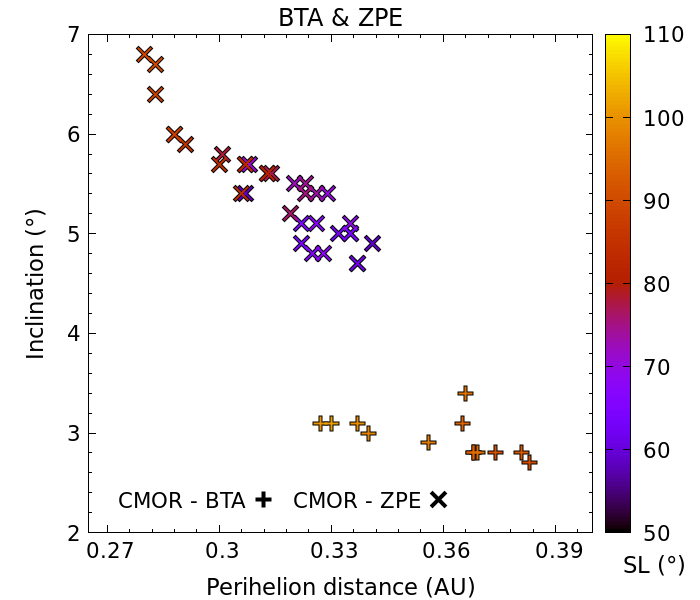}
	\caption{\label{fig:qi} Perihelion distance and inclination of the NTA (left), STA (middle), BTA and ZPE (right) as measured by CAMS, CMOR (wavelet analysis) or selected by \protect\cite{Svoren2011} from the IAU MDC database. }
\end{figure}

\textbf{C3.3 Orbital variation of $a$, $e$ in CMOR data} \\

The following figures compare the correlation of $a$ vs $e$ measured for CMOR individual orbits extracted with the Convex Hull method (see C2.2) with the wavelet analysis results. \\

\begin{figure}[!ht]
    \centering
    \includegraphics[width=.49\textwidth]{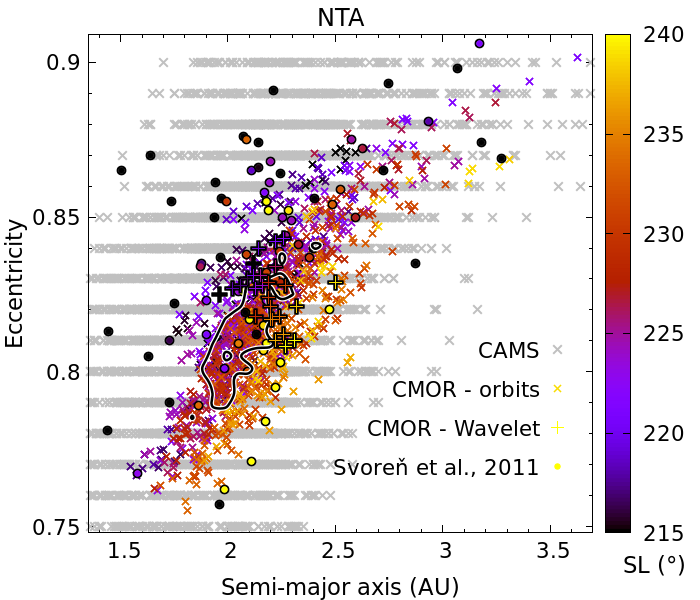}
    \includegraphics[width=.49\textwidth]{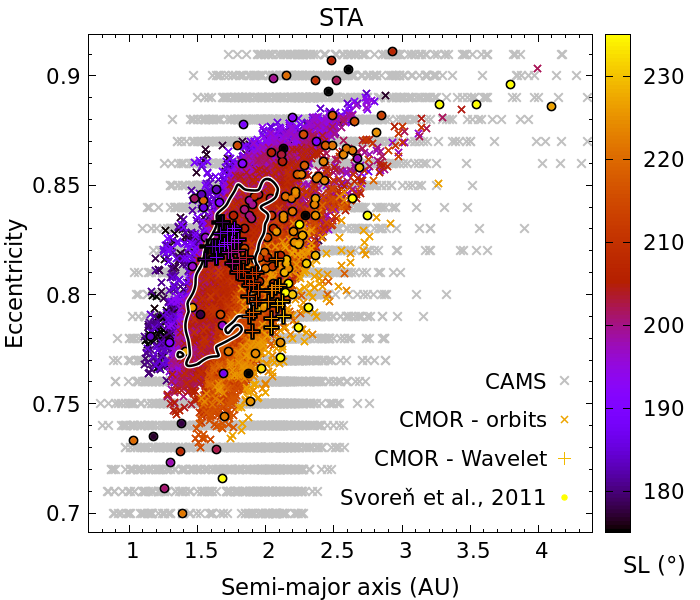} \\
    \includegraphics[width=.49\textwidth]{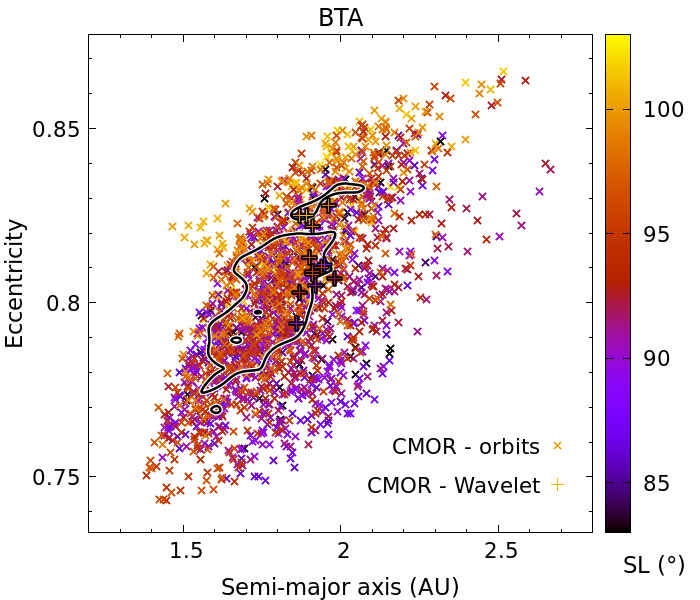}
    \includegraphics[width=.49\textwidth]{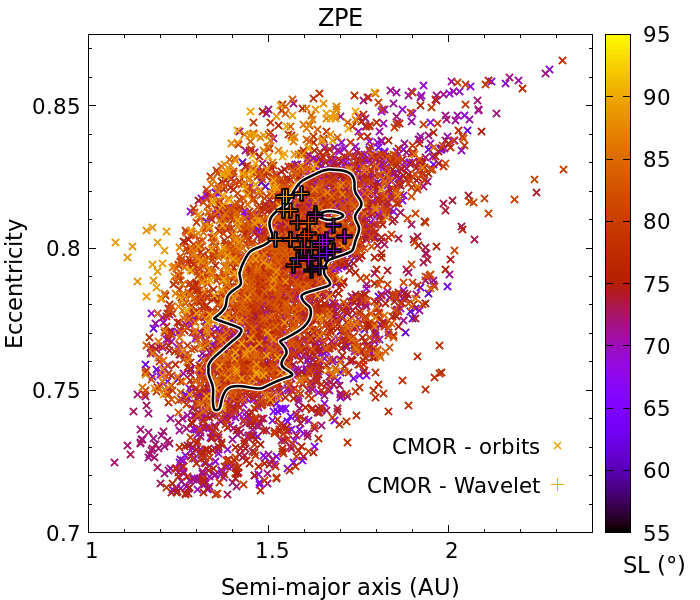}
    \caption{Eccentricity and semi-major axis of the NTA, STA, BTA and ZPE as measured by CMOR. Measurements are colour-coded by solar longitude. Individual radar orbits are represented by the oblique crosses (x) while the results of the wavelet analysis are plotted with vertical crosses (+). NTA and STA meteors recorded by CAMS are represented with grey points. The black contours encompass the regions of highest density of meteors extracted with the Convex Hull approach, corresponding to the red and yellow areas in the maps of section C2.2. }
    \label{fig:ae_CMOR}
\end{figure}

\newpage
\textbf{C3.4 Orbital variation of $i$, $\omega$} \\

\begin{figure}[!ht]
    \centering
    \includegraphics[width=.49\textwidth]{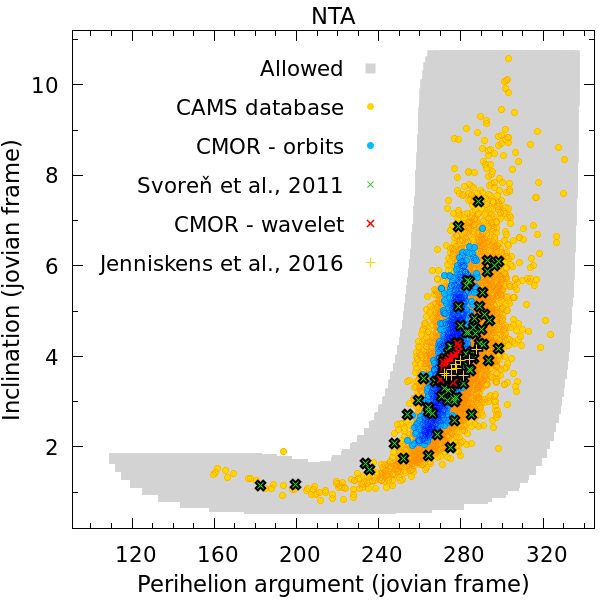}
    \includegraphics[width=.49\textwidth]{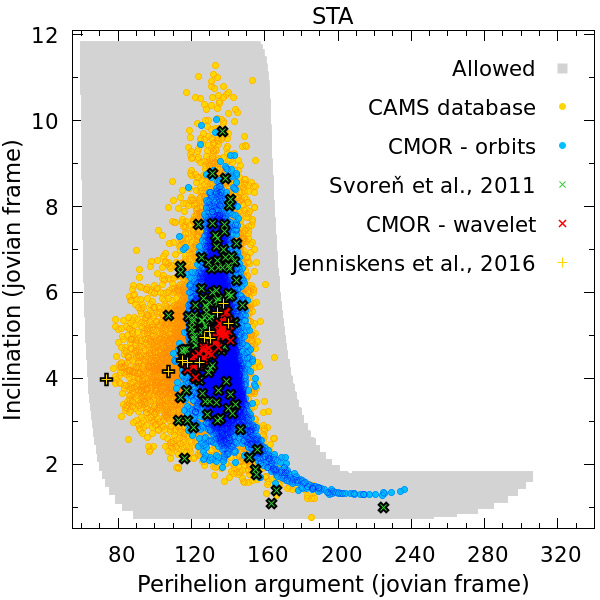}
    \includegraphics[width=.49\textwidth]{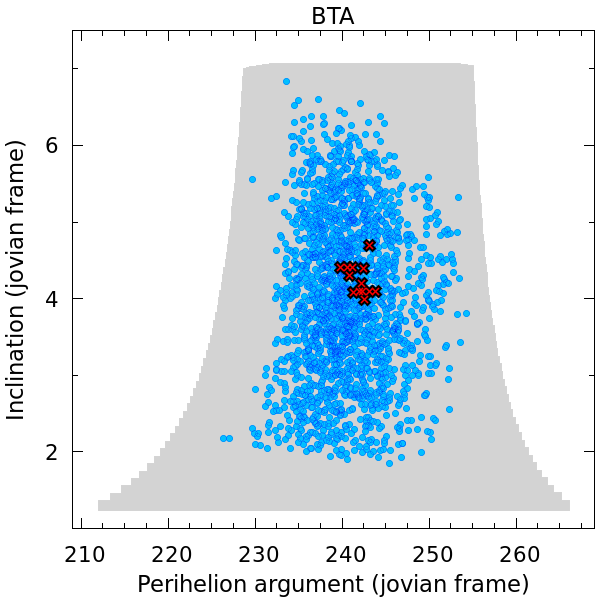}
    \includegraphics[width=.49\textwidth]{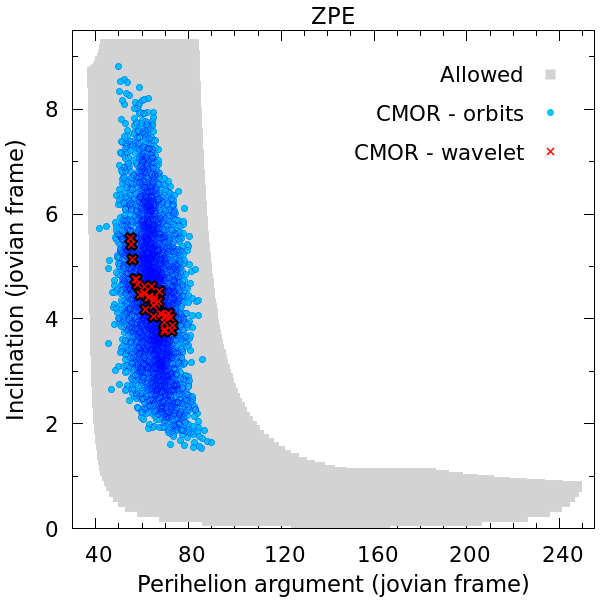}
    \caption{\label{fig:iw_jovian} Inclination and perihelion argument (reffered to Jupiter's orbital plane) of the observed NTA, STA, BTA and ZPE as measured by CAMS, CMOR (wavelet analysis) or selected by \protect\cite{Svoren2011} and \protect\cite{Jenniskens2016}. The grey dots represent synthetic meteoroids, selected to fill the observed orbital range of each shower while being detectable from Earth. The light blue dots represent individual orbits recorded by the CMOR radar, extracted using the Convex Hull method (BTA \& ZPE) -- see text for details.}
\end{figure}

In this section, we revisit the analysis of \cite{Steel1991} in order to analyze the $i-\omega$ evolution of each branch in the Jovian frame. The meteoroids orbital elements in the Jovian frame were obtained by rotating the meteoroids state vector $X$ (in ecliptic coordinates) into Jupiter's orbital plane ($X'$)  with the relation 
\begin{equation}
    X'=[R_x(i).R_z(\Omega)]X
\end{equation}
with $i=1.303\degree$ the inclination and $\Omega=100.464\degree$ the node longitude of Jupiter. Figure \ref{fig:iw_jovian} presents the obtained values of $i$ and $\omega$ (referred to Jupiter's orbit) computed from CMOR (wavelet: dark blue, individual orbits: cyan), CAMS (orange), and the selected meteors of \cite{Svoren2011} (red) and \cite{Jenniskens2016} (green).  

Following the methodology of \cite{Steel1991}, we also plot in Figure \ref{fig:iw_jovian} the range of $i-\omega$ distribution allowing a meteoroid to intersect Earth's orbit. We first generate a sample of synthetic orbits filling the observational phase of the four Taurid showers. Then, we verify that each orbit included has a Minimum Orbit Intersecting Distance (MOID) with Earth below 0.05 AU. The resulting orbits, converted into the Jovian frame, are represented by the grey circles in Figure \ref{fig:iw_jovian}. 
To obtain these "allowed" orbital regions, we varied the meteoroids inclination from 0$\degree$ to 11$\degree$ with a step of 0.2$\degree$ (for the NTA and STA) or 0.1$\degree$ (for the daytime showers). The perihelion argument and longitude of the ascending node varied by 5$\degree$ and 3$\degree$ respectively, over the ranges: \\[0.2cm]

\begin{itemize}
    \item $\omega \in [270,340]$, $\Omega \in [130,250]$ for the NTA
    \item $\omega \in [60,155]$, $\Omega \in [-60,60]$ for the STA
    \item $\omega \in [233,252]$, $\Omega \in [260,295]$ for the BTA
    \item $\omega \in [38,76]$, $\Omega \in [50,95]$ for the ZPE
\end{itemize}
The semi-major axis ($a$) and the eccentricity ($e$) of the meteoroids, that do not contribute significantly to the ($i-\omega$) distribution, were taken to vary from 1 AU to 4 AU and 0.7 to 1.0 with a step of 0.5 AU and 0.2 respectively.  

\newpage
\clearpage
\bibliographystyle{mnras}
\bibliography{references} 